%% file: main.tex
\documentclass[lettersize,journal]{IEEEtran}
\usepackage{amsmath,amsfonts}
\usepackage{algorithmic}
\usepackage{array}
\usepackage[normalem]{ulem}
\usepackage{textcomp}
\usepackage{stfloats}
\usepackage{url}
\usepackage{verbatim}
\usepackage{amsmath,bm}
\usepackage{mathrsfs}
\usepackage{amssymb}
\usepackage{booktabs}
\usepackage{svg}
\usepackage{siunitx}
\usepackage[ruled,lined]{algorithm2e}
\usepackage{makecell}
\usepackage{multirow}
\usepackage{arydshln}
\usepackage{stackengine}
\usepackage{graphicx}
\usepackage{xspace}
\usepackage[export]{adjustbox}
\usepackage{subcaption}
\usepackage{titling}

\newbox{\bigpicturebox}

\makeatletter
\DeclareRobustCommand\onedot{\futurelet\@let@token\@onedot}
\def\@onedot{\ifx\@let@token.\else.\null\fi\xspace}

\def\eg{\emph{e.g}\onedot} 
\def\ie{\emph{i.e}\onedot}

\def\etal{\emph{et al}\onedot}
\makeatother

\usepackage[pagebackref,breaklinks]{hyperref}
\usepackage[capitalize]{cleveref}
\crefname{section}{Sec.}{Secs.}
\crefname{section}{Section}{Sections}
\crefname{algorithm}{Alg.}{Algs.}

\usepackage[switch]{lineno}

\def\BibTeX{{\rm B\kern-.05em{\sc i\kern-.025em b}\kern-.08em
    T\kern-.1667em\lower.7ex\hbox{E}\kern-.125emX}}
\usepackage{balance}

\makeatletter
\def\endthebibliography{%
	\def\@noitemerr{\@latex@warning{Empty `thebibliography' environment}}%
	\endlist
}
\makeatother

\begin{document}
\title{Towards Robust and Generalizable Lensless Imaging with Modular Learned Reconstruction}
\author{Eric Bezzam, Yohann Perron, and Martin Vetterli}
\date{}

\markboth{}
{Bezzam
\MakeLowercase{\textit{(et~al.)}: 
Towards Robust and Generalizable Lensless Imaging with Modular Learned Reconstruction}}

\maketitle


\begin{abstract}
Lensless cameras disregard the conventional design that imaging should mimic the human eye. This is done by replacing the lens with a thin mask, and moving image formation to the digital post-processing. State-of-the-art lensless imaging techniques use learned approaches that combine physical modeling and neural networks. However, these approaches make simplifying modeling assumptions for ease of calibration and computation. Moreover, the generalizability of learned approaches to lensless measurements of new masks has not been studied. To this end, we utilize a modular learned reconstruction in which a key component is a pre-processor prior to image recovery. We theoretically demonstrate the pre-processor's necessity for standard image recovery techniques (Wiener filtering and iterative algorithms), and through extensive experiments show its effectiveness for multiple lensless imaging approaches and across datasets of different mask types (amplitude and phase). We also perform the first generalization benchmark across mask types to evaluate how well reconstructions trained with one system generalize to others. Our modular reconstruction enables us to use pre-trained components and transfer learning on new systems to cut down weeks of tedious measurements and training. As part of our work, we open-source four datasets, and software for measuring datasets and for training our modular reconstruction.
\end{abstract}


\input{sec/1_introduction}  
\input{sec/2_related}
\input{sec/3_proposed}
\input{sec/4_experiments}

\input{sec/6_conclusion}

\section*{Acknowledgment}

The authors would like to thank Jonathan Dong for discussions and helpful feedback.

\bibliographystyle{IEEEtran}
\bibliography{main}

\input{sec/supplementary}

\end{document}

%% file: sec/1_introduction.tex
\section{Introduction}
\label{sec:introduction}

\IEEEPARstart{L}{ensless} imaging has emerged as a promising alternative to traditional optical systems, circumventing the rigid requirements of lens-based designs.
By substituting a lens with a thin modulating mask, an imaging system can achieve compactness, lower cost, and enhanced visual privacy~\cite{boominathan2022recent}.
Conventional imaging relies on lenses to establish a direct one-to-one mapping between scene points and sensor pixels. 
In contrast, lensless imaging employs an optical element to create a one-to-many encoding by modulating the phase~\cite{Antipa:18,Monakhova:19,phlatcam,9239993,Lee:23} and/or amplitude~\cite{flatcam,wu2020single} of incident light.
With a sufficient understanding of these one-to-many mappings, \ie the point spread functions (PSFs), 
computational algorithms can be used to reconstruct viewable images from these multiplexed measurements.

Despite advancements that combine physical modeling with deep learning~\cite{Monakhova:19,9239993},
lensless imaging systems face challenges in robustness and generalizability. 
Robustness issues arise from approximations when modeling the imaging system, such as the linear shift-invariance (LSI) assumption, which can degrade reconstruction quality due to model mismatch~\cite{9546648}.
Additionally, learned reconstruction approaches are often trained under a specific signal-to-noise ratio (SNR), 
resulting in performance degradation to SNR changes at inference~\cite{Rego2021,Perron2023}.
With regards to generalizability, few studies have evaluated the transferability of learned reconstructions to different masks or PSFs from those used during training. 
While slight manufacturing variations have been explored~\cite{Lee:23},
generalizability to significant PSF changes remains underexplored. 
Rego \etal~\cite{Rego2021} train models on measurements from multiple PSFs but only evaluate on simulations/measurements with the same PSFs seen at training,
leaving the generalizability to unseen PSFs unclear.
As current methods rely on supervised training with paired lensless-lensed datasets for a given PSF,
the scalability of high-quality lensless imaging is limited.
Robustness to different settings and generalizability to PSF changes would make lensless imaging system much more practical, \ie cutting down weeks of measurement/training time and enabling improved reconstruction when data collection is difficult or impossible, \eg \textit{in-vivo} or due to privacy constraints.

\begin{figure}[t!]
    \centering
\includegraphics[width=\linewidth]{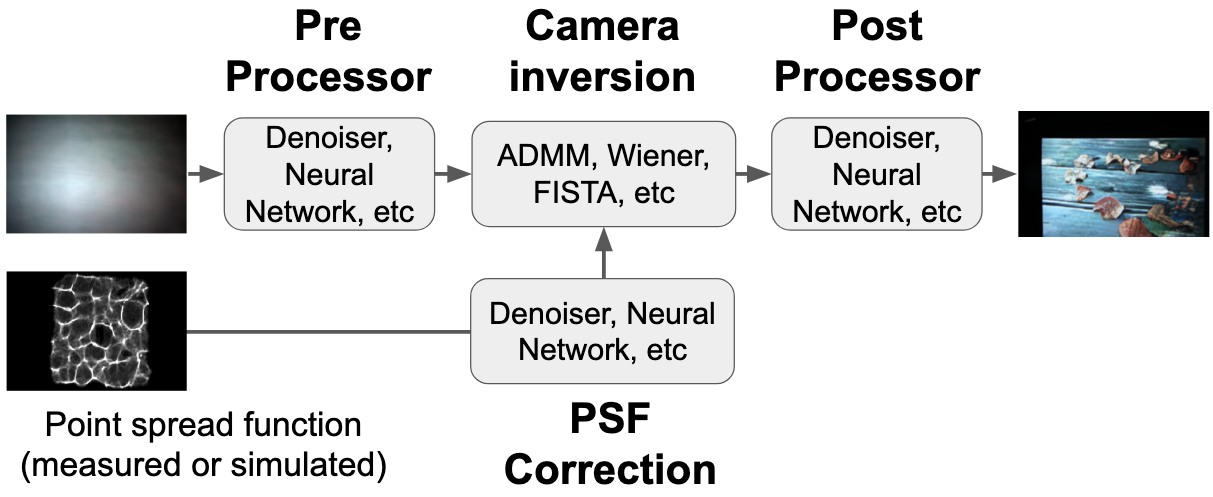}
\caption{Modular lensless imaging pipeline. Pre- and post-processors and PSF correction are optional.}
  \label{fig:pipeline}
\end{figure}

This work addresses these gaps by advancing the robustness and generalizability of lensless imaging recovery. Our contributions include:
\begin{itemize}
    \item \textit{Versatile Modular Reconstruction}: We propose and apply a modular reconstruction framework, as shown in \cref{fig:pipeline} to multiple imaging systems and previously-proposed camera inversion approaches. The framework extends our previous work that introduced a pre-processor~\cite{Perron2023}.
\item \textit{Robustness Analysis and Experiments:} We motivate the pre-processor and our modular framework by showing how camera inversion methods amplify input noise and introduce error terms due to inevitable model mismatch in lensless imaging. With our modular approach, we experimentally show improved robustness to varying input SNR and model mismatch.
    \item \textit{Benchmarking and Improving Generalizability}: We conduct the first benchmark across multiple mask patterns and types, assessing how well reconstruction approaches trained on one system generalize to others. With our modular reconstruction, we explore techniques to improve
    generalization to unseen PSFs.
    \item \textit{Hardware Prototype}: We introduce \textit{DigiCam}, a programmable-mask system that is $30\times$ cheaper than existing alternatives, and enables convenient evaluation across multiple masks/PSFs.
\end{itemize}
For reproducibility and to encourage further research, we open-source:
\begin{itemize}
    \item \textit{Datasets}: Four public datasets, including the first multi-mask dataset with 100 unique masks and 250 measurements per mask~\cite{tapecam,digicam_celeba,digicam_single,digicam_multi}.
    \item \textit{Code}: Reconstruction and training implementations, including that of baseline algorithms.
    \item \textit{Tooling}: Scripts for dataset collection using the Raspberry Pi HQ sensor~\cite{rpi_hq} and tools for uploading datasets to Hugging Face.
\end{itemize}
All resources are integrated into a documented toolkit for lensless imaging hardware and software~\cite{Bezzam2023}.\footnote{\href{https://lensless.readthedocs.io}{lensless.readthedocs.io}}

%% file: sec/2_related.tex
\section{Related Work}

\noindent In this section, we give an overview of lensless cameras, image recovery techniques, and previous work that addresses robustness and generalizability in lensless imaging.

\subsection{Lensless Cameras}

\noindent The earliest cameras, such as the \textit{camera obscura} and the pinhole camera, were inherently lensless, though they required long exposure times due to their limited light throughput. 
The introduction of lenses with larger apertures resolved this limitation, by allowing shorter exposures while producing sharp, in-focus images.
Mask-based lensless imaging found its first notable applications beyond the visible spectrum, namely in astronomy, where X-rays and gamma rays cannot be easily focused with conventional lenses or mirrors. Instead, increasing the number of apertures enabled better signal collection and imaging capabilities~\cite{dicke1968scatter,caroli1987coded}.
The commoditization of digital sensors paved the way for lensless imaging in the visible spectrum.
Camera miniaturization and advancements in compressive sensing enabled the shift of image formation from traditional optics to digital post-processing. 
Ultra-compact lensless imaging systems can be fabricated to sub-mm thickness using scalable lithography techniques~\cite{Lee:23,flatcam}, 
while the multiplexing property of lensless cameras allows higher-dimensional quantities to be recovered from 2D measurements: refocusable/3D imaging~\cite{Antipa:18,sweepcam2020,zheng2021programmable3dcam}, hyperspectral~\cite{Monakhova:20}, and videos~\cite{antipa2019video}.
The compact design can also enable \textit{in-vivo} imaging of hard-to-reach areas in biological systems, as demonstrated in calcium imaging of live mouse cortices~\cite{adams2022vivo}.

Lensless cameras replace traditional lenses with masks that modulate the phase~\cite{Antipa:18,Monakhova:19,phlatcam,9239993,Lee:23} and/or amplitude~\cite{flatcam,wu2020single} of incident light.
Off-the-shelf materials such as diffusers~\cite{Antipa:18} or even double-sided tape~\cite{Bezzam2023,diffusercam_tut} can be used as a static mask, or it can be fabricated with photolithography for a desired structure/PSF~\cite{phlatcam,Lee:23,flatcam,wu2020single}.
For reconfigurable systems, spatial light modulators (SLMs)~\cite{sweepcam2020,zheng2021programmable3dcam,10.1117/1.OE.54.2.023102} or liquid crystal displays (LCDs)~\cite{4472247,huang2013,zomet2006}  can be used as a programmable mask.
If design constraints permit, phase masks are preferred for superior light efficiency and concentration, as this leads to higher-quality reconstructions~\cite{boominathan2022recent}.

\subsection{Lensless Image Recovery}

\noindent Lensless image recovery is inherently an ill-posed inverse problem due to the multiplexing nature of such cameras.
To solve this, an optimization framework is typically employed, consisting of: (1) a data fidelity term that ensures consistency with the measurements through a forward model, and (2) regularization term(s) that incorporate prior knowledge about the desired image.
In certain cases, such as with $\ell_2$ regularization~\cite{flatcam} or Wiener filtering~\cite{Li:23}, 
the problem can be solved in closed-form, offering computational efficiency. 
More expressive priors, like non-negativity constraints or total variation (TV) minimization~\cite{Antipa:18,phlatcam} require iterative solvers, such as the fast iterative shrinkage-thresholding algorithm (FISTA)~\cite{beck2009fast} or the alternating direction method of multipliers (ADMM)~\cite{ADMM}.
While such solvers are slower due to multiple iterations needed for convergence,
they generally perform much better than closed-form approaches.

Incorporating deep learning can accelerate image formation time and enhance performance.
Unrolling iterative solvers is a notable approach that combines the strengths of deep learning with traditional optimization methods and physical modeling,
as a fixed number of iterations of an iterative solver are represented as layers of a neural network. 
Each layer is parameterized with its own learnable hyperparameters, such as step sizes, and these are optimized end-to-end using backpropagation~\cite{lista}.
Unrolled algorithms can significantly reduce convergence time by learning optimal hyperparameters for fewer iterations.
In lensless imaging, Monakhova \etal~\cite{Monakhova:19} demonstrated that only five iterations of unrolled ADMM with learned hyperparameters achieved similar performance to 100 iterations with manually-selected fixed parameters.
Furthermore, they incorporated a learned denoiser, a U-Net architecture with approximately 10M parameters~\cite{unet}, at the output to further improve reconstruction quality.
Combining deep learning with physical priors produces state-of-the-art results with fewer hallucinations and improved interpretability~\cite{9239993,Perron2023}.
Unlike purely data-driven approaches, these hybrid methods leverage both the underlying physics of the imaging system and the representational power of neural networks, achieving accurate reconstructions with less data.

\subsection{Robust Lensless Imaging}

\noindent Lensless imaging recovery is a challenging ill-posed inverse problem, as the highly-multiplexed nature makes it sensitive to model mismatch in the forward modeling within the data fidelity term.
A common assumption is linear shift invariance (LSI), which approximates off-axis PSFs as lateral shifts of the on-axis PSF.
This simplifies calibration and reduces computational complexity when computing the forward model.
However, this approximation introduces errors, particularly when iterative solvers like ADMM are used, as these errors accumulate over multiple iterations~\cite{9546648}.

To address model mismatch, Zeng \etal~\cite{9546648} proposed a neural network-based compensation branch that uses intermediate outputs from unrolled ADMM iterations to reduce the error resulting from model mismatch.
Other works attempt to reduce the model mismatch itself, \eg by fine-tuning the on-axis PSF~\cite{9239993,Kingshott:22} or applying transformations to it~\cite{Li:23}.
However, these methods still operate within the constraints of the LSI assumption.
While LSI can be valid for certain systems (\eg for DiffuserCam where off-axis and on-axis PSFs have at least \SI{75}{\percent} similarity within a \SI{37.5}{\degree} field of view~\cite{Antipa:18}),
some level of model mismatch is inevitable.
More advanced models, such as spatially-varying forward models~\cite{Yanny:22,cai2024phocolens}, aim to relax the LSI assumption.
However, these approaches can still suffer from inaccuracies if the locally shift-invariant regions are not well-parameterized.

Incorporating deep learning into reconstruction introduces additional challenges, as performance can degrade when test data deviates from the training distribution~\cite{9782500}.
In lensless imaging, there are several factors that could change between training and inference:
\eg scene content, positioning, lighting, SNR, and the imaging system's mask/PSF.
Existing methods demonstrate robustness to scene variations~\cite{Monakhova:19,9239993,Lee:23},
while our previous work~\cite{Perron2023} showed improved robustness to SNR variations by incorporating a pre-processor.

\subsection{Generalizable Lensless Imaging}

\noindent While existing methods demonstrate robustness to scene variations,
few address generalization to changes in the mask/PSF of the imaging system.
Lee \etal~\cite{Lee:23} evaluated robustness to minor manufacturing variations in masks,
but these are very minimal changes to the PSF.
Rego \etal~\cite{Rego2021} formulated lensless imaging as a blind deconvolution problem, such that the PSF is not needed during inference, but all possible PSFs (and measurements or simulations with them) are seen during training.
Collecting extensive datasets for each PSF is impractical, given the already time-consuming nature of acquiring data for a single PSF.
Untrained networks~\cite{Monakhova:21} eliminate the need for labeled datasets but require impractically long reconstruction times (\eg, several hours).
A lack of generalizability studies is largely due to the dearth of publicly-available lensless datasets: DiffuserCam (25K examples)~\cite{Monakhova:19}, FlatCam (10K examples)~\cite{9239993}, PhlatCam (10K examples)~\cite{9239993}, and SweepCam (380 examples)~\cite{zheng2021programmable3dcam}.

Adapting pre-trained models offers another avenue for generalization. 
Gilbert \etal~\cite{9477112} proposed methods to adapt a network trained on one forward model to a new one, but their results are limited to simple blur kernels ($7\times7$ pixels), which are far smaller and less complex than the PSFs encountered in lensless imaging.

Reconstruction methods that are robust (to model mismatch and noise) and that generalize to unseen PSFs are crucial for advancing lensless imaging. 
Such methods would reduce the need for exhaustive dataset collection, 
making lensless imaging more practical, 
particularly in scenarios where data acquisition is infeasible due to privacy concerns or inaccessibility. By open-sourcing four large datasets and leveraging a modular reconstruction pipeline, 
we aim to improve the generalizability and usability of lensless imaging systems.

%% file: sec/3_proposed.tex
\section{Sensitivity to Model Mismatch}

\noindent In this section, we present the physical modeling of lensless imaging, and mathematically demonstrate the sensitivity of common image recovery techniques to model mismatch.
This theoretical analysis helps to explain the empirical success of previous work that apply post-processors~\cite{Monakhova:19,9239993}, PSF fine-tuning~\cite{9239993}, and PSF correction~\cite{Li:23}.
Moreover, it motivates our use of a pre-processor in minimizing the input noise amplified by the inevitable model mismatch.

\subsection{Forward Modeling}

\noindent Assuming a desired scene is comprised of point sources that are incoherent with each other, a lensless imaging system
can be modeled as a linear matrix-vector multiplication with the system matrix $\bm{H}$:
\begin{align}
    \label{eq:forward_gen}
    \bm{y} = \bm{H}\bm{x} + \bm{n},
\end{align}
where $\bm{y}$ and $\bm{x}$ are the vectorized lensless measurement and scene intensity respectively, and $\bm{n}$ is the measurement noise.
Due to the highly multiplexed characteristic of lensless cameras, image recovery amounts to a large-scale deconvolution problem where the kernel $\bm{H}$ is a very dense matrix.
Each column of $\bm{H}$ is a PSF, mapping a single point in the scene to a response at the measurement plane.

As obtaining $\bm{H}$ would require an expensive calibration, the PSFs in lensless imaging are approximated as shift-invariant, \ie off-axis PSFs are assumed to be lateral shifts of the on-axis PSF.
This approximation allows $\bm{H}$ to take on a Toeplitz structure, such that the forward operation can be written as a 2D convolution with the on-axis PSF.
Using the convolution theorem, we can write the forward operation as a point-wise multiplication in the frequency domain:
\begin{align}
\label{eq:lsi_forward}
    \bm{Y} = \bm{P} \odot \bm{X} + \bm{N},
\end{align}
where $\{\bm{Y}, \bm{P}, \bm{X}, \bm{N}\} \in \mathbb{C}^{N_x \times N_y}$ are 2D Fourier transforms of the measurement, the on-axis PSF, the scene, and the noise respectively, and $\odot$ is point-wise multiplication.
The on-axis PSF can be either measured, \eg with a white LED at far-field in a dark room, or simulated if the mask structure is known~\cite{9239993,Li:23}.
As well as requiring less calibration measurements and storage, the above convolution can be computed efficiently with the fast Fourier transform (FFT).

\subsection{Consequences of Model Mismatch}
\label{sec:model_mismatch}

\noindent Whether or not we assume shift-invariance, there will be model mismatch with the true system matrix $\bm{H}$. 
Either the measurement of the on-axis PSF will be noisy, its simulation will make simplifying assumptions, or the LSI modeling is too simplistic.
In other words, the forward modeling can impact the amount of model mismatch.

In the most general case, \ie not assuming shift-invariance, we can denote our estimate system matrix as $\bm{\hat{H}}=(\bm{H}+\bm{\Delta}_H)$ where the deviation from the true system matrix is $\bm{\Delta}_H$.
Our forward model from \cref{eq:forward_gen} can then be written as:
\begin{align}
    \label{eq:mismatch_forward}
    \bm{y} = \bm{H}\bm{x} + \bm{n} = (\bm{\hat{H}} - \bm{\Delta}_H)\bm{x} + \bm{n}.
\end{align}
The quality of the sensor and the optical components can influence the amount of mismatch $\bm{\Delta}_H$ and of measurement noise $\bm{n}$.
As both are inevitable, 
image recovery approaches
yield a noisy estimate of the form:
\begin{align}
    \label{eq:mismatch_general}
    \bm{\hat{x}}^{\text{noisy}} = \bm{\hat{x}} + \underbrace{f(\bm{x}, \bm{\Delta}_H)}_{\text{model mismatch}} +  \underbrace{g(\bm{n}, \bm{\Delta}_H)}_{\text{noise amplification}},
\end{align}
where $\bm{\hat{x}}$ is the estimate when $\bm{\Delta}_H=0$ and $\bm{n}=0$, 
the model mismatch perturbation $f(\bm{x}, \bm{\Delta}_H)$ depends on the target image $\bm{x}$ and $\bm{\Delta}_H$, and the noise amplification $g(\bm{n}, \bm{\Delta}_H)$ depends on $\bm{n}$ and $\bm{\Delta}_H$.

The breakdown in \cref{eq:mismatch_general} provides insight to motivate our use of a pre-processor and the modular framework as a whole, \ie to minimize measurement noise and model mismatch before and after inevitable amplification by camera inversion. 
This motivation is further discussed in \cref{sec:modular}.
Below we demonstrate this breakdown for common image recovery approaches for lensless cameras.
Detailed derivations can be found in \cref{app:mismatch}.

\subsubsection{Direct inversion}

\noindent Assuming the system is invertible and with spectral radius $\rho(\bm{H}) < 1$, using the estimate $\bm{\hat{H}}$ for direct inversion yields~\cite{9546648,9157433}:
\begin{align}
   \label{eq:inversion_terms}
   \bm{\hat{x}} &= \bm{x} - \underbrace{\bm{H}^{-1}\bm{\Delta}_H \bm{x}}_{\text{model mismatch}} + \underbrace{(\bm{I} - \bm{H}^{-1}\bm{\Delta}_H)\bm{H}^{-1}\bm{n}}_{\text{noise amplification}} + \mathcal{O}(\| \bm{\Delta}_H\|_F^2).
\end{align}
In \cref{eq:inversion_terms}, we observe how noise and model mismatch are amplified, particularly if $\bm{H}$ is ill-conditioned as $\bm{H}^{-1}$ could be very large.

\subsubsection{Wiener filtering} From the point-wise forward model in \cref{eq:lsi_forward}, minimizing the mean squared error yields the classic Wiener filtering estimate:
\begin{align}
    \label{eq:wiener}
    \bm{\hat{X}} = \dfrac{\bm{P}^*  \odot \bm{Y}}{ |\bm{P}|^2 + \bm{R}} =  \dfrac{\bm{P}^* \odot (\bm{P} \odot\bm{X} + \bm{N})}{ |\bm{P}|^2 + \bm{R}},
\end{align}
where all operations are point-wise,
the noise $\bm{N}$ is assumed to be independent to $\bm{X}$, and $\bm{R}\in \mathbb{R}^{N_x \times N_y}$ is the \textit{inverse} of the SNR at each frequency.

If we use a mismatched version of the on-axis PSF's Fourier transform, \ie $\bm{\hat{P}} = (\bm{P}+ \bm{\Delta}_P)$, 
our Wiener-filtered estimate of the scene becomes:
\begin{align}
    \bm{\hat{X}}^{\text{noisy}} &= \dfrac{\bm{\hat{P}}^* \odot \bm{Y}}{ |\bm{\hat{P}}|^2 + \bm{R}} \nonumber\\
    &= \bm{\hat{X}} + \underbrace{\bm{M} \odot \bm{P} \odot  \bm{X}}_{\text{model mismatch}} + \underbrace{\bm{M} \odot \bm{N}}_{\text{noise amplification}}, \label{eq:noisy_wiener}
\end{align}
where:
\begin{align}
\bm{M} &= \dfrac{\bm{\Delta}_P^*}{\bm{B}} - \dfrac{\bm{\Delta}_B \odot (\bm{P}^* + \bm{\Delta}_P^*)}{\bm{B}^2 + \bm{B} \odot \bm{\Delta}_B}, \\
\bm{B} &= |\bm{P}|^2 + \bm{R},\\
\bm{\Delta}_B &= |\bm{\Delta}_P|^2 + \bm{\Delta}_P^* \odot \bm{P} + \bm{P}^*\odot\bm{\Delta}_P.
\end{align}
While Wiener filtering avoids adverse amplification, \ie with $\bm{H}^{-1}$ as in direct inversion,
model mismatch still leads to similar error terms as shown in \cref{eq:mismatch_general}.

\subsubsection{Iterative solvers} 
\label{sec:error_iterative}

A common approach to avoid amplification with $\bm{H}^{-1}$ is to cast image recovery as a regularized optimization problem:
\begin{align}
\label{eq:opt_gen}
   \bm{\hat{x}} = \arg \min_{\bm{x}} \frac{1}{2} ||\bm{H}\bm{x} - \bm{y}||_2^2 + \lambda \mathcal{R}(\bm{x}),
\end{align}
where $\mathcal{R}(\cdot)$ is a regularization function on the estimate image.
In lensless imaging, it is common to apply non-negativity and sparsity constraints in the TV space~\cite{Antipa:18,phlatcam}.
When the regularization function uses the $l_1$ norm, an iterative solver is needed to optimize \cref{eq:opt_gen}.
A common approach is ADMM, for which we can obtain a similar decomposition of model mismatch and noise amplification at \textit{each iteration}.
Zeng \etal~\cite{9546648} show how model mismatch leads to an accumulation of mismatch errors over multiple ADMM iterations, but they do not show noise amplification.
From Eq.~15 of~\cite{9546648}, by expanding the terms from the previous iteration ($\bm{\epsilon}^{(k-1)}$)  that depend on the model mismatch, we obtain:
\begin{align}
&\bm{\hat{x}}^{(k),\text{noisy}} = \bm{\hat{x}}^{(k)}  + \underbrace{\bm{W}_4 \bm{W}_2\bm{C}^T \bm{n}}_{\text{noise amplification}} \nonumber \\
\label{eq:admm_mismatch} 
&\underbrace{+ \bm{W}_1^{-1}\rho_x \delta_{\bm{H}} \bm{\hat{x}}^{(k)} + \bm{W}_4 \bm{W}_2\left(\bm{C}^T\bm{C} \bm{H}\bm{x} + \bm{\gamma}^{(k-1)} \right) + \bm{W}_4\bm{W}_3}_{\text{model mismatch}}, 
\end{align}
where:
\begin{align}
\delta_{\bm{H}} &= \left( \bm{\Delta}_H^T\bm{H} + \bm{\hat{H}}^T \bm{\Delta}_H \right),\\
\bm{W}_1 &= \rho_x \bm{\hat{H}}^T \bm{\hat{H}} + \rho_z \bm{C}^T\bm{C} + \rho_y \bm{I}, \\
\bm{W}_2 &= (\bm{W}_1 + \rho_x \delta_{\bm{H}})^{-1} \Delta_{\bm{H}}^T \rho_x (\bm{C}^T\bm{C} + \rho_x \bm{I})^{-1}, \\
\bm{W}_3 &= \left( \bm{W}_1 + \rho_x \delta_{\bm{H}} \right)^{-1} \bm{\hat{H}}^T \rho_x^2 \Delta_{\bm{H}} \bm{\hat{x}}^{(k-1)},\\
\bm{W}_4 &= (\bm{I} + \bm{W}_1^{-1} \rho_x \delta_{\bm{H}}),
\end{align}
$\{\rho_x, \rho_y, \rho_z\}$ are positive penalty parameters, $\bm{C}$ crops the image to the sensor size~\cite{Antipa:18}, and $\bm{\gamma}^{(k-1)}$ contains terms from the previous iterations that do not depend on model mismatch.
Similar to Wiener filtering, while there is no amplification with $\bm{H}^{-1}$,
model mismatch leads to noise amplification and error terms,
as shown in \cref{eq:mismatch_general},
at each iteration.

\section{Methodology}
\label{sec:methodology}

\subsection{Modular Reconstruction}
\label{sec:modular}

\noindent As shown in \cref{sec:model_mismatch}, there are typically two noise sources in lensless imaging: (1) at measurement~$\bm{n}$ and (2) model mismatch $\bm{\Delta}_H$.
Our modular reconstruction pipeline, 
as shown in~\cref{fig:pipeline},
can address these perturbations and their consequences
for multiple camera inversion approaches.
While previous work has proposed various lensless recovery approaches that jointly train camera inversion with post-processors~\cite{Monakhova:19, 9239993},
they do not address noise amplification by camera inversion, nor do they theoretically motivate the use of a post-processor 
(apart from~\cite{9546648} for ADMM).

One of our contributions is to introduce a pre-processor to minimize the inevitably-amplified noise,
as shown with $g(\bm{n}, \bm{\Delta}_H)$ in \cref{eq:mismatch_general}.
While simply using a post-processor could address $g(\bm{n}, \bm{\Delta}_H)$, a low SNR may result in a poor camera inversion output, which can be challenging for post-processing alone to effectively address. 
Similarly, if noise amplification is non-linear or leads to clipping, post-processing alone may struggle.
Therefore, pre-processing can alleviate the task of the post-processor by denoising \textit{prior} to camera inversion,
such that the latter component's output is easier for the post-processor to enhance.
In our experiments, we evaluate on both low and high SNRs to demonstrate the added benefit from the pre-processor.

Moreover, the insight from \cref{sec:model_mismatch} better motivates the design choices of previous work, which have otherwise relied on experimental results to support their design choices.
Firstly, learned camera inversions that attempt to reduce model mismatch~\cite{9239993,Li:23,Kingshott:22} can reduce both perturbation terms in \cref{eq:mismatch_general},
and like the pre-processor, reduce the effort needed by the post-processor in treating the camera inversion output.
In our modular framework shown in~\cref{fig:pipeline}, we incorporate a PSF correction component to reduce model mismatch in the PSF prior to camera inversion. Moreover, while pre-processing and PSF correction can be incorporated to directly address $\bm{n}$ and $\bm{\Delta}_H$,
there will be residual error.
The post-processor can address the now simpler denoising task, while also performing perceptual enhancements.

To train our modular reconstruction and to demonstrate the effectiveness of our proposed pre-processor,
we need a sufficient amount of data.
Another one of our contributions is collecting and open-sourcing four lensless datasets,
which use a variety of masks/PSFs to demonstrate the effectiveness for different imaging systems.
These datasets are summarized in \cref{tab:datasets} and are further explained in \cref{sec:experiments}.
We also open-source the tooling for others to more conveniently collect and share their own datasets.\footnote{\href{https://lensless.readthedocs.io/en/latest/measurement.html}{lensless.readthedocs.io/en/latest/measurement.html}}

\subsubsection{Pre- and Post-Processor Design}
\label{sec:processors}

From a single measurement, it is difficult to obtain meaningful information about $\bm{\Delta}_H$ and/or $\bm{n}$ to design the pre- and post-processors.
Moreover, 
as shown in \cref{sec:model_mismatch}, 
each inversion approach amplifies the input noise in a unique manner.
To this end, our solution is to train both processors and the camera inversion end-to-end, such that the appropriate processing can be learned from measurements rather than heuristically-designed processing.
As each camera inversion approach results in a unique amplification of the model mismatch and noise,
the learned pre- and post-processors are trained for a specific inversion approach,
\ie their ability to transfer between camera inversion approaches is not guaranteed. 

Similar to previous work, we use a loss function that is a sum of the mean-squared error (MSE) and a perceptual loss between the reconstruction output $\bm{\hat{x}}$ and the ground-truth $\bm{x}$:
\begin{equation}
    \label{eq:loss_mse_lpips}
    \mathscr{L}\left(\bm{x},\bm{\hat{x}}\right) = \mathscr{L}_{\text{MSE}}\left(\bm{x},\bm{\hat{x}}\right) + \mathscr{L}_{\text{LPIPS}}\left(\bm{x},\bm{\hat{x}}\right).
\end{equation}
We use the Learned Perceptual Image Patch Similarity (LPIPS) metric~\cite{zhang2018perceptual} as a perceptual loss, which promotes photo-realistic images at a patch level, rather than pixel-wise as MSE.

For the pre- and post-processors in \cref{fig:pipeline},
we use a denoising residual U-Net (DRUNet) architecture  shown to be very effective for denoising, deblurring, and super-resolution tasks~\cite{zhang2021plug}.
The DRUNet architecture is presented in \cref{sec:drunet}.
Transformers~\cite{Zamir2021Restormer,Pan:22} or diffusion models~\cite{cai2024phocolens} could also be applied as pre- and post-processors,
but this work concentrates on DRUNet as it is does not require many parameters (unlike transformers) nor many iteration steps (as diffusion models).

\begin{table}[t!]
\renewcommand{\arraystretch}{1.3}
\caption{Comparison of trainable camera inversion approaches.}
\label{tab:compare_inv}
\centering
\resizebox{\columnwidth}{!}{
\begin{tabular}{c||c|c|c}
\hline
& \makecell{Noisy estimate\\due to mismatch} & \makecell{\# trainable\\parameters} & \makecell{PSF\\correction} \\
\hline\hline
Unrolled ADMM~\cite{Monakhova:19}  &  \makecell{\cref{eq:admm_mismatch}} & $10^1 - 10^2$ & No \\
\hline
\makecell{Trainable inversion~\cite{9239993}}  & 
\makecell{\cref{eq:inversion_terms}} & $10^4 - 10^6$ & Yes \\
\hline
\makecell{Unrolled ADMM\\with model-mismatch\\compensation network~\cite{9546648}}    &  \makecell{\cref{eq:admm_mismatch}} & $10^6-10^7$  & No \\
\hline
\makecell{Multi-Wiener deconvolution\\ network (MWDN)~\cite{Li:23}}  & 
\makecell{\cref{eq:noisy_wiener}} & $10^6-10^7$ & Optional \\
\hline
\end{tabular}
}
\end{table}

\subsubsection{Camera Inversion Approaches}

\noindent We investigate four trainable camera inversion approaches proposed by previous work: unrolled ADMM~\cite{Monakhova:19}, trainable inversion~\cite{9239993}, unrolled ADMM with a model mismatch compensation network (MMCN)~\cite{9546648},
and multi-Wiener deconvolution network (MWDN)~\cite{Li:23}.
The architecture of all camera inversion approaches are visualized in \cref{app:inversion},
and their characteristics are compared in \cref{tab:compare_inv}.

As shown in \cref{fig:pipeline},
the input to camera inversion is either the raw lensless measurement or the output of a pre-processor,
while the output of camera inversion can be optionally fed to a post-processor.
For promoting measurement consistency,
the camera inversion can take as input the on-axis PSF,
which can be fine-tuned~\cite{9239993} or corrected with neural networks~\cite{Li:23}.
Fine-tuning may be preferable if reconstruction is only expected for a single PSF/imaging system,
but the learned adjustments cannot be transferred if there are changes in the imaging system,
in which case a correction network may be preferable.
As we investigate the transferability of learned components, 
we optionally add a DRUNet for correcting the PSF.

\subsection{DigiCam: Hardware and Modeling}
\label{sec:digicam}

\noindent For our generalizability experiments, we propose a programmable-mask system entitled \textit{DigiCam}. 
As a programmable mask, we use a low-cost LCD~\cite{adafruitlcd}.
While previous work with LCDs requires multiple measurements~\cite{4472247,huang2013}
or only uses a few LCD pixels to simply control the aperture~\cite{zomet2006},
we are the first to apply an LCD for single-shot lensless imaging.
Moreover, LCDs are significantly cheaper than SLMs: our component is around $150\times$ cheaper than the SLM used in~\cite{sweepcam2020,zheng2021programmable3dcam}.
A reconfigurable system is an extremely convenient way to experimentally evaluate generalizability to different PSFs, as the mask pattern can be simply reprogrammed to have an imaging system with a different PSF.
Moreover, as the mask structure is known, the PSF can be simulated for calibration-free imaging (after an initial alignment).
While a programmable mask cannot represent all possible lensless imaging PSFs, it can provide useful insight into the generalizability of learned reconstruction approaches.
Below we describe the hardware, and detail the wave-propagation modeling needed for simulating the PSF.
More modeling details can be found in \cref{app:psf_modeling,app:mask_modeling}.

\subsubsection{Hardware Prototype}

\begin{figure}[t!]
		\centering
		\includegraphics[width=0.7\linewidth]{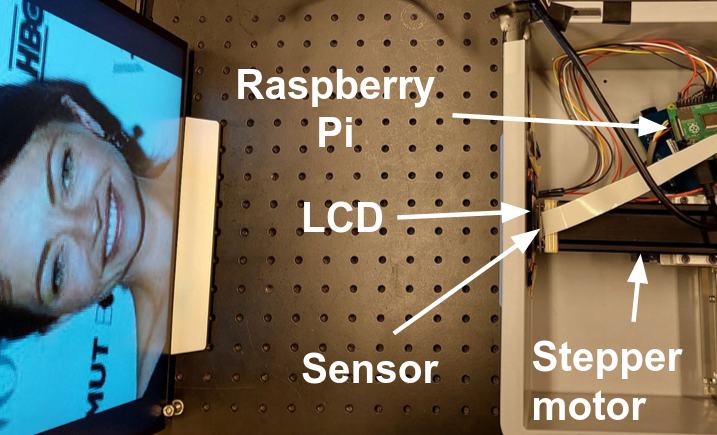}
	\caption{DigiCam prototype and measurement setup.}
	\label{fig:prototype_labeled}
\end{figure}

\noindent A programmable mask serves as the only optical component, 
specifically an off-the-shelf LCD driven by the ST7735R device,
which can be purchased for $20$ USD~\cite{adafruitlcd}.
The LCD component was selected because it has a higher spatial resolution than other off-the-shelf LCDs,
and has a Python API to set pixel values.
The LCD has an interleaved pattern of red, blue, and green sub-pixels, 
but a monochrome programmable mask with sufficient spatial resolution could also be used.
Our experimental prototype can be seen in~\cref{fig:prototype_labeled}.
The LCD is wired to a Raspberry Pi (RPi) ($35$~USD) with the RPi High Quality (HQ) $12.3$ MP Camera~\cite{rpi_hq} ($50$~USD) as a sensor,
totaling our design to just $105$~USD. 
This is significantly cheaper than other programmable mask-based prototypes that make use of an SLM~\cite{sweepcam2020,zheng2021programmable3dcam}, 
which can cost a few thousand USD.
Our prototype includes an optional stepper motor for programmatically setting the distance between the LCD and the sensor.

\subsubsection{Wave-Based Modeling}

\noindent Accurately simulating the PSF is crucial for the reconstruction quality,
and to minimize model mismatch and its consequences.
One advantage of using a programmable mask is that it has a well-defined structure, 
which allows us to model propagation through the programmable mask to simulate the PSF.
A simulation based on wave optics (as opposed to ray optics) may be necessary to account for diffraction due to the small apertures of the mask and for wavelength-dependent propagation. 
The Fresnel number $ N_F $ can be used to determine whether a wave-optics simulation is necessary, with ray optics generally requiring $N_F\gg 1$~\cite{boominathan2022recent}.
The Fresnel number is given by $ N_F = a^2 / d\lambda $, where $ a $ is the size of the mask's open apertures, $ d $ the propagation distance, and $ \lambda $ the wavelength. 
For our prototype, $ a = \SI{0.06}{\milli\meter} $, $ d = \SI{2}{\milli\meter}$ between the mask and sensor, and $ \lambda \in [\SI{450}{\nano\meter}, \SI{750}{\nano\meter}] $ (visible light),
such that $ N_F \in [2.4, 4] $. 
As $N_F$ is not significantly greater than one, diffraction effects need to be accounted for.

We model the PSF similar to~\cite{sitzmann2018e2e}, \ie as spherical waves up to the optical element followed by free-space propagation to the sensor.
The point-source wave field at the sensor for a given wavelength $\lambda$
can be written as:
\begin{align}
	\label{eq:wavefield}
	&u(\bm{r}; d_1, d_2, \lambda) = \nonumber \\ 
	&\mathcal{F}^{-1}\Big(\mathcal{F} \Big( m(\bm{r}; \lambda) \underbrace{e^{j \frac{2\pi}{\lambda} \sqrt{\|\bm{r}\|_2^2 +  d_1^2}}}_{\text{spherical waves}}
	\Big) \times h(\bm{u}; z=d_2, \lambda) \Big),
\end{align}
where 
$d_1$ is the distance from the point source to the optical element,
$d_2$ is the distance from the optical element to the sensor,
$ h(\bm{u}; z, \lambda)$ is the free-space propagation frequency response, and
$\bm{u} \in \mathbb{R}^2$ are the spatial frequencies of $\bm{r} \in \mathbb{R}^2$.
For the free-space propagation kernel, we use bandlimited angular spectrum (BLAS)~\cite{Matsushima2009}.

As the illumination is incoherent, PSFs from different scene points will add in intensity at the sensor~\cite{Goodman2004}.
Therefore, we take the squared amplitude of \cref{eq:wavefield} for the intensity PSF~\cite{Goodman2004}.
As our sensor measures RGB, the PSF of each color channel $c \in \{R,G,B\}$ should account for its wavelength sensitivity.
Similar to~\cite{sitzmann2018e2e}, we assume a narrowband around the RGB wavelengths, and compute the PSFs for $ c\in\{R,G,B\} $ for the respective wavelengths of (\SI{640}{\nano\meter}, \SI{550}{\nano\meter}, \SI{460}{\nano\meter}):
\begin{equation}
	\label{eq:intensity_psf_simple}
	p(\bm{r}; d_1, d_2, c) = | u(\bm{r}; d_1, d_2, \lambda_c)|^2.
\end{equation}
\cref{eq:wavefield} defines the response for an arbitrary optical encoder $m(\bm{r}; \lambda)$.
A programmable mask like that of \textit{DigiCam} can be modeled as a superposition of apertures for each adjustable RGB sub-pixel in $ \bm{r} \in \mathbb{R}^2 $:
\begin{align}
	\label{eq:mask_simple}
	m(\bm{r}; \lambda=\lambda_c) &= \sum_{k_c \in K_c}
	w_{k_c}
	a(\bm{r} -\bm{r}_{k_c}), \quad c\in\{R,G,B\},
\end{align}
where we again assume a narrowband around the RGB wavelengths,
$K_c$ is the set of pixel corresponding to the color filter $c$,
$w_{k_c} \in [0, 1]$ are the mask weights for each sub-pixel, $ \{(\bm{r}_{k_c})\}_{k_c=1}^{K_c} $ are the centers of each sub-pixel, 
and the aperture function $a(\cdot)$ is modeled as a rectangle of size $\SI{0.06}{\milli\meter}\times\SI{0.18}{\milli\meter}$ (the dimensions of each sub-pixel).
\cref{eq:mask_simple} accounts for the mask \textit{deadspace} (occluding regions that are not controllable due to circuitry) and \textit{pixel pitch} (distance between pixels) by setting the appropriate centers $ \{(\bm{r}_{k_c})\}_{k_c=1}^{K_c} $.
An alternative approach to account for pixel pitch is to modify the wave propagation model to include higher-order diffraction and attenuation~\cite{Gopakumar:21}, but this approach does not account for the deadspace.

\newcommand{\figsizepsf}{0.21}
\newcommand{\figsizepsfdigi}{0.21}
\begin{figure*}[t!]
    \centering
    \begin{subfigure}{\figsizepsf\linewidth}
		\centering
		\includegraphics[width=0.99\linewidth]{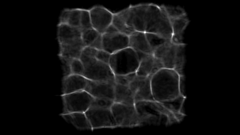} 
		\caption{DiffuserCam~\cite{Monakhova:19} \textit{(meas)}.}
		\label{fig:diffusercam}
	\end{subfigure}
    \begin{subfigure}{\figsizepsf\linewidth}
		\centering
		\includegraphics[width=0.99\linewidth]{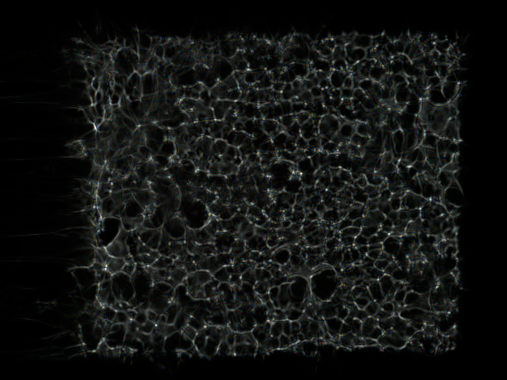} 
		\caption{TapeCam \textit{(meas)}.}
		\label{fig:tapecam}
	\end{subfigure}
 \begin{subfigure}{\figsizepsf\linewidth}
		\centering
		\includegraphics[width=0.99\linewidth]{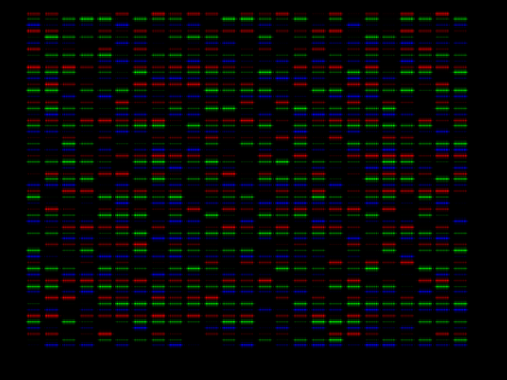} 
		\caption{DigiCam-CelebA \textit{(sim)}.}
		\label{fig:digicam_celeba}
	\end{subfigure}
 \begin{subfigure}{\figsizepsf\linewidth}
		\centering
  \includegraphics[width=0.99\linewidth]{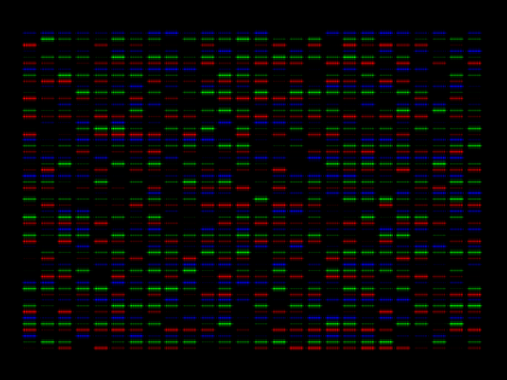} 
		\caption{DigiCam-Single \textit{(sim)}.}
		\label{fig:digicam_mirflickr}
	\end{subfigure}
 \\[5pt]
 \begin{subfigure}{\figsizepsfdigi\linewidth}
		\centering
		\includegraphics[width=\linewidth]{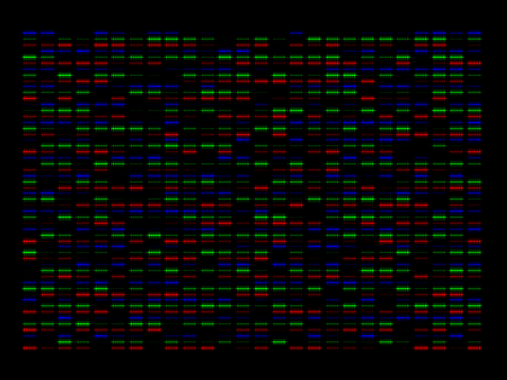} 
		\caption{Seed=1 \textit{(sim)}.}
		\label{fig:multi1}
	\end{subfigure}
	\begin{subfigure}{\figsizepsfdigi\linewidth}
		\centering 
  \includegraphics[width=\linewidth]{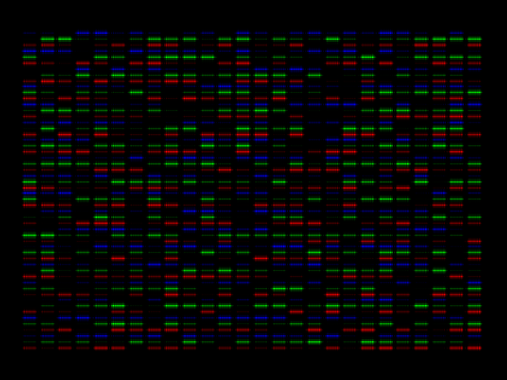} 
		\caption{Seed=2 \textit{(sim)}.}
		\label{fig:multi2}
	\end{subfigure}
    \begin{subfigure}{\figsizepsfdigi\linewidth}
		\centering
		\includegraphics[width=\linewidth]{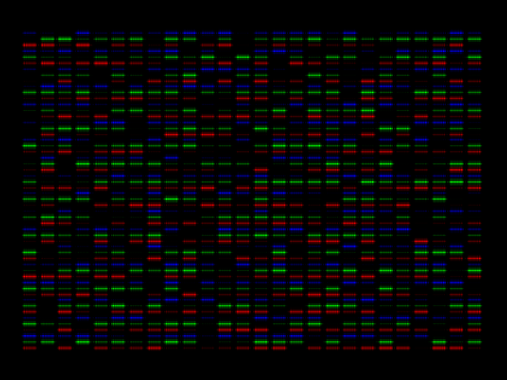} 
		\caption{Seed=3 \textit{(sim)}.}
		\label{fig:multi3}
	\end{subfigure}
    \begin{subfigure}{\figsizepsfdigi\linewidth}
		\centering
		\includegraphics[width=\linewidth]{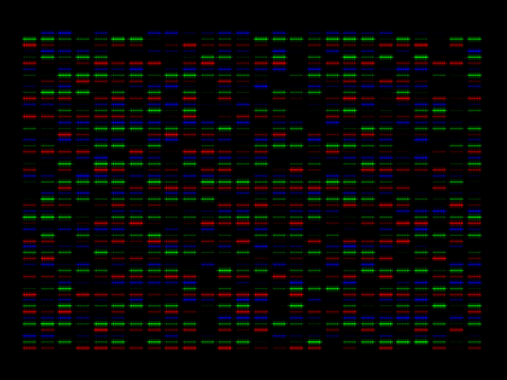} 
		\caption{Seed=4 \textit{(sim)}.}
		\label{fig:multi4}
	\end{subfigure}
	\caption{Point spread functions of datasets used in this work, where \textit{(meas)} refers to a measured PSF and \textit{(sim)} refers to simulated. \cref{fig:multi1,fig:multi2,fig:multi3,fig:multi4} are four of 100 simulated PSFs of the mask patterns used in measuring the \textit{DigiCam-Multi} dataset.}
	\label{fig:compared_psfs_crop}
\end{figure*}

\cref{fig:digicam_mirflickr,fig:multi1,fig:multi2,fig:multi3,fig:multi4} show simulated \textit{DigiCam} PSFs.
In \cref{app:compare_psf}, a measured PSF is compared with various simulation approaches with regards to reconstruction performance.
Wave-based propagation and programmable-mask modeling (with PyTorch support) is made available in \textit{waveprop}~\cite{waveprop}.\footnote{\href{https://github.com/ebezzam/waveprop}{github.com/ebezzam/waveprop}}

\subsection{Improving Generalizability}
\label{sec:improve_gen}

\noindent Learned reconstructions for lensless imaging face generalizability issues because they are typically not exposed to measurements and PSFs from different systems during training.
With our \textit{DigiCam} system, we can conveniently collect measurements from multiple mask patterns by programmatically setting the LCD, and can use \cref{eq:intensity_psf_simple,eq:mask_simple} to simulate the corresponding PSF. 
Consequently, a multi-mask dataset can be collected to train a reconstruction approach that generalizes to unseen \textit{DigiCam} patterns.
Our modular reconstruction, as shown in \cref{fig:pipeline}, can learn pre-processing, PSF correction, and post-processing that generalizes to measurements from unseen mask patterns.

Whether or not a programmable-mask system is used, transfer learning can be applied between lensless imaging systems.
This can be done by fine-tuning a learned reconstruction (trained with real measurements with one system) on simulations with a \textit{new system's} PSF, \ie by convolving ground-truth data with the new PSF.
While this requires training with the new PSF, it can avoid the need to collect a dataset which may not be possible.
Moreover, we can exploit modular components that have been trained with real measurements by other imaging systems, such as the pre-processor.
Training on simulations may not always generalize to real measurements.
It depends on the validity of the modeling assumptions, \eg the validity of LSI in \cref{eq:lsi_forward} for a wide-enough FOV.
If this width of LSI validity is too narrow or if there are significant differences in coloring, training with simulated data may not generalize to measured data.

%% file: sec/4_experiments.tex
\begin{table*}[t!]
\renewcommand{\arraystretch}{1.3}
\caption{Summary of datasets.}
\label{tab:datasets}
\centering
\begin{tabular}{c||c|c|c|c|c|c}
\hline
\textit{Dataset} & Source data & Mask & Sensor & PSF(s) & $\#$ train & $\#$ test\\
\hline\hline
DiffuserCam~\cite{Monakhova:19} & MirFlickr~\cite{huiskes2008mir} & Random diffuser~\cite{luminit} & Basler daA1920-30uc~\cite{basler} & \cref{fig:diffusercam} &24K & 1K \\
\hline
TapeCam  & MirFlickr~\cite{huiskes2008mir} & Double-sided tape as in~\cite{diffusercam_tut} & Raspberry Pi HQ~\cite{rpi_hq} & \cref{fig:tapecam} & 21.25K & 3.75K \\
\hline
DigiCam-Single & MirFlickr~\cite{huiskes2008mir} & Random LCD pattern~\cite{adafruitlcd} & Raspberry Pi HQ~\cite{rpi_hq} & \cref{fig:digicam_mirflickr} & 21.25K & 3.75K \\
\hline
DigiCam-Multi  & MirFlickr~\cite{huiskes2008mir} & 
100 random LCD patterns~\cite{adafruitlcd} & Raspberry Pi HQ~\cite{rpi_hq}
& \eg \cref{fig:digicam_mirflickr,fig:multi1,fig:multi2,fig:multi3,fig:multi4} &
\makecell{21.25K\\(85 masks)}
& 
\makecell{3.75K\\(15 masks)}
\\
\hline
DigiCam-CelebA  & CelebA~\cite{liu2015faceattributes} & Random LCD pattern~\cite{adafruitlcd} & Raspberry Pi HQ~\cite{rpi_hq} & \cref{fig:digicam_celeba}  &22.1K & 3.9K \\
\hline
\end{tabular}
\end{table*}

\section{Experiments and Results}
\label{sec:experiments}

\noindent We perform the following experiments:
\begin{enumerate}
    \item Show the strength of our modular approach and the benefit of using a pre-processor across different imaging systems and reconstruction approaches (\cref{sec:exp_preproc}).
    \item Demonstrate the robustness of our modular approach by digitally adding noise and model mismatch (\cref{sec:robustness_exp}).
    \item Evaluate the performance of a learned reconstruction on a system \textit{different} than the one it was trained for (\cref{sec:exp_eval_gen}).
    \item Utilize our modular reconstruction for improving the generalizability to measurements of PSFs not seen during training (\cref{sec:exp_improve_gen}).
\end{enumerate}

\subsection{Experimental Setup}

\subsubsection{Datasets}

\noindent Our experiments make use of five datasets to evaluate performance and generalizability.
The datasets are summarized in \cref{tab:datasets}. 
Apart from \textit{DiffuserCam}~\cite{Monakhova:19}, all datasets have been collected as part of this work and have a resolution of $(380\times507)$.
They are available on Hugging Face~\cite{tapecam,digicam_celeba,digicam_single,digicam_multi}, which provides a visualization of the measurements and a Python interface for downloading each dataset.
With our datasets, the goal is to use low-cost and accessible materials to demonstrate the potential for scalable, cost-effective lensless imaging.
For this reason, we use the RPi HQ sensor~\cite{rpi_hq}, double-sided tape as a phase mask (\textit{TapeCam)}, and an LCD as a reconfigurable amplitude mask (\textit{DigiCam}).
For the \textit{DigiCam} datasets, a random pattern of size $(3\times18\times26) = 1404$ pixels is generated using a uniform distribution, with 100 randomly generated patterns used for the multimask dataset (\textit{DigiCam-Multi}).
The training set measurements of \textit{DigiCam-Multi} use 85 different random mask patterns, while the test set uses another 15 random mask patterns, 
and there are 250 measurements per mask.
All other datasets use the same mask for the training and test set measurements.
The scene of interest, \ie an image displayed at a pre-defined resolution on a computer monitor as shown in \cref{fig:prototype_labeled}, is \SI{30}{\centi\meter} from the camera, while the mask is roughly \SI{2}{\milli\meter} from the sensor.
For our datasets, the ground-truth image, which is needed to compute the loss in \cref{eq:loss_mse_lpips}, is obtained by reshaping the image to the same resolution when displayed on the screen, and again reshaping to the corresponding region-of-interest (ROI) in the reconstruction.
The ROI is the region of the lensless reconstruction that corresponds to the object of interest, such that we remove black regions before computing the loss/metrics.
The loss is then computed between the reshaped ground-truth image and the extracted ROI from the reconstruction.

For \textit{DiffuserCam}~\cite{Monakhova:19}, lensed images are simultaneously captured with a beamsplitter,
and we downsample both lensless and lensed by $2\times$ to a resolution of $(135\times240)$.
Both lensed and lensless images are captured with Basler Dart (daA1920-30uc) sensors, which is more than $3\times$ the cost of the RPi HQ sensor used to collect our datasets~\cite{basler}.

\subsubsection{Models}

\noindent Equating the number of model parameters between reconstruction approaches has not been done in previous works, and is needed to fairly compare reconstruction approaches.
To this end,
we parameterize the number of feature representation channels between the four downsampling/upsampling scales in the DRUNet architecture,
such that an approximately equal number of model parameters (around 8.2M) are distributed between the pre-processor, camera inversion, and post-processor.
Unless noted otherwise, we consider three different sizes for the processors:
(1) around 8.2M parameters that increases the number of feature representation channels when downscaling from 32 to 256 according to (32, 64, 128, 256),
(2) around 4.1M parameters with (32, 64, 116, 128) feature representation channels, 
and (3) around 2M parameters with (16, 32, 64, 128) feature representation channels.
When upscaling back to the image shape, the number of feature representation channels decreases symmetrically.
Pre- and post-processor are denoted as $\textit{Pre}_{X}$ and $\textit{Post}_{X}$ respectively,
where $X$ refers to the number of parameters in millions.
For example, $\textit{Pre}_{4}$ refers to a pre-processor with around 4.1M parameters.
We use \textit{ADMMX} to refer to conventional ADMM with fixed hyperparameters and \textit{X} iterations,
and  \textit{LeADMMX} to denote unrolled ADMM with \textit{X} unrolled layers.
For each unrolled layer there are four hyperparameters.
Trainable inversion is denoted as \textit{TrainInv}.
As MMCN and MWDN use neural network components, the number of feature representation channels between the downsampling/upsampling scales are parameterized to maintain a similar number of model parameters as the other reconstruction approaches.
To this end, $\textit{MMCN}_{4}$ uses (24, 64, 128, 256, 400) feature representation channels 
for around 4.1M total parameters (original network~\cite{9546648} used (24, 64, 128, 256, 512) channels), 
and $\textit{MWDN}_{8}$ uses (32, 64, 128, 256, 436) feature representation channels 
for around 8.2M total parameters (original network~\cite{Li:23} used (64, 128, 256, 512, 1024) channels).

\subsubsection{Training and Evaluation Details}

PyTorch~\cite{Paszke2017} is used for training and evaluation. 
Unless noted otherwise, all learned methods are trained with the Adam optimizer with a learning rate of $10^{-4}$, $\beta_1=0.9$, $\beta_2=0.999$ for $25$ epochs and a batch size of $4$.
Training is done on an Intel Xeon E5-2680 v3 \SI{2.5}{\giga\hertz} CPU
and 4$\times$ Nvidia Titan X Pascal GPUs.
Three metrics are used: (1) peak signal-to-noise ratio (PSNR) which operates pixel-wise and higher is better, 
(2) structural similarity index measure (SSIM) which analyzes local regions and higher is better within $[-1,1]$, 
and (3) LPIPS which uses pre-trained VGG neural networks as feature extractors to compute similarity and lower is better within $[0,1]$.
Measurement and training scripts are made available in \textit{LenslessPiCam}~\cite{Bezzam2023}.

\subsection{Benefit of Pre-Processor}
\label{sec:exp_preproc}

\begin{table*}[!t]
	\renewcommand{\arraystretch}{1.3}
	\caption{Average image quality metrics (PSNR $\uparrow$ / SSIM $\uparrow$ / LPIPS $\downarrow$) for reconstructions on the test set of various measured datasets. Bold is used to denote the best performance across reconstruction methods (along columns). For the \textit{DiffuserCam} dataset, the number of parameters for \textit{TrainInv} differs as the PSF (which itself is a parameter) has a different resolution as another sensor is used, \ie 8.3M parameters for (\textit{TrainInv+}$\textit{Post}_{8}$) and 8.2M parameters for ($\textit{Pre}_{4}$\textit{+TrainInv+}$\textit{Post}_{4}$).}
	\label{tab:exp1_benchmark}
	\centering
\begin{tabular}{c|c|c||c|c|c|c}
		\hline
		Method & \makecell{\# learnable \\parameters} & \makecell{Inference\\time [ms]} & DiffuserCam & TapeCam & DigiCam-Single  & DigiCam-CelebA \\
		\hline\hline
		ADMM100~\cite{Antipa:18} & - & 771 & 15.0 / 0.457 / 0.511 & 10.2 / 0.234 / 0.720 &  10.6 / 0.291 / 0.751  & 10.1 / 0.352 / 0.737 \\
		\hline
        \hline
		TrainInv+$\text{Post}_{8}$~\cite{9239993} & 8.7M & 29.6 & 21.5 / 0.748 / 0.252 & 16.2 / 0.411 / 0.565 & 17.7 / 0.470 / 0.517 & 20.1 / 0.643 / 0.321 \\
		\hline
		$\text{MMCN}_{4}$+$\text{Post}_{4}$~\cite{9546648}  & 8.2M & 73.9 & 
  22.9 / 0.786 / 0.210
 & 16.7 / 0.483 / 0.505
  & 
  16.9 / 0.477 / 0.538
  & 
  18.0 / 0.614 / 0.363
  \\
  \hline
  $\text{Pre}_{8}$+LeADMM5 & 8.2M & 67.7 & 18.9 / 0.662 / 0.284 & 16.2 / 0.352 / 0.576 & 15.8 / 0.297 / 0.578 & 16.9 / 0.525 / 0.407 \\
		\hline
  LeADMM5+$\text{Post}_{8}$~\cite{Monakhova:19} & 8.2M & 67.6 & 23.8 / 0.806 / 0.202 & 18.6 / 0.505 / 0.478 & 19.1 / 0.515 / 0.469 & 20.9 / 0.667 / 0.296 \\
  		\hline
		$\text{MWDN}_{8}$~\cite{Li:23} & 8.1M & 20.2 & 24.2 / 0.797 / 0.206 & 16.5 / 0.480 / 0.541
  & 18.1 / 0.501 / 0.531 & 16.3 / 0.549 / 0.449 \\
		\hline
  \hline
		$\text{Pre}_{4}$+TrainInv+$\text{Post}_{4}$ & 8.7M & 49.2 & 23.5 / 0.794 / 0.214 & 19.3 / 0.555 / 0.461 & 19.9 / 0.525 / 0.454 & 22.1 / 0.696 / 0.265 \\
		\hline
		$\text{Pre}_{2}$+$\text{MMCN}_{4}$+$\text{Post}_{2}$  & 8.2M & 72.9 &  
 22.4 / 0.801 / 0.199
  & 
  18.0 / 0.518 / 0.484
  & 
  17.3 / 0.509 / 0.521
  & 
  19.2 / 0.631 / 0.346
  \\
		\hline
  $\text{Pre}_{4}$+LeADMM5+$\text{Post}_{4}$ & 8.1M & 88.1 & 25.3 / 0.838 / 0.171 & 19.7 / 0.564 / 0.441 & 19.6 / 0.531 / 0.449 & 22.5 / 0.703 / 0.263 \\
		\hline
        \hline
        $\text{Pre}_{4}$+LeADMM10+$\text{Post}_{4}$ & 8.1M & 129 & 26.1 / 0.851 / 0.160 & 19.8 / 0.560 / 0.441 & \textbf{20.1} / 0.551 / 0.440 & \textbf{23.0} / \textbf{0.709} / \textbf{0.262} \\
		\hline
        \makecell{$\text{Pre}_{4}$+LeADMM5+$\text{Post}_{4}$\\with PSF correction} & 8.1M & 93.9 & \textbf{26.4} / \textbf{0.857} / \textbf{0.154}  & \textbf{20.2} / \textbf{0.575} / \textbf{0.426}  & \textbf{20.1} / \textbf{0.552} / \textbf{0.439}  & 22.3 / 0.704 / 0.263 \\
		\hline
	\end{tabular}
\end{table*}

\begin{figure*}[t!]
    \centering
    \begin{subfigure}{0.32\linewidth}
		\centering
		\includegraphics[width=0.99\linewidth]{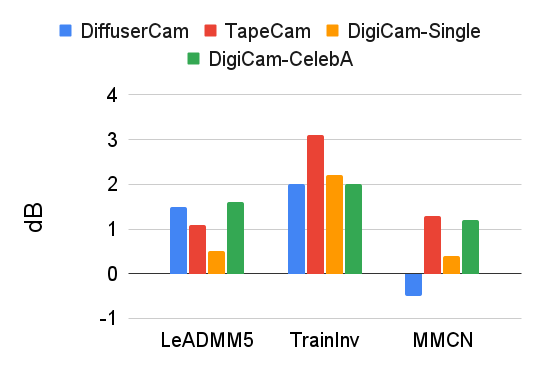} 
		\caption{PSNR improvement (dB).}
		\label{fig:exp1_psnr}
	\end{subfigure}
    \begin{subfigure}{0.32\linewidth}
		\centering
		\includegraphics[width=0.99\linewidth]{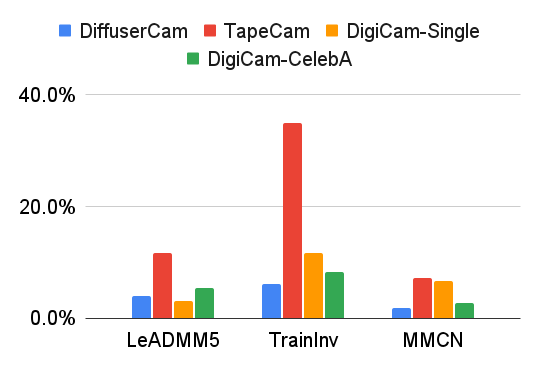} 
		\caption{SSIM relative improvement.}
		\label{fig:exp1_ssim}
	\end{subfigure}
 \begin{subfigure}{0.32\linewidth}
		\centering
		\includegraphics[width=0.99\linewidth]{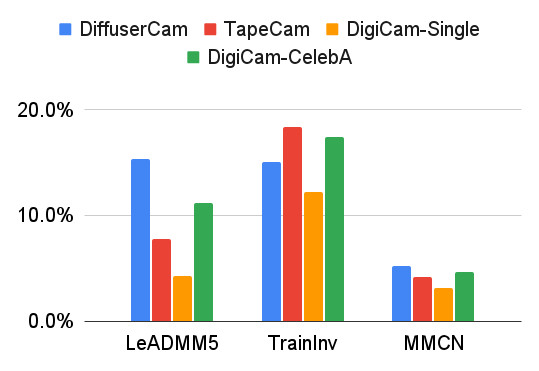}
		\caption{LPIPS relative improvement.}
		\label{fig:exp1_lpips}
	\end{subfigure}
 	\caption{Visualization of improvement in image quality metrics when splitting the number of model parameters between pre- and post-processors, rather than only using a post-processor.}
	\label{fig:exp1_improvement_viz}
\end{figure*}

\newcommand{\figsizegen}{0.10}
\newcommand{\figsizegendiffusercam}{0.13}
\newcommand{\figsizegenceleba}{0.068}
\newcommand{\newlinegen}{15pt}
\begin{figure*}[t!]
\centering
	\renewcommand{\arraystretch}{1} 
	\setlength{\tabcolsep}{0.06em} 
	\begin{tabular}{ccccccccc}
    &   
    \multicolumn{2}{c}{DiffuserCam~\cite{Monakhova:19}}
    & \multicolumn{2}{c}{TapeCam}
    & \multicolumn{2}{c}{DigiCam-Single}
    & \multicolumn{2}{c}{DigiCam-CelebA}
    \\ 
    \cmidrule(r){2-3} \cmidrule(r){4-5} \cmidrule(r){6-7} \cmidrule(r){8-9}
    \makecell{Raw data +\\Ground-truth}
&
\includegraphics[width=\figsizegendiffusercam\linewidth,valign=m]{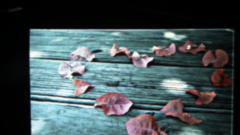}
\llap{\shortstack[l]{%
		\hspace{-1.9cm}\includegraphics[scale=.09]{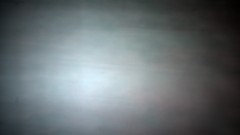}\\
		\rule{0ex}{0.13in}%
	}
	\rule{0.2in}{0ex}}
&
\includegraphics[width=\figsizegendiffusercam\linewidth,valign=m]{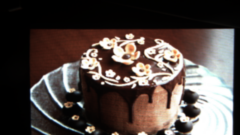}
\llap{\shortstack[l]{%
		\hspace{-1.9cm}\includegraphics[scale=.09]{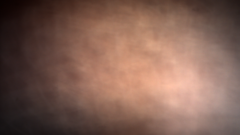}\\
		\rule{0ex}{0.13in}%
	}
	\rule{0.2in}{0ex}}
&
\includegraphics[width=\figsizegen\linewidth,valign=m]{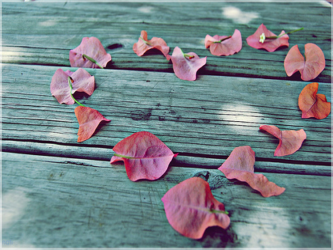}
\llap{\shortstack[l]{%
		\hspace{-1.35cm}\includegraphics[scale=.035]{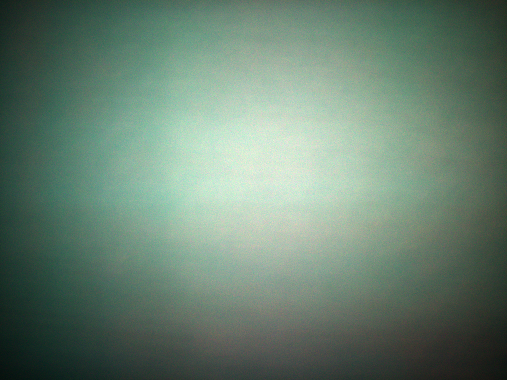}\\
		\rule{0ex}{0.13in}%
	}
	\rule{0.2in}{0ex}}
&
\includegraphics[width=\figsizegen\linewidth,valign=m]{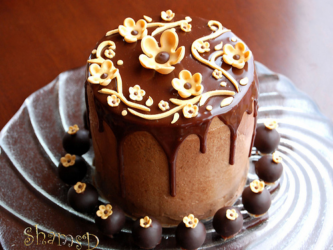}
\llap{\shortstack[l]{%
		\hspace{-1.35cm}\includegraphics[scale=.035]{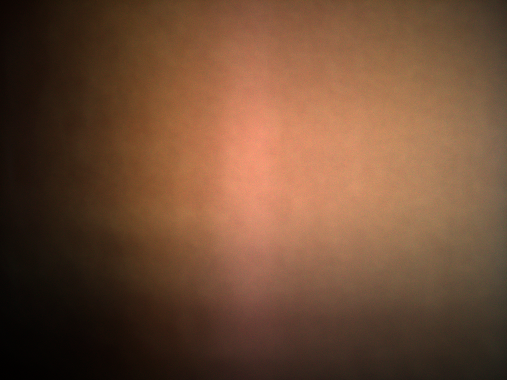}\\
		\rule{0ex}{0.13in}%
	}
	\rule{0.2in}{0ex}}
&
\includegraphics[width=\figsizegen\linewidth,valign=m]{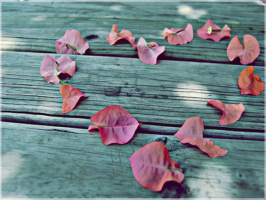}
\llap{\shortstack[l]{%
		\hspace{-1.35cm}\includegraphics[scale=.035]{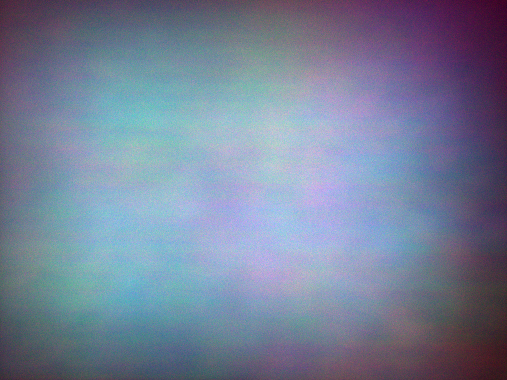}\\
		\rule{0ex}{0.13in}%
	}
	\rule{0.2in}{0ex}}
&
\includegraphics[width=\figsizegen\linewidth,valign=m]{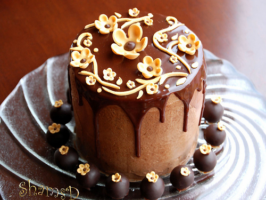}
\llap{\shortstack[l]{%
		\hspace{-1.35cm}\includegraphics[scale=.035]{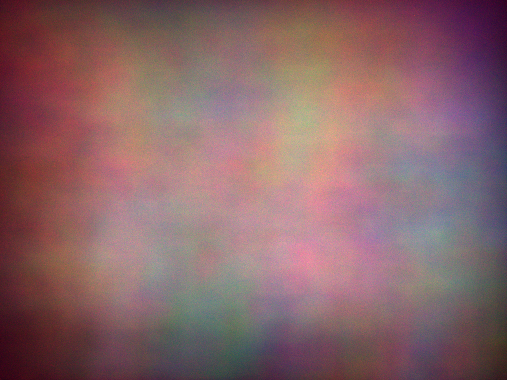}\\
		\rule{0ex}{0.13in}%
	}
	\rule{0.2in}{0ex}}
&
\includegraphics[width=\figsizegenceleba\linewidth,valign=m]{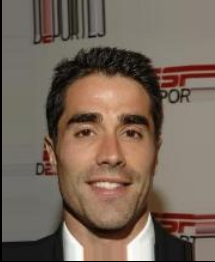}
\llap{\shortstack[l]{%
		\hspace{-0.8cm}\includegraphics[scale=.035]{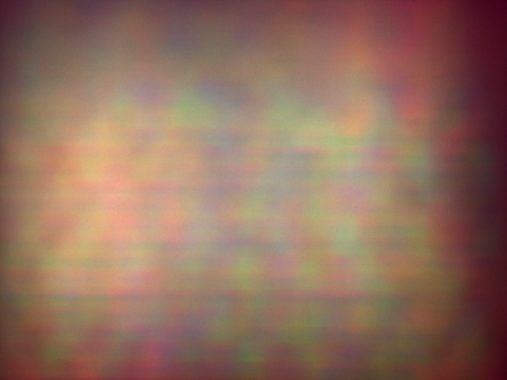}\\
		\rule{0ex}{0.15in}%
	}
	\rule{0.2in}{0ex}}
&
\includegraphics[width=\figsizegenceleba\linewidth,valign=m]{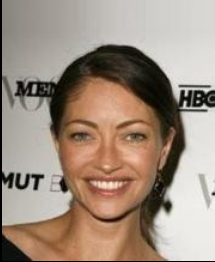}
\llap{\shortstack[l]{%
		\hspace{-0.8cm}\includegraphics[scale=.035]{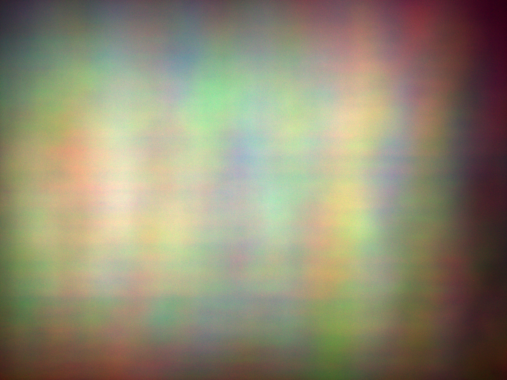}\\
		\rule{0ex}{0.15in}%
	}
	\rule{0.2in}{0ex}}
    \\[\newlinegen]
    \hline
    ADMM100
&\hspace{-0.35em}\includegraphics[width=\figsizegendiffusercam\linewidth,valign=m]{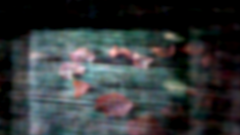}
&\hspace{-0.35em}\includegraphics[width=\figsizegendiffusercam\linewidth,valign=m]{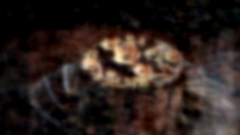}
&\hspace{-0.35em}\includegraphics[width=\figsizegen\linewidth,valign=m]{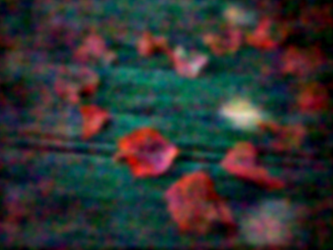}
&\hspace{-0.35em}\includegraphics[width=\figsizegen\linewidth,valign=m]{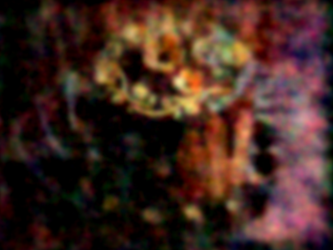}
&\hspace{-0.35em}\includegraphics[width=\figsizegen\linewidth,valign=m]{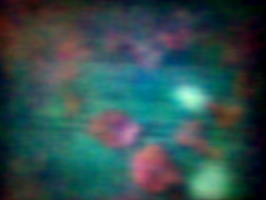}
&\hspace{-0.35em}\includegraphics[width=\figsizegen\linewidth,valign=m]{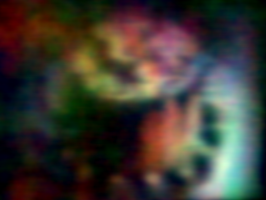}
&\hspace{-0.35em}\includegraphics[width=\figsizegenceleba\linewidth,valign=m]{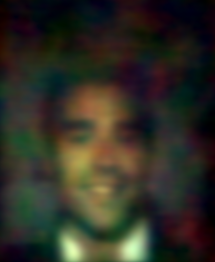}
&\hspace{-0.35em}\includegraphics[width=\figsizegenceleba\linewidth,valign=m]{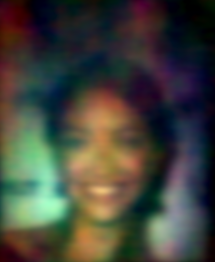}
    \\[\newlinegen]
\makecell{TrainInv\\+$\text{Post}_{8}$~\cite{9239993}}
&\hspace{-0.35em}\includegraphics[width=\figsizegendiffusercam\linewidth,valign=m]{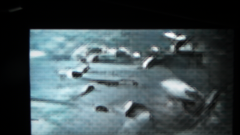}
&\hspace{-0.35em}\includegraphics[width=\figsizegendiffusercam\linewidth,valign=m]{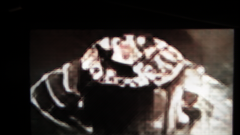}
&\hspace{-0.35em}\includegraphics[width=\figsizegen\linewidth,valign=m]{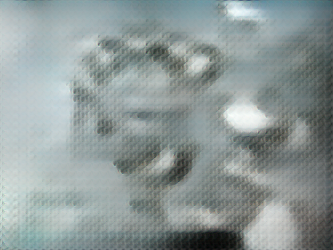}
&\hspace{-0.35em}\includegraphics[width=\figsizegen\linewidth,valign=m]{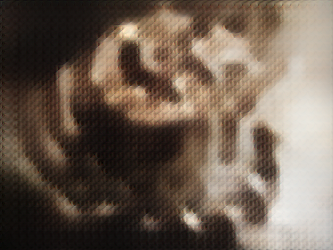}
&\hspace{-0.35em}\includegraphics[width=\figsizegen\linewidth,valign=m]{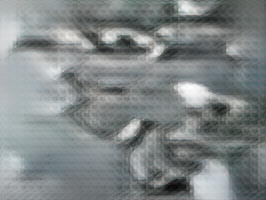}
&\hspace{-0.35em}\includegraphics[width=\figsizegen\linewidth,valign=m]{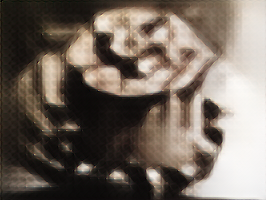}
&\hspace{-0.35em}\includegraphics[width=\figsizegenceleba\linewidth,valign=m]{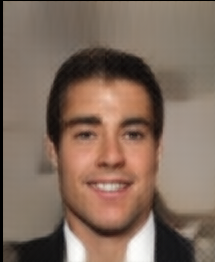}
&\hspace{-0.35em}\includegraphics[width=\figsizegenceleba\linewidth,valign=m]{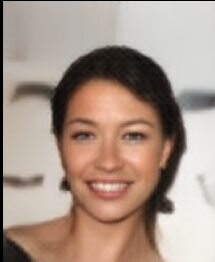}
    \\[\newlinegen]
\makecell{$\text{MMCN}_{4}$\\+$\text{Post}_{4}$~\cite{9546648}}
&\hspace{-0.35em}\includegraphics[width=\figsizegendiffusercam\linewidth,valign=m]{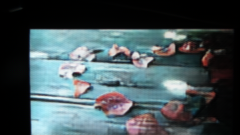}
&\hspace{-0.35em}\includegraphics[width=\figsizegendiffusercam\linewidth,valign=m]{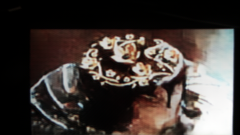}
&\hspace{-0.35em}\includegraphics[width=\figsizegen\linewidth,valign=m]{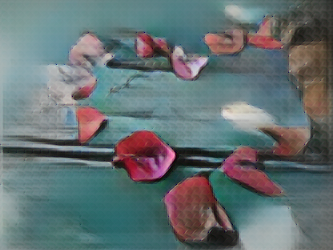}
&\hspace{-0.35em}\includegraphics[width=\figsizegen\linewidth,valign=m]{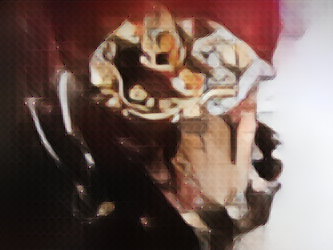}
&\hspace{-0.35em}\includegraphics[width=\figsizegen\linewidth,valign=m]{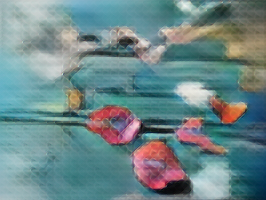}
&\hspace{-0.35em}\includegraphics[width=\figsizegen\linewidth,valign=m]{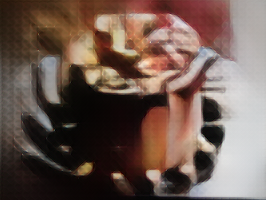}
&\hspace{-0.35em}\includegraphics[width=\figsizegenceleba\linewidth,valign=m]{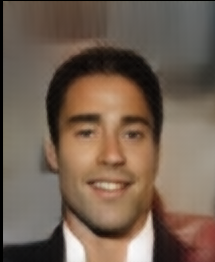}
&\hspace{-0.35em}\includegraphics[width=\figsizegenceleba\linewidth,valign=m]{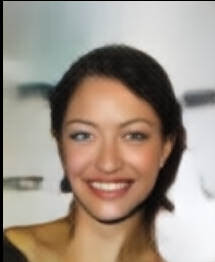}
    \\[\newlinegen]

\makecell{$\text{Pre}_{8}$+\\LeADMM5}
&\hspace{-0.35em}\includegraphics[width=\figsizegendiffusercam\linewidth,valign=m]{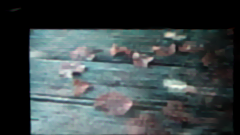}
&\hspace{-0.35em}\includegraphics[width=\figsizegendiffusercam\linewidth,valign=m]{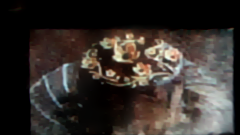}
&\hspace{-0.35em}\includegraphics[width=\figsizegen\linewidth,valign=m]{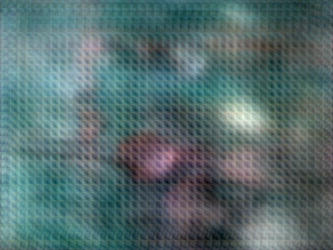}
&\hspace{-0.35em}\includegraphics[width=\figsizegen\linewidth,valign=m]{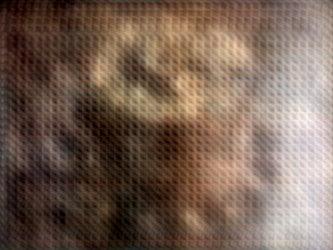}
&\hspace{-0.35em}\includegraphics[width=\figsizegen\linewidth,valign=m]{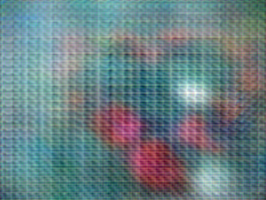}
&\hspace{-0.35em}\includegraphics[width=\figsizegen\linewidth,valign=m]{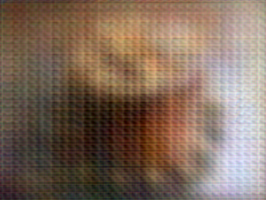}
&\hspace{-0.35em}\includegraphics[width=\figsizegenceleba\linewidth,valign=m]{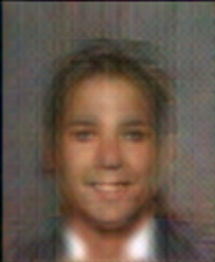}
&\hspace{-0.35em}\includegraphics[width=\figsizegenceleba\linewidth,valign=m]{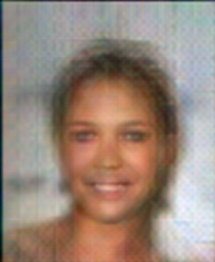}
    \\[\newlinegen]
    
\makecell{LeADMM5\\+$\text{Post}_{8}$~\cite{Monakhova:19}}
&\hspace{-0.35em}\includegraphics[width=\figsizegendiffusercam\linewidth,valign=m]{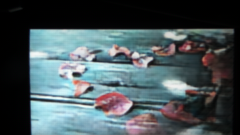}
&\hspace{-0.35em}\includegraphics[width=\figsizegendiffusercam\linewidth,valign=m]{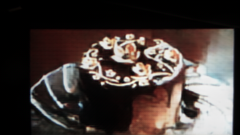}
&\hspace{-0.35em}\includegraphics[width=\figsizegen\linewidth,valign=m]{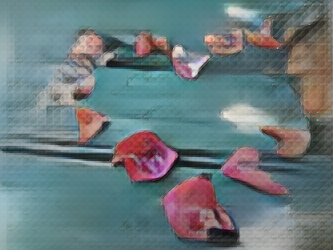}
&\hspace{-0.35em}\includegraphics[width=\figsizegen\linewidth,valign=m]{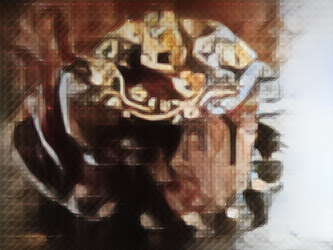}
&\hspace{-0.35em}\includegraphics[width=\figsizegen\linewidth,valign=m]{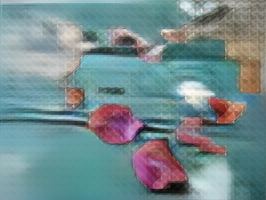}
&\hspace{-0.35em}\includegraphics[width=\figsizegen\linewidth,valign=m]{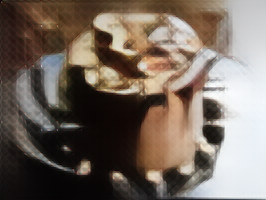}
&\hspace{-0.35em}\includegraphics[width=\figsizegenceleba\linewidth,valign=m]{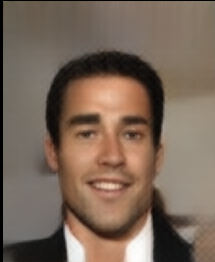}
&\hspace{-0.35em}\includegraphics[width=\figsizegenceleba\linewidth,valign=m]{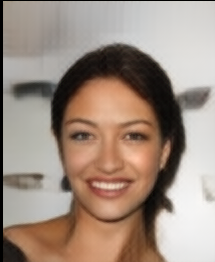}
    \\[\newlinegen]
    
\makecell{$\text{MWDN}_{8}$~\cite{Li:23}}
&\hspace{-0.35em}\includegraphics[width=\figsizegendiffusercam\linewidth,valign=m]{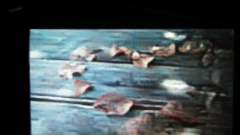}
&\hspace{-0.35em}\includegraphics[width=\figsizegendiffusercam\linewidth,valign=m]{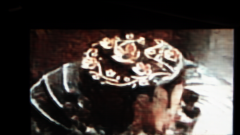}
&\hspace{-0.35em}\includegraphics[width=\figsizegen\linewidth,valign=m]{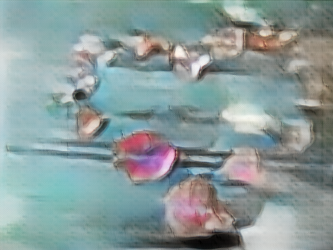}
&\hspace{-0.35em}\includegraphics[width=\figsizegen\linewidth,valign=m]{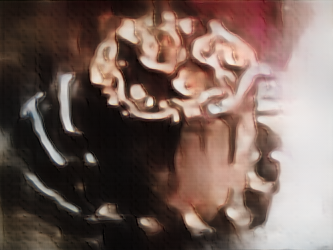}
&\hspace{-0.35em}\includegraphics[width=\figsizegen\linewidth,valign=m]{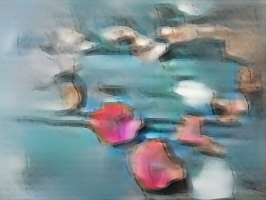}
&\hspace{-0.35em}\includegraphics[width=\figsizegen\linewidth,valign=m]{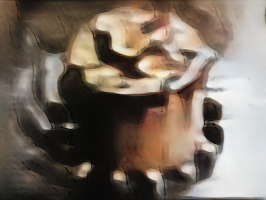}
&\hspace{-0.35em}\includegraphics[width=\figsizegenceleba\linewidth,valign=m]{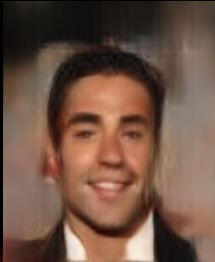}
&\hspace{-0.35em}\includegraphics[width=\figsizegenceleba\linewidth,valign=m]{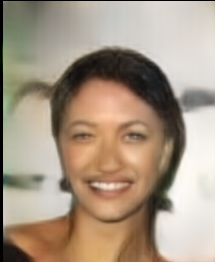}
    \\[\newlinegen]
    
\hline
\makecell{$\text{Pre}_{4}$+\\TrainInv\\+$\text{Post}_{4}$}
&\hspace{-0.35em}\includegraphics[width=\figsizegendiffusercam\linewidth,valign=m]{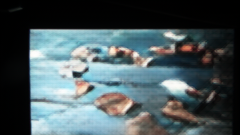}
&\hspace{-0.35em}\includegraphics[width=\figsizegendiffusercam\linewidth,valign=m]{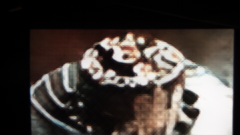}
&\hspace{-0.35em}\includegraphics[width=\figsizegen\linewidth,valign=m]{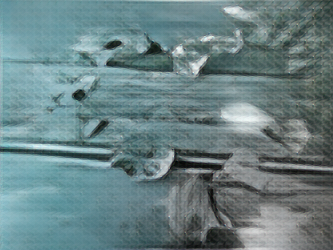}
&\hspace{-0.35em}\includegraphics[width=\figsizegen\linewidth,valign=m]{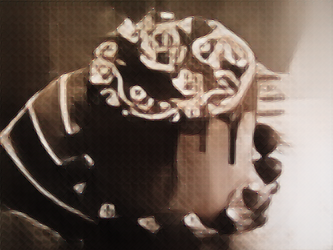}
&\hspace{-0.35em}\includegraphics[width=\figsizegen\linewidth,valign=m]{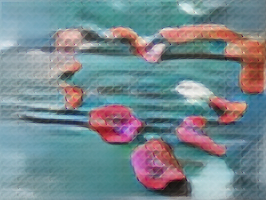}
&\hspace{-0.35em}\includegraphics[width=\figsizegen\linewidth,valign=m]{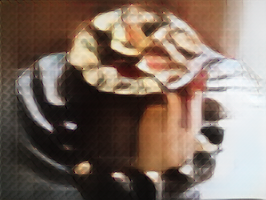}
&\hspace{-0.35em}\includegraphics[width=\figsizegenceleba\linewidth,valign=m]{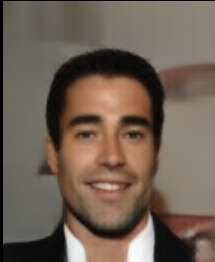}
&\hspace{-0.35em}\includegraphics[width=\figsizegenceleba\linewidth,valign=m]{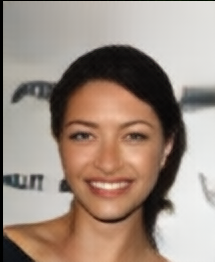}
    \\[\newlinegen]
\makecell{$\text{Pre}_2$+\\$\text{MMCN}_{4}$\\+$\text{Post}_2$}
&\hspace{-0.35em}\includegraphics[width=\figsizegendiffusercam\linewidth,valign=m]{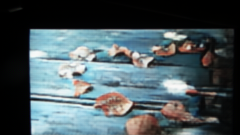}
&\hspace{-0.35em}\includegraphics[width=\figsizegendiffusercam\linewidth,valign=m]{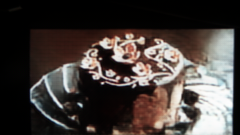}
&\hspace{-0.35em}\includegraphics[width=\figsizegen\linewidth,valign=m]{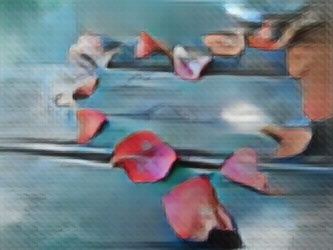}
&\hspace{-0.35em}\includegraphics[width=\figsizegen\linewidth,valign=m]{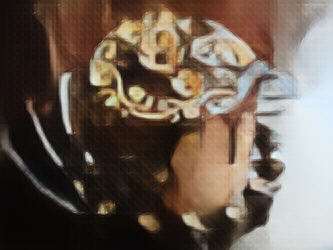}
&\hspace{-0.35em}\includegraphics[width=\figsizegen\linewidth,valign=m]{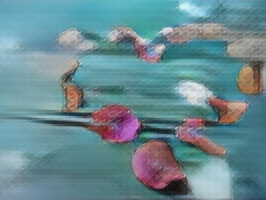}
&\hspace{-0.35em}\includegraphics[width=\figsizegen\linewidth,valign=m]{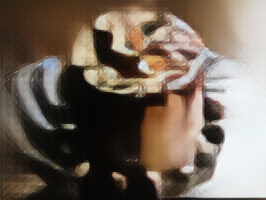}
&\hspace{-0.35em}\includegraphics[width=\figsizegenceleba\linewidth,valign=m]{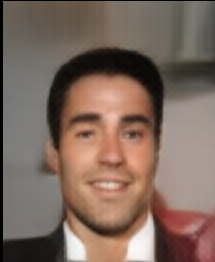}
&\hspace{-0.35em}\includegraphics[width=\figsizegenceleba\linewidth,valign=m]{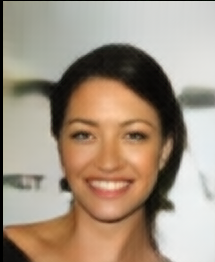}
    \\[\newlinegen]
\makecell{$\text{Pre}_{4}$+\\LeADMM5\\+$\text{Post}_{4}$}
&\hspace{-0.35em}\includegraphics[width=\figsizegendiffusercam\linewidth,valign=m]{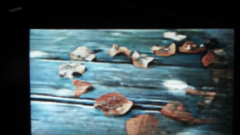}
&\hspace{-0.35em}\includegraphics[width=\figsizegendiffusercam\linewidth,valign=m]{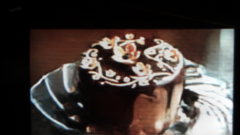}
&\hspace{-0.35em}\includegraphics[width=\figsizegen\linewidth,valign=m]{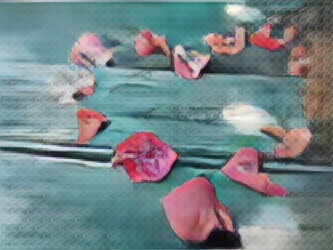}
&\hspace{-0.35em}\includegraphics[width=\figsizegen\linewidth,valign=m]{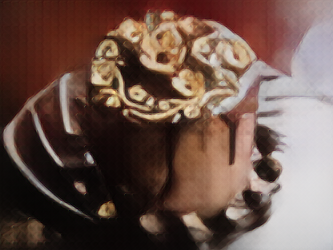}
&\hspace{-0.35em}\includegraphics[width=\figsizegen\linewidth,valign=m]{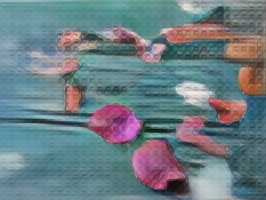}
&\hspace{-0.35em}\includegraphics[width=\figsizegen\linewidth,valign=m]{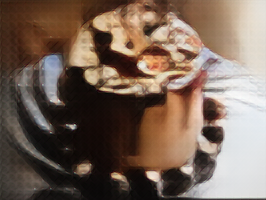}
&\hspace{-0.35em}\includegraphics[width=\figsizegenceleba\linewidth,valign=m]{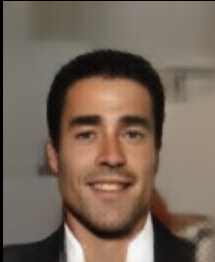}
&\hspace{-0.35em}\hspace{-0.5em}\includegraphics[width=\figsizegenceleba\linewidth,valign=m]{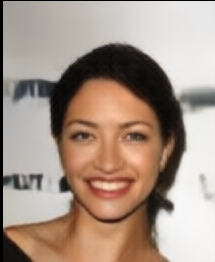}
    \\
	\end{tabular}
	\caption{Visual comparison of reconstructions on test set examples of datasets of different mask types (amplitude and phase). Models are trained on the corresponding training set of each dataset/system.}
    \label{fig:exp1_visual_comparison}
\end{figure*}

\noindent In this experiment, we demonstrate the benefit of the pre-processor for multiple camera inversion techniques and across multiple datasets of three different imaging systems.
We compare three camera inversion approaches with and without a pre-processor: ADMM with 5 unrolled layers (\textit{LeADMM5})~\cite{Monakhova:19}, trainable inversion (\textit{TrainInv})~\cite{9239993}, and ADMM with 5 unrolled layers and a model mismatch compensation network ($\textit{MMCN}_{4}$)~\cite{9546648}.
For multi-Wiener deconvolution network with PSF correction ($\textit{MWDN}_{8}$)~\cite{Li:23}, 
we do not add a pre- nor post-processor as its architecture already contains convolutional layers before and after (multiple) camera inversions.

\cref{tab:exp1_benchmark} presents image quality metrics for all reconstruction approaches and across four datasets.
For all approaches (\textit{LeADMM5}, \textit{TrainInv}, $\textit{MMCN}_{4}$) and across all datasets, we see improved performance when splitting the parameters between the pre- and post-processors.
This improvement is visualized and quantified in \cref{fig:exp1_improvement_viz}.
There is a slight decrease in PSNR for \textit{DiffuserCam} with $\textit{MMCN}_{4}$ but both SSIM and LPIPS improve.
We observe significant improvement when using \textit{TrainInv} for camera inversion.
This is confirmed by looking at a few outputs from the test sets in \cref{fig:exp1_visual_comparison}.
With just a post-processor, (\textit{TrainInv}+$\textit{Post}_{8}$) has difficulty in faithfully recovering the colors of the original image, but adding a pre-processor helps to reproduce the original colors.

Equating the number of parameters across models helps to identify which techniques lead to improved performance.
For example, we observe that ($\textit{MMCN}_{4}$+$\textit{Post}_{4}$) is worse than \textit{uncompensated} unrolled ADMM (\textit{LeADMM5}+$\textit{Post}_{8}$).
This indicates that using a more performant post-processor is better at handling model mismatch
than adding a compensation network.
With ($\textit{Pre}_{8}$+\textit{LeADMM5}), we put all the neural network parameters in the pre-processor.
While better than \textit{ADMM100}, it lacks the denoising and perceptual enhancements that a post-processor can offer.
For $\textit{MWDN}_{8}$, with respect to approaches that \textit{do not} use a pre-processor, we only observe improved performance for the \textit{DiffuserCam} dataset, as shown by the original authors~\cite{Li:23}.
For other datasets, $\textit{MWDN}_{8}$ is noticeably worse; likely because it is more sensitive to noise from the low-cost hardware of our datasets, and due to multiple camera inversions in its approach (see \cref{fig:multiwiener}).

When performing imaging for a specific type of data content, \eg reconstruction faces with \textit{DigiCam-CelebA} rather than general-purpose imaging with \textit{DigiCam-Single}, we observe a significant improvement in performance: \SI{2.9}{\decibel} improvement in PSNR, and \SI{32}{\percent} and \SI{41}{\percent} relative improvement in SSIM and LPIPS respectively (for $\textit{Pre}_{4}$+\textit{LeADMM5}+$\textit{Post}_{4}$).

\subsubsection{Improving Camera Inversion} Using unrolled ADMM for camera inversion has more flexibility when it comes to improving the camera inversion capabilities.
Improving the capacity of MMCN and MWDN requires introducing a large amount of model parameters, while unrolled ADMM only requires four more hyperparameters per unrolled layer.
By simply adding five more unrolled layers (\ie~just 20 more parameters) for \textit{LeADMM10}, we can further improve results ($\textit{Pre}_{4}$+\textit{LeADMM10}+$\textit{Post}_{4}$ row of \cref{tab:exp1_benchmark}).
This however comes at a cost in inference time.

To directly address model mismatch,
we can add a \textit{PSF correction} network as shown in~\cref{fig:pipeline}.
A similar approach is used by MWDN~\cite{Li:23}, in which the input PSF is fed to a downscaling network (see \cref{fig:multiwiener}).
For the last row in \cref{tab:exp1_benchmark},
we feed the PSF to a DRUNet with (4, 8, 16, 32) feature representation channels (128K parameters) and slightly decrease the pre-processor size to (32, 64, 112, 128) channels (3.9M parameters).
Intermediate outputs, \ie after the pre-processor,
camera inversion, and PSF correction, can be found in \cref{app:intermediate}.

\subsubsection{Inference Time} In \cref{tab:exp1_benchmark}, we report average inference time computed over 100 trials on an Intel Xeon E5-2680 v3 \SI{2.5}{\giga\hertz} CPU with a single Nvidia Titan X Pascal GPU.
Using unrolled ADMM and MMCN is significantly slower due to multiple iterations, while approaches based on inverse/Wiener filtering (\textit{TrainInv} and MWDN) are much faster.

\subsection{Improved Robustness}
\label{sec:robustness_exp}

\noindent In these experiments, we demonstrate the improved robustness of our modular approach by numerically varying the noise sources:
(1) the measurement noise~$\bm{n}$ and (2) the model mismatch $\bm{\Delta}_H$.
We perform these experiments on the \textit{DiffuserCam} dataset.
For both experiments, intermediate outputs can be found in \cref{app:intermediate}.

\subsubsection{Shot Noise}
\label{sec:shot_noise_exp}
During training, we add shot noise (\ie~signal-dependent noise following a Poisson distribution) at an SNR of \SI{10}{\decibel}, which is representative of a low-light/photon scenario. 
We evaluate at different SNRs to determine robustness to variations of the input SNR.
\cref{tab:exp2_robustness} shows average test set metrics,
and \cref{fig:robustness_10db} shows example outputs.
The model that does not use a pre-processor (\textit{LeADMM5}+$\textit{Post}_{8}$) is unable to recover high frequency details,
and the image quality metrics are significantly worse (\cref{tab:exp2_robustness}).
Incorporating a pre-processor is capable of recovering such details ($\textit{Pre}_{4}$+\textit{LeADMM5}+$\textit{Post}_{4}$), 
and is robust to SNRs lower than the one used at training.

\subsubsection{Model Mismatch}
\label{sec:mismatch_exp}
To evaluate robustness to model mismatch,
we digitally add Gaussian noise to \textit{DiffuserCam}'s PSF at multiple SNRs,
as shown in the first row of \cref{fig:robustness_psf_err}.
The remaining rows show example outputs,
and \cref{tab:exp2_robustness_psf} presents average test set metrics on the clean \textit{DiffuserCam} dataset.
Using both a pre- and post-processor is more robust to the increasing mismatch in the PSF than just using a post-processor,
and re-allocating some of the pre-processor parameters to PSF correction further improves performance.

\begin{table}[!t]
	\renewcommand{\arraystretch}{1.3}
	\caption{Average image quality metrics (PSNR $\uparrow$ / SSIM $\uparrow$ / LPIPS $\downarrow$) on models (each column) that have been trained on the \textit{DiffuserCam} dataset with a signal-to-noise ratio (SNR) of \SI{10}{\decibel} (digitally-added Poisson noise). At test time, Poisson noise is added according to the SNR in the left-most column.}
	\label{tab:exp2_robustness}
	\centering
	\begin{tabular}{c||c|c}
		\hline
		 Test SNR & LeADMM5+$\text{Post}_{8}$~\cite{Monakhova:19} & $\text{Pre}_{4}$+LeADMM5+$\text{Post}_{4}$ \\
		\hline\hline
		 \SI{0}{\decibel} & 16.4 / 0.569 / 0.345 &  \textbf{20.0} / \textbf{0.755} / \textbf{0.230} \\
         \hline
		\SI{5}{\decibel} & 18.2 / 0.627 / 0.316 &  \textbf{23.9} / \textbf{0.818} / \textbf{0.186} \\
		\hline
		 \makecell{\SI{10}{\decibel}\\(Train SNR)} & 19.4 / 0.672 / 0.290 & \textbf{24.6} / \textbf{0.827}  / \textbf{0.176}  \\
		\hline
        \SI{15}{\decibel} & 19.7 / 0.687 / 0.282  & \textbf{23.7} / \textbf{0.820} / \textbf{0.184} \\
		\hline
        \SI{20}{\decibel} & 19.7 / 0.690 / 0.281  & \textbf{21.6} / \textbf{0.784} / \textbf{0.215} \\
		\hline
	\end{tabular}
\end{table}

\begin{table}[!t]
	\renewcommand{\arraystretch}{1.2}
	\caption{Average image quality metrics (PSNR $\uparrow$ / SSIM $\uparrow$ / LPIPS $\downarrow$) on models (each column) that have been trained on the \textit{DiffuserCam} dataset with Gaussian noise added to the PSF (according to SNR in left-most column). Corrupted PSFs can be seen in~\cref{fig:robustness_psf_err}.}
	\label{tab:exp2_robustness_psf}
	\centering
	\scalebox{0.85}{\begin{tabular}{c||c|c|c}
		\hline
		 PSF SNR & \makecell{LeADMM5\\+$\text{Post}_{8}$~\cite{Monakhova:19}} & \makecell{$\text{Pre}_{4}$+LeADMM5\\+$\text{Post}_{4}$} & \makecell{$\text{Pre}_{4}$+LeADMM5\\+$\text{Post}_{4}$ (PSF correction)} \\
		\hline\hline
		 Clean & 23.8 / 0.806 / 0.202 &  25.3 / 0.838 / 0.171 & \textbf{26.4} / \textbf{0.857} / \textbf{0.154}  \\
         \hline
		\SI{0}{\decibel} & 23.1 / 0.781 / 0.222 & 24.7 / 0.827 / 0.181 & \textbf{26.2} / \textbf{0.853} / \textbf{0.159} \\
		\hline
        \SI{-10}{\decibel} & 22.3 / 0.750 / 0.250 &  24.4 / 0.818 / 0.193 & \textbf{25.7} / \textbf{0.849} / \textbf{0.164} \\
		\hline
        \SI{-20}{\decibel} & 20.2 / 0.673 / 0.297 & 23.4 / 0.790 / 0.215 & \textbf{26.2} / \textbf{0.858} / \textbf{0.155} \\
		\hline
	\end{tabular}}
\end{table}

\newcommand{\figsizeood}{0.135}
\newcommand{\newlineood}{14pt}
\begin{figure*}[t!]
\centering
	\begingroup
	\renewcommand{\arraystretch}{1} 
	\setlength{\tabcolsep}{0.08em} 
	\begin{tabular}{c cccccc}
		  &\multicolumn{2}{c}{\SI{0}{\decibel}} & \multicolumn{2}{c}{\makecell{\SI{10}{\decibel}\\(train SNR)}}  & \multicolumn{2}{c}{\SI{20}{\decibel}}\\

          \cmidrule(r){2-3} \cmidrule(r){4-5} \cmidrule(r){6-7} 
    
\makecell{Raw data} &
\includegraphics[width=\figsizeood\linewidth,valign=m]{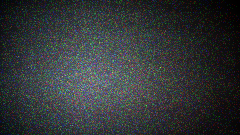}
  & 
  \includegraphics[width=\figsizeood\linewidth,valign=m]{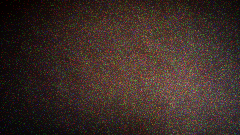}

  &\includegraphics[width=\figsizeood\linewidth,valign=m]{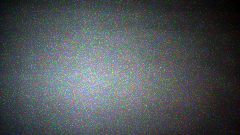}
  & 
  \includegraphics[width=\figsizeood\linewidth,valign=m]{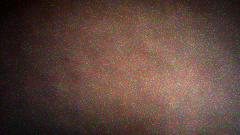}

  & \includegraphics[width=\figsizeood\linewidth,valign=m]{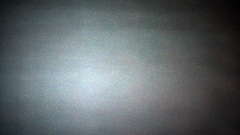}
  &\includegraphics[width=\figsizeood\linewidth,valign=m]{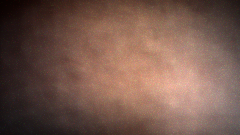}
\\[\newlineood]

\hline

\makecell{LeADMM5\\+$\text{Post}_{8}$~\cite{Monakhova:19}} &
\includegraphics[width=\figsizeood\linewidth,valign=m]{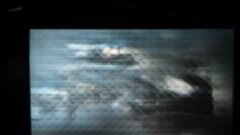}
& 
\includegraphics[width=\figsizeood\linewidth,valign=m]{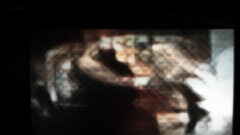}
& \includegraphics[width=\figsizeood\linewidth,valign=m]{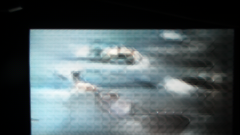}
& \includegraphics[width=\figsizeood\linewidth,valign=m]{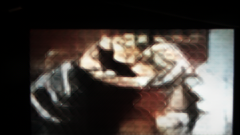}
  & \includegraphics[width=\figsizeood\linewidth,valign=m]{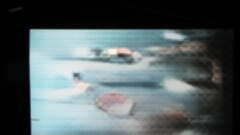}
  &\includegraphics[width=\figsizeood\linewidth,valign=m]{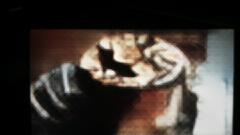}
\\[\newlineood]

\makecell{$\text{Pre}_{4}$\\+LeADMM5\\+$\text{Post}_{4}$} & \includegraphics[width=\figsizeood\linewidth,valign=m]{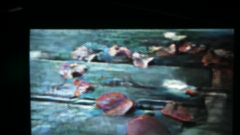}
& \includegraphics[width=\figsizeood\linewidth,valign=m]{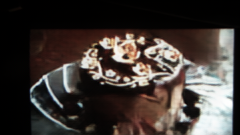}
& \includegraphics[width=\figsizeood\linewidth,valign=m]{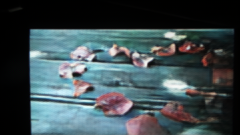}
& \includegraphics[width=\figsizeood\linewidth,valign=m]{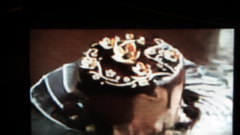}
  & \includegraphics[width=\figsizeood\linewidth,valign=m]{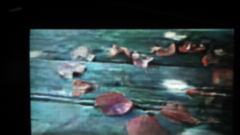}
  &\includegraphics[width=\figsizeood\linewidth,valign=m]{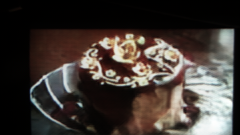}
\\
	\end{tabular}
	\endgroup
	\caption{Example outputs of applying (\textit{LeADMM5+$\text{Post}_{8}$}) and (\textit{$\text{Pre}_{4}$+LeADMM5+$\text{Post}_{4}$}) at various signal-to-noise ratios (SNRs). Both approaches are trained with lensless measurements where Poisson noise is added according to an SNR \SI{10}{\decibel}.}
  \label{fig:robustness_10db}
\end{figure*}

\newcommand{\figsizepsferr}{0.135}
\begin{figure*}[t!]
\centering
	\begingroup
	\renewcommand{\arraystretch}{1} 
	\setlength{\tabcolsep}{0.08em} 
	\begin{tabular}{ccccccc}
          & \multicolumn{2}{c}{\SI{0}{\decibel}} & \multicolumn{2}{c}{\makecell{\SI{-10}{\decibel}}} & \multicolumn{2}{c}{\SI{-20}{\decibel}}\\

          \cmidrule(r){2-3} \cmidrule(r){4-5} \cmidrule(r){6-7} 
    
\makecell{PSF} 
  & 
  \multicolumn{2}{c}{\includegraphics[width=0.18\linewidth,valign=m]{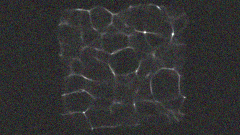}}
&\multicolumn{2}{c}{\includegraphics[width=0.18\linewidth,valign=m]{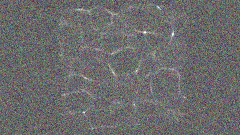}}
  & \multicolumn{2}{c}{\includegraphics[width=0.18\linewidth,valign=m]{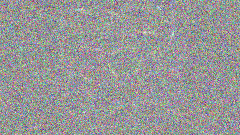}}
\\[20pt]

\hline

\makecell{LeADMM5\\+$\text{Post}_{8}$~\cite{Monakhova:19}} 
& \includegraphics[width=\figsizepsferr\linewidth,valign=m]{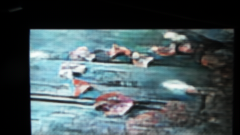}
& 
\includegraphics[width=\figsizepsferr\linewidth,valign=m]{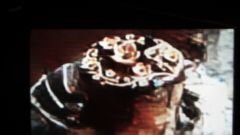}
& \includegraphics[width=\figsizepsferr\linewidth,valign=m]{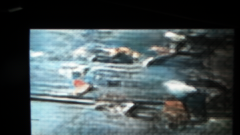}
& 
\includegraphics[width=\figsizepsferr\linewidth,valign=m]{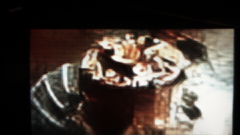}
& \includegraphics[width=\figsizepsferr\linewidth,valign=m]{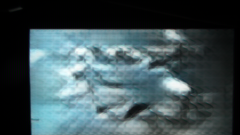}
& 
\includegraphics[width=\figsizepsferr\linewidth,valign=m]{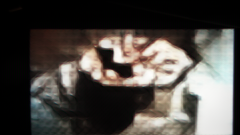}
\\[\newlineood]

\makecell{$\text{Pre}_{4}$+LeADMM5\\+$\text{Post}_{4}$} 
& \includegraphics[width=\figsizepsferr\linewidth,valign=m]{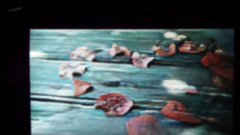}
& 
\includegraphics[width=\figsizepsferr\linewidth,valign=m]{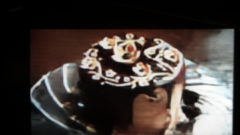}
& \includegraphics[width=\figsizepsferr\linewidth,valign=m]{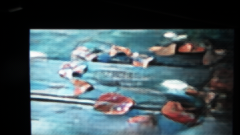}
& 
\includegraphics[width=\figsizepsferr\linewidth,valign=m]{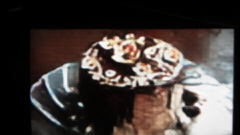}
& \includegraphics[width=\figsizepsferr\linewidth,valign=m]{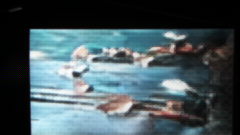}
& 
\includegraphics[width=\figsizepsferr\linewidth,valign=m]{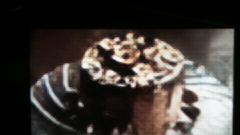}
\\[\newlineood]

\makecell{$\text{Pre}_{4}$+LeADMM5\\+$\text{Post}_{4}$ (PSF corr.)} 
& \includegraphics[width=\figsizepsferr\linewidth,valign=m]{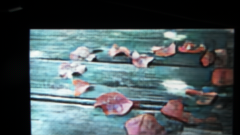}
& 
\includegraphics[width=\figsizepsferr\linewidth,valign=m]{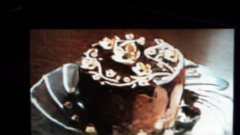}
& \includegraphics[width=\figsizepsferr\linewidth,valign=m]{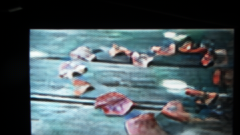}
& 
\includegraphics[width=\figsizepsferr\linewidth,valign=m]{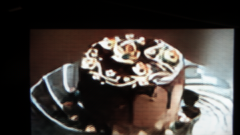}
& \includegraphics[width=\figsizepsferr\linewidth,valign=m]{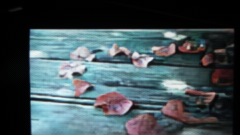}
& 
\includegraphics[width=\figsizepsferr\linewidth,valign=m]{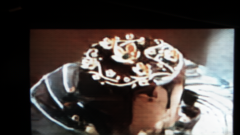}
\\
	\end{tabular}
	\endgroup
	\caption{Example outputs of various reconstruction approach that have been trained on digitally-corrupted PSFs at various signal-to-noise ratios (SNRs) to evaluate robustness to model mismatch.}
  \label{fig:robustness_psf_err}
\end{figure*}

\subsection{Evaluating Generalizability to PSF Changes}
\label{sec:exp_eval_gen}

\newcommand{\figsizegentrans}{0.083}
\newcommand{\figsizegencelebatrans}{0.07}
\begin{figure*}[t!]
\centering
	\renewcommand{\arraystretch}{1} 
	\setlength{\tabcolsep}{0.1em} 
	\begin{tabular}{c|cc|cc|cc||cc|cc}
    \makecell{\textit{Train set} $\rightarrow$\\\textit{Test set} $\downarrow$}
    &   
    \multicolumn{2}{c|}{DiffuserCam}
    & \multicolumn{2}{c|}{TapeCam}
    & \multicolumn{2}{c||}{DigiCam-Single}
    & \multicolumn{2}{c|}{\makecell{ADMM100\\(no training)}}
    & 
    \multicolumn{2}{c}{Ground-truth}
    \\
\hline 
\makecell{DiffuserCam}
&\includegraphics[width=\figsizegentrans\linewidth,valign=m]{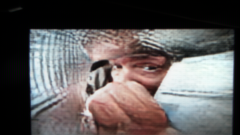}
&\includegraphics[width=\figsizegentrans\linewidth,valign=m]{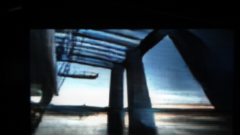}
&\includegraphics[width=\figsizegentrans\linewidth,valign=m]{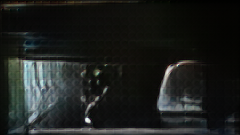}
&\includegraphics[width=\figsizegentrans\linewidth,valign=m]{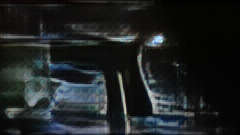}
&\includegraphics[width=\figsizegentrans\linewidth,valign=m]{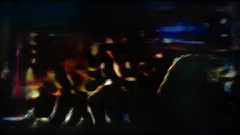}
&\includegraphics[width=\figsizegentrans\linewidth,valign=m]{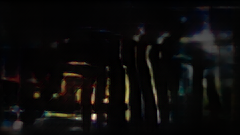}
&\includegraphics[width=\figsizegentrans\linewidth,valign=m]{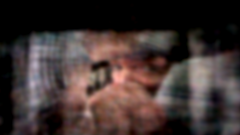}
 
&\includegraphics[width=\figsizegentrans\linewidth,valign=m]{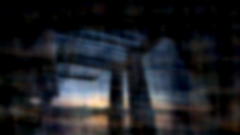}

&\includegraphics[width=\figsizegentrans\linewidth,valign=m]{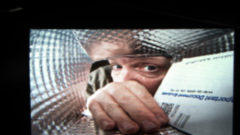}
&\includegraphics[width=\figsizegentrans\linewidth,valign=m]{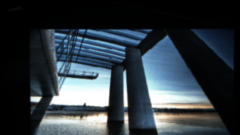}
    \\[6pt]
\hline 
\makecell{TapeCam}
&\includegraphics[width=\figsizegentrans\linewidth,valign=m]{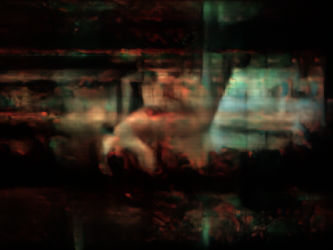}
&\includegraphics[width=\figsizegentrans\linewidth,valign=m]{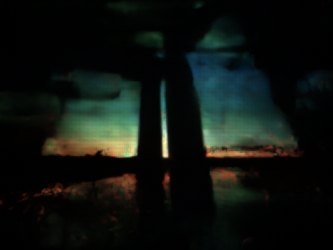}
&\includegraphics[width=\figsizegentrans\linewidth,valign=m]{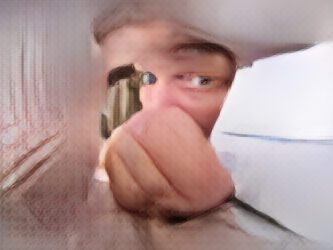}
&\includegraphics[width=\figsizegentrans\linewidth,valign=m]{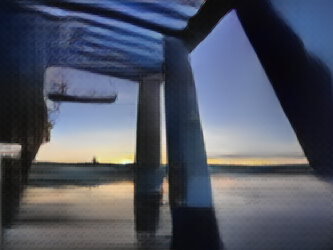}
&\includegraphics[width=\figsizegentrans\linewidth,valign=m]{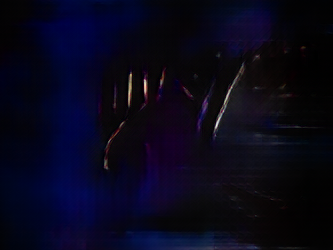}
&\includegraphics[width=\figsizegentrans\linewidth,valign=m]{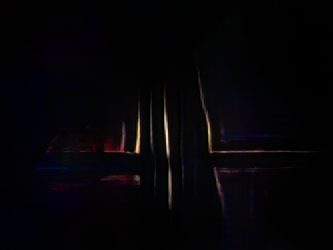}
&\includegraphics[width=\figsizegentrans\linewidth,valign=m]{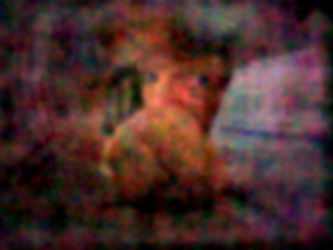}
&\includegraphics[width=\figsizegentrans\linewidth,valign=m]{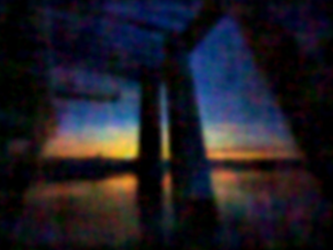}

&\includegraphics[width=\figsizegentrans\linewidth,valign=m]{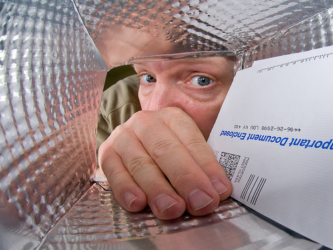}
&\includegraphics[width=\figsizegentrans\linewidth,valign=m]{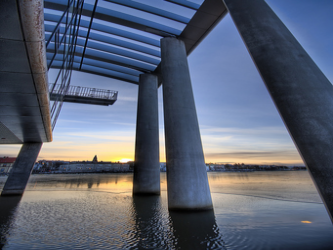}
\\[10pt]
\hline
\makecell{DigiCam\\-Single}
&\includegraphics[width=\figsizegentrans\linewidth,valign=m]{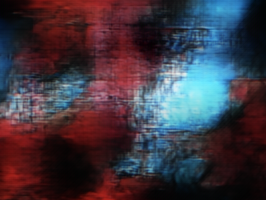}
&\includegraphics[width=\figsizegentrans\linewidth,valign=m]{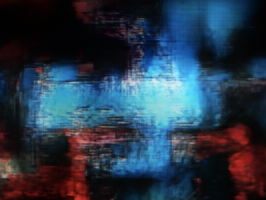}
&\includegraphics[width=\figsizegentrans\linewidth,valign=m]{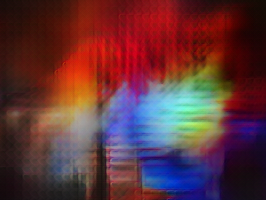}
&\includegraphics[width=\figsizegentrans\linewidth,valign=m]{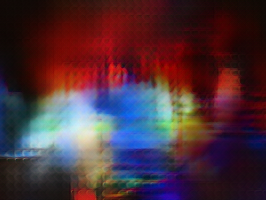}
&\includegraphics[width=\figsizegentrans\linewidth,valign=m]{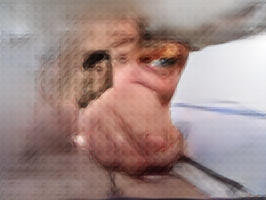}
&\includegraphics[width=\figsizegentrans\linewidth,valign=m]{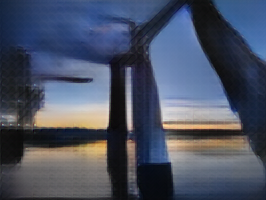}
&\includegraphics[width=\figsizegentrans\linewidth,valign=m]{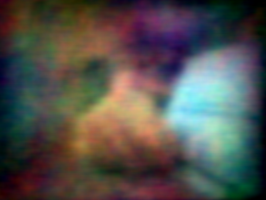}
&\includegraphics[width=\figsizegentrans\linewidth,valign=m]{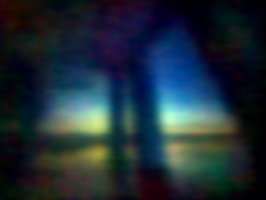}

&\includegraphics[width=\figsizegentrans\linewidth,valign=m]{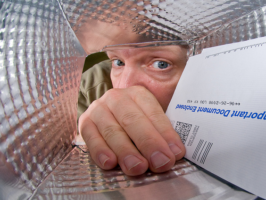}
&\includegraphics[width=\figsizegentrans\linewidth,valign=m]{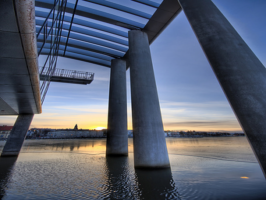}
\\[10pt]
\hline
\hline
\makecell{DigiCam\\-Multi}
&\includegraphics[width=\figsizegentrans\linewidth,valign=m]{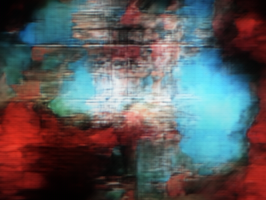}
&\includegraphics[width=\figsizegentrans\linewidth,valign=m]{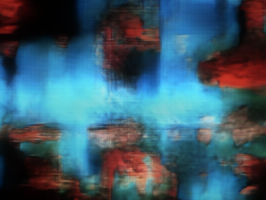}
&\includegraphics[width=\figsizegentrans\linewidth,valign=m]{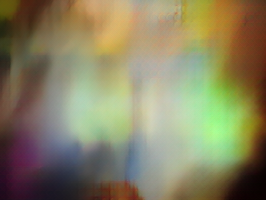}
&\includegraphics[width=\figsizegentrans\linewidth,valign=m]{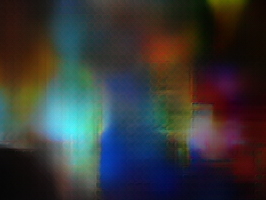}
&\includegraphics[width=\figsizegentrans\linewidth,valign=m]{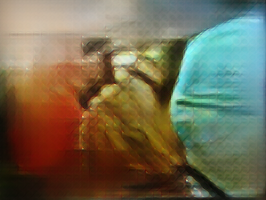}
&\includegraphics[width=\figsizegentrans\linewidth,valign=m]{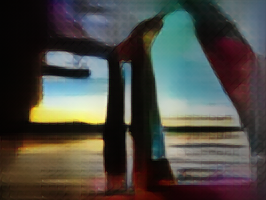}
&\includegraphics[width=\figsizegentrans\linewidth,valign=m]{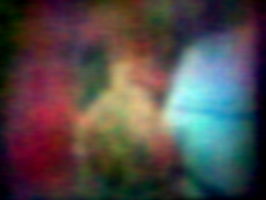}
&\includegraphics[width=\figsizegentrans\linewidth,valign=m]{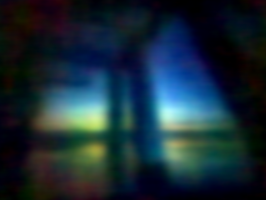}

&\includegraphics[width=\figsizegentrans\linewidth,valign=m]{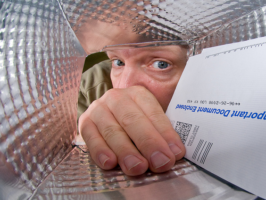}
& \hspace{-0.5em}\includegraphics[width=\figsizegentrans\linewidth,valign=m]{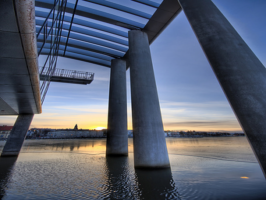}

\\

	\end{tabular}
	\caption{Example outputs of ($\textit{Pre}_{4}$\textit{+LeADMM5+}$\textit{Post}_{4}$) trained on the system/dataset indicated along the columns, and evaluated on the system/dataset indicated along the rows.}
    \label{fig:exp2_visual_comparison}
\end{figure*}

\noindent While the results of learned reconstruction approach can improve significantly from classical techniques such as ADMM, it has not been studied how such approaches generalize to other systems, \eg if the PSF changes.
Before applying techniques for improving the generalizability to measurements of unseen PSFs (\cref{sec:exp_improve_gen}), in this section we first benchmark the learned reconstructions from the previous section.

In \cref{fig:exp2_visual_comparison}, ($\textit{Pre}_{4}$\textit{+LeADMM5+}$\textit{Post}_{4}$) that was trained on measurements from one system is evaluated on measurements from other systems.
Along the ``diagonal'' of the first three rows are reconstructions where the system/PSF is identical during training and testing, \ie what previous work normally performs as an evaluation.
Off-diagonal and in the last row (unseen mask patterns of \textit{DigiCam)} are evaluations on system changes.
In general, we observe that the performance of learned reconstructions significantly deteriorate when evaluated on another system.
Similar results are observed for the models with PSF correction in \cref{app:gen_benchmark_psf_corr}.

When testing on phase masks (first two rows of \cref{fig:exp2_visual_comparison}), we observe slightly discernible content with systems trained on another phase mask (\textit{DiffuserCam} and \textit{TapeCam}),
but it is not as good as simply using \textit{ADMM100}.
The model trained on \textit{DigiCam-Single} (third column) fails to recover meaningful outputs from \textit{DiffuserCam} and \textit{TapeCam} measurements.
When testing on \textit{DigiCam-Single} and \textit{DigiCam-Multi} (amplitude masks, last two rows),
the reconstruction approaches trained on phase masks perform very poorly.
The model trained on \textit{DigiCam-Single} is able to better generalize to \textit{DigiCam-Multi} as the mask structure is similar, but there are significant coloring artifacts as the learned reconstruction fails to generalize to measurements of the unseen masks in \textit{DigiCam-Multi}.

\subsection{Generalizing to Measurements of a New PSF}
\label{sec:exp_improve_gen}

\newcommand{\figsizedigicamgennew}{0.105}
\newcommand{\newlinedigicamgennew}{18pt}
\begin{figure*}[t!]
	\centering
	\renewcommand{\arraystretch}{1} 
	\setlength{\tabcolsep}{0.1em} 
	\begin{tabular}{c cc|ccc|ccc}
     &   
    \multicolumn{2}{c}{DigiCam-Single (test set)}
    & \multicolumn{6}{c}{DigiCam-Multi (test set)}
    \\ 
    \cmidrule(r){2-3} \cmidrule(r){4-9} 
Ground-truth  & \makecell{Single-Mask\\(PSF corr.)} & \makecell{Multi-Mask\\(PSF corr.)} & 
\makecell{Single-Mask} & \makecell{With\\PSF corr.} & \makecell{PSF corr.\\and P\&P} & 
\makecell{Multi-Mask} &
\makecell{With\\PSF corr.} & \makecell{PSF corr.\\and P\&P}\\

\includegraphics[width=\figsizedigicamgennew\linewidth,valign=m]{figs/benchmark_digicam_mirflickr/GROUND_TRUTH/1.png}
    &
\includegraphics[width=\figsizedigicamgennew\linewidth,valign=m]{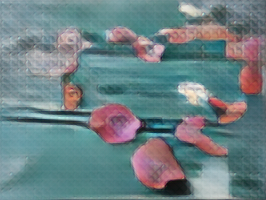}
		& 
\includegraphics[width=\figsizedigicamgennew\linewidth,valign=m]{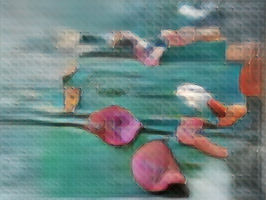}
&
\includegraphics[width=\figsizedigicamgennew\linewidth,valign=m]{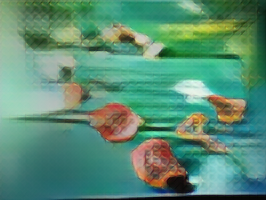}
&
\includegraphics[width=\figsizedigicamgennew\linewidth,valign=m]{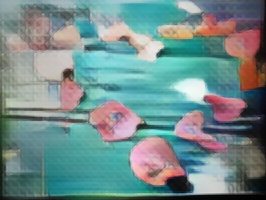}
&
\includegraphics[width=\figsizedigicamgennew\linewidth,valign=m]{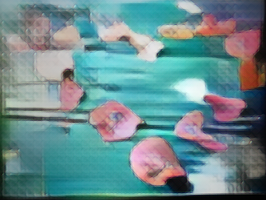}
& 
\includegraphics[width=\figsizedigicamgennew\linewidth,valign=m]{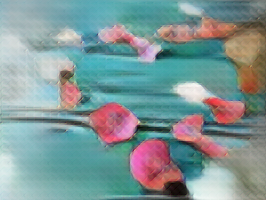}
&
\includegraphics[width=\figsizedigicamgennew\linewidth,valign=m]{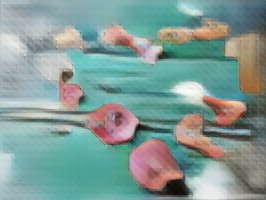}
&
\includegraphics[width=\figsizedigicamgennew\linewidth,valign=m]{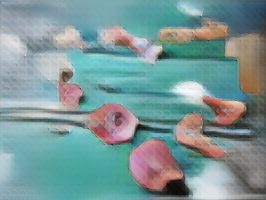}
\\[\newlinedigicamgennew]

\includegraphics[width=\figsizedigicamgennew\linewidth,valign=m]{figs/benchmark_digicam_mirflickr/GROUND_TRUTH/2.png}
    &
\includegraphics[width=\figsizedigicamgennew\linewidth,valign=m]{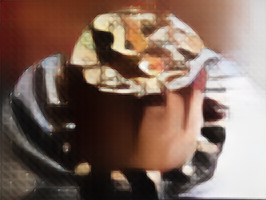}
		& 
\includegraphics[width=\figsizedigicamgennew\linewidth,valign=m]{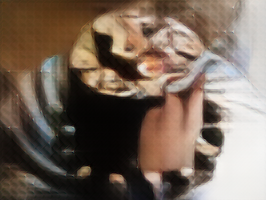}
&
\includegraphics[width=\figsizedigicamgennew\linewidth,valign=m]{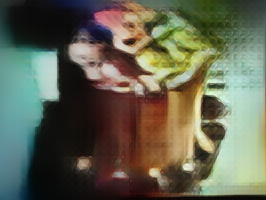}
&
\includegraphics[width=\figsizedigicamgennew\linewidth,valign=m]{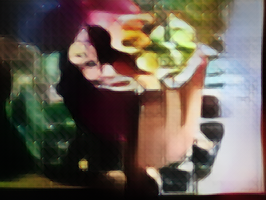}
&
\includegraphics[width=\figsizedigicamgennew\linewidth,valign=m]{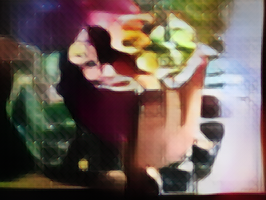}
& 
\includegraphics[width=\figsizedigicamgennew\linewidth,valign=m]{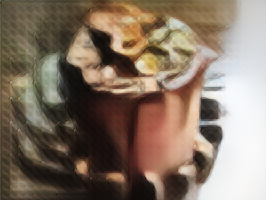}
&
\includegraphics[width=\figsizedigicamgennew\linewidth,valign=m]{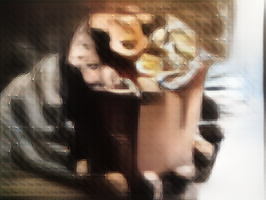}
&
\includegraphics[width=\figsizedigicamgennew\linewidth,valign=m]{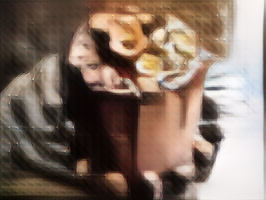}
\\[\newlinedigicamgennew]

\includegraphics[width=\figsizedigicamgennew\linewidth,valign=m]{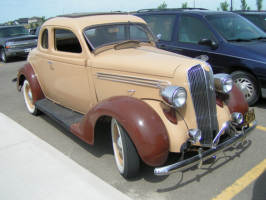}
    &
\includegraphics[width=\figsizedigicamgennew\linewidth,valign=m]{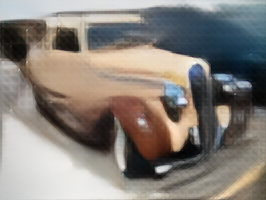}
		& 
\includegraphics[width=\figsizedigicamgennew\linewidth,valign=m]{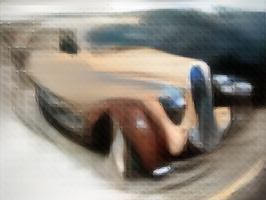}
&
\includegraphics[width=\figsizedigicamgennew\linewidth,valign=m]{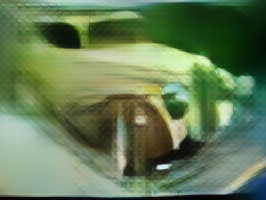}
&
\includegraphics[width=\figsizedigicamgennew\linewidth,valign=m]{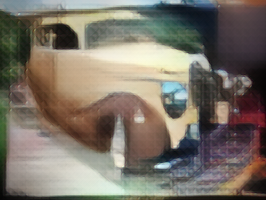}
&
\includegraphics[width=\figsizedigicamgennew\linewidth,valign=m]{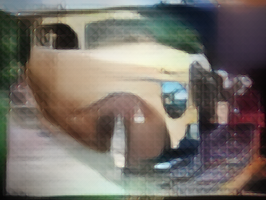}
& 
\includegraphics[width=\figsizedigicamgennew\linewidth,valign=m]{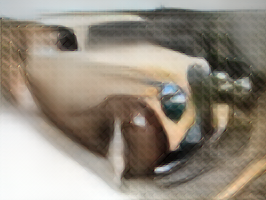}
&
\includegraphics[width=\figsizedigicamgennew\linewidth,valign=m]{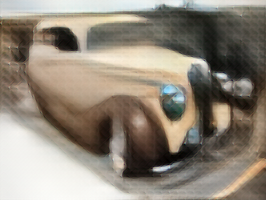}
&
\includegraphics[width=\figsizedigicamgennew\linewidth,valign=m]{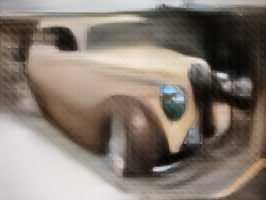}
\\
	\end{tabular}
  \caption{Example reconstructions on the \textit{DigiCam-Single} and \textit{DigiCam-Multi} test sets, as indicated by upper-most column labels.\textit{Single-Mask} and \textit{Multi-Mask} refer to the training dataset for ($\textit{Pre}_{4}$\textit{+LeADMM5+}$\textit{Post}_{4}$), where \textit{PSF corr.} incorporates a PSF correction module and \textit{P\&P} applies model adaptation with parameterize-and-perturb~\cite{9477112}.}
	\label{fig:digicam_gen}
\end{figure*}

\newcommand{\figsizedirect}{0.32}
\newcommand{\newlinedirect}{23pt}
\begin{figure}[t!]
	\centering
	\renewcommand{\arraystretch}{1} 
	\setlength{\tabcolsep}{0.2em} 
	\begin{tabular}{ccc}
		ADMM100
		& Single-Mask
		& 
  Multi-Mask
		\\\includegraphics[width=\figsizedirect\linewidth,valign=m]{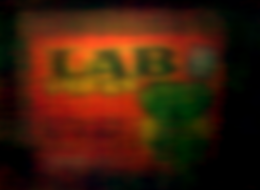}
		&\includegraphics[width=\figsizedirect\linewidth,valign=m]{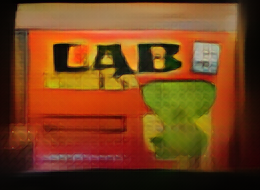}
		&  \includegraphics[width=\figsizedirect\linewidth,valign=m]{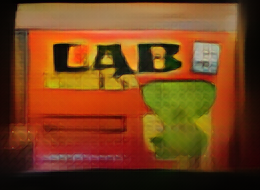}
		\\[\newlinedirect]
		\includegraphics[width=\figsizedirect\linewidth,valign=m]{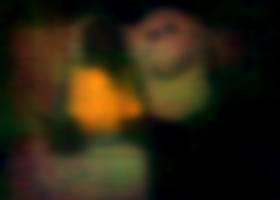}
&\includegraphics[width=\figsizedirect\linewidth,valign=m]{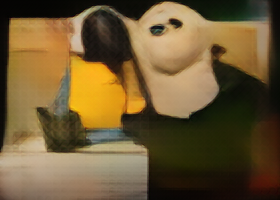}&\includegraphics[width=\figsizedirect\linewidth,valign=m]{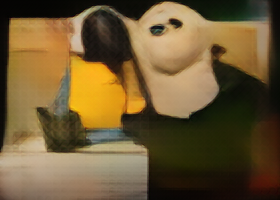}
		\\
	\end{tabular}
	\caption{Direct-capture reconstructions with \textit{DigiCam}, \ie real objects instead of images displayed on a computer monitor. Second and third columns apply ($\textit{Pre}_{4}$\textit{+LeADMM5+}$\textit{Post}_{4}$) trained on  \textit{DigiCam-Single} (measurements with a single mask) and \textit{DigiCam-Multi} (measurements with multiple masks).}
	\label{fig:direct_capture}
\end{figure}

\noindent As shown in \cref{fig:exp2_visual_comparison}, it is not possible to use a reconstruction trained on one system and expect high-quality image recovery with measurements from another system.
However, 
it would be desirable to exploit the perceptual improvements of learned approaches.
In this experiment, we explore (1) multi-mask training to improve the generalizability of \textit{DigiCam} to mask variations, (2) model adaptation as proposed by~\cite{9477112}, and (3) transfer learning to exploit the trained-components from one system for a completely new system.
For all approaches, we make use of our modular reconstruction architecture, namely ($\textit{Pre}_{4}$\textit{+LeADMM5+}$\textit{Post}_{4}$).

\begin{table}[!t]
	\renewcommand{\arraystretch}{1.3}
	\caption{Average image quality metrics (PSNR $\uparrow$ / SSIM $\uparrow$ / LPIPS $\downarrow$) on measurements of \textit{DigiCam} mask patterns not seen during training, \ie the \textit{DigiCam-Multi} test set. \textit{Single-mask} denotes training ($\textit{Pre}_{4}$\textit{+LeADMM5+}$\textit{Post}_{4}$) on \textit{DigiCam-Single}, and \textit{Multi-mask} denotes training on \textit{DigiCam-Multi}.}
	\label{tab:exp3_gen_digicam}
	\centering
	\begin{tabular}{c||c|c}
		\hline
		 & DigiCam-Single & DigiCam-Multi \\
		\hline\hline
		ADMM100 &  10.6 / 0.291 / 0.751 & 10.6 / 0.301 / 0.760  \\
		\hline
		\hline
		Single-Mask & 19.6 / 0.531 / 0.449 & 13.6 / 0.368 / 0.646  \\
        \hline
		Multi-Mask & 17.4 / 0.474 / 0.492 & 18.1 / 0.498 / 0.489  \\
        \hline
        \hline
        \makecell{Single-Mask\\with PSF corr.} &  \textbf{20.1} / \textbf{0.552} / \textbf{0.439} & 15.1 / 0.421 / 0.571  \\
		\hline
		\makecell{Multi-Mask\\with PSF corr.} & 17.7 / 0.484 / 0.484 &  \textbf{18.5} / \textbf{0.509} / \textbf{0.477}  \\
		\hline
	\end{tabular}
\end{table}

\begin{table}[!t]
	\renewcommand{\arraystretch}{1.3}
	\caption{Average image quality metrics on measurements of \textit{DigiCam} mask patterns not seen during training, \ie the \textit{DigiCam-Multi} test set. \textit{Single-mask} denotes training ($\textit{Pre}_{4}$\textit{+LeADMM5+}$\textit{Post}_{4}$) on \textit{DigiCam-Single}, and \textit{Multi-mask} denotes training on \textit{DigiCam-Multi}. Parameterize-and-perturb (P\&P)~\cite{9477112} is used to adapt the model weights for each test example.}
	\label{tab:exp_digicam_gilton}
	\centering
	\begin{tabular}{c||c|c|c|c}
		\hline
		 & PSNR $\uparrow$ & SSIM $\uparrow$ & LPIPS $\downarrow$ & Data Fidelity $\downarrow$\\
		\hline\hline
		ADMM100 & 10.6 & 0.301 & 0.760 & \textbf{0.00575} \\
		\hline
        \hline
        \makecell{Single-Mask\\with PSF corr.} & 15.1 & 0.421 & 0.571 & 0.0138 \\
		\hline
        \makecell{Adapted\\with P\&P} & 14.6 & 0.404 & 0.593  & 0.00962 \\
		\hline
        \hline
        \makecell{Multi-Mask\\with PSF corr.} & \textbf{18.5} & \textbf{0.509} & \textbf{0.477} & 0.0178 \\
		\hline
        \makecell{Adapted\\with P\&P} & 17.9 & 0.495 & 0.497  & 0.0131 \\
		\hline
	\end{tabular}
\end{table}

\subsubsection{Multi-Mask Training}

For multi-mask training, we use the \textit{DigiCam-Multi} dataset, whose training set has measurements from 85 different masks (250 measurements per mask).
During training, the corresponding PSF is passed to \textit{LeADMM5} such that all masks share the learned pre-processor, unrolled ADMM, and post-processor parameters.
We also add the PSF correction network, to learn processing that is common across mask patterns. 

In \cref{tab:exp3_gen_digicam},
the \textit{DigiCam-Multi} column evaluates various reconstructions on measurements from 15 masks not seen during training.
The model trained with multiple mask patterns (\textit{Multi-mask}) generalizes better to the unseen mask patterns than the model trained on a single mask pattern (\textit{Single-Mask}).
Incorporating PSF correction improves the performance for both:
with training on multiple masks still significantly better.
\cref{fig:digicam_gen} shows example outputs on the \textit{DigiCam-Multi} test set.
\textit{Single-Mask} has significant coloring artifacts.
PSF correction can improve these color artifacts but is still present, \eg with the cake. 
\textit{Multi-Mask}, on the other hand, has more consistent coloring with respect to the ground-truth.

While better generalizability to mask pattern changes can be achieved with multi-mask training, 
there is a degradation with respect to optimizing for a fixed mask, \ie first column of \cref{tab:exp3_gen_digicam} where we evaluate on \textit{DigiCam-Single} (the same mask whose measurements are used to train \textit{Single-Mask}).
The second and third columns of \cref{fig:digicam_gen} show performance on the \textit{DigiCam-Single} test set.
While the metrics are different, the reconstructed outputs are of similar quality.
Moreover, multi-mask training to achieve generalizability altogether removes the need for (1) measuring datasets with new mask patterns and (2) training new models, cutting down several weeks of development.
\cref{fig:direct_capture} shows results on real objects, \ie not displayed on the screen and using the mask pattern of \textit{DigiCam-Single},
and the \textit{Single-Mask} and \textit{Multi-Mask} models perform similarly.

\subsubsection{Model Adaptation}

Gilton \etal~\cite{9477112} proposed multiple approaches to adapt a learned reconstruction from one forward model to another.
Without the need for ground-truth data,
all approaches minimize the data fidelity.
For example, parameterize-and-perturb (P\&P) minimizes the following for each measurement-PSF pair $\{\bm{y}_i, \bm{p}_i\}_{i=0}^{N}$:
\begin{align}
\label{eq:pnp}
    \min_{\bm{\theta}} ||\bm{p}_i \ast r(\bm{y}_i; \bm{\theta}, \bm{p}_i) - \bm{y}_i||_2^2 + \mu ||\bm{\theta} - \bm{\theta}_0||_2^2,
\end{align}
where $\bm{\theta}$ are the retrained model parameters, $\bm{\theta}_0$ are the original model parameters, $r(\bm{y}; \bm{\theta}, \bm{p})$ recovers an image given a set of parameters and a PSF, and $\mu > 0$ controls the regularization on the retrained parameters.

\cref{tab:exp_digicam_gilton} compares models with and without P\&P to evaluate generalizability to measurements from unseen mask patterns in the \textit{DigiCam-Multi} test set.
For P\&P, \cref{eq:pnp} is minimized with stochastic gradient descent for 10 iterations with a learning rate of $3 \cdot 10^{-3}$ and $\mu = 10^{-3}$.
While P\&P reduces the data fidelity for both the \textit{Single-Mask} and \textit{Multi-Mask} models,
the other image quality metrics deteriorate.
This is consistent with the findings of \textit{DiffuserCam}~\cite{Monakhova:19}, namely for lensless imaging there is a trade-off between image quality and matching the
imaging model due the imperfect forward modeling. 
\cref{fig:digicam_gen} shows example outputs of the model parameters adapted with P\&P,
which are very similar to outputs from the original model.

\subsubsection{Transfer Learning}

Another approach for generalizing to new PSFs is to apply transfer learning, 
\eg fine-tuning a model that has been trained on one system to a new system.
Fine-tuning still requires data of the new system, 
but this data can be simulated to avoid measuring a dataset.
In the previous experiment,
we observed that the commonly-used shift-invariant (convolutional) model is not suitable for minimizing data fidelity for model adaption~\cite{9477112}.
In this experiment,
we investigate whether it is sufficient for simulating data for fine-tuning.
We perform our experiments with \textit{DiffuserCam} (whose on-axis PSF has a high degree of similarity with off-axis PSFs~\cite{Antipa:18}),
by convolving the lensed data with the PSF and adding Poisson noise with an SNR of \SI{40}{\decibel}.

\cref{tab:exp3_gen_phase} quantifies performance on the \textit{DiffuserCam} test set and \cref{fig:diffuser_gen} shows example outputs.
Our baselines are (1) \textit{ADMM100} (no training required) and (2) \textit{DiffuserCam-Sim} (trained from scratch with simulated data).
While \textit{DiffuserCam-Sim} performs worse than \textit{ADMM100} in terms of metrics and has grainier outputs, 
it is better at recovering finer details  (\eg lines on the butterfly wings in \cref{fig:diffuser_gen}).
Moreover, it performs better than models trained on other datasets, \ie \textit{TapeCam} and \textit{DigiCam-Multi}.

\begin{table}[!t]
	\renewcommand{\arraystretch}{1.3}
	\caption{Average image quality metrics of reconstructions on the \textit{DiffuserCam} test set. No model is trained with the measured lensless data from \textit{DiffuserCam}.}
	\label{tab:exp3_gen_phase}
	\centering
	\begin{tabular}{c||c|c|c}
		\hline
	 & PSNR $\uparrow$ & SSIM $\uparrow$ & LPIPS $\downarrow$   \\
		\hline\hline
		ADMM100 & 15.0 & 0.457 & 0.511   \\
  \hline
        \makecell{DiffuserCam-Sim} & 13.6 & 0.389 & 0.525  \\
		\hline
		\hline
		TapeCam & 10.7 & 0.217 & 0.556 \\
		\hline
		\makecell{Fine-tune ($\text{Pre}_{4}$+LeADMM5+$\text{Post}_{4}$)}  & 15.3 & 0.563 & 0.337 \\
  \hline
  \makecell{Fine-tune ($\text{Pre}_{4}$+LeADMM5)} & 16.1 & 0.516 & 0.350  \\
\hline
  \makecell{Fine-tune (LeADMM5+$\text{Post}_{4}$)} & \textbf{16.2} & \textbf{0.604} & \textbf{0.305}  \\
  \hline
  \hline
  \makecell{DigiCam-Multi} &  10.2 & 0.330 & 0.542  \\
		\hline
  \makecell{Fine-tune ($\text{Pre}_{4}$+LeADMM5+$\text{Post}_{4}$)}  & 15.0 & 0.569 & 0.327  \\
		\hline
     \makecell{Fine-tune ($\text{Pre}_{4}$+LeADMM5)} & \textbf{16.2} & 0.506 & 0.368   \\
    \hline
    \makecell{Fine-tune (LeADMM5+$\text{Post}_{4}$)} & 15.9 & \textbf{0.589} & \textbf{0.324} \\
\hline
	\end{tabular}
\end{table}

By fine-tuning \textit{TapeCam} and \textit{DigiCam-Multi} with \textit{DiffuserCam} simulations,
we can obtain approaches that surpass the performance of \textit{ADMM100} and \textit{DiffuserCam-Sim}.
Fine-tuning exploits the modular components that have been learned on real measurements from other systems to generalize to the new \textit{DiffuserCam} system.
We fine-tune various components with a smaller learning rate of $10^{-5}$, and find that freezing the pre-processor yields the best results, 
\ie indicating that the pre-processor generalizes to measurements of other systems.
While training on actual measurement is significantly better (see \cref{tab:exp1_benchmark}), fine-tuning learned reconstructions on simulated data can exploit learnings from the original system and remove the need of collecting a lensed-lensless dataset.

\newcommand{\figsizediffusergen}{0.135}
\newcommand{\newlinediffusergen}{12pt}
\begin{figure*}[t!]
	\centering
	\renewcommand{\arraystretch}{1} 
	\setlength{\tabcolsep}{0.1em} 
	\begin{tabular}{ccccccc}
        Ground-truth
        & ADMM100
		& \makecell{DiffuserCam\\-Sim}
		& TapeCam
        & \makecell{TapeCam\\(fine-tuned)}
        & DigiCam-Multi
        & \makecell{DigiCam-Multi\\(fine-tuned)}
		\\

\includegraphics[width=\figsizediffusergen\linewidth,valign=m]{figs/benchmark_diffusercam/GROUND_TRUTH/3.png}

  &\includegraphics[width=\figsizediffusergen\linewidth,valign=m]{figs/benchmark_diffusercam/ADMM/100/3.png}
		&\includegraphics[width=\figsizediffusergen\linewidth,valign=m]{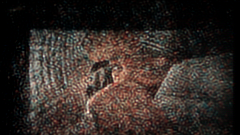}
		& \includegraphics[width=\figsizediffusergen\linewidth,valign=m]{figs/benchmark_diffusercam/hf_tapecam_mirflickr_Unet4M+U5+Unet4M/3.png}

  & \includegraphics[width=\figsizediffusergen\linewidth,valign=m]{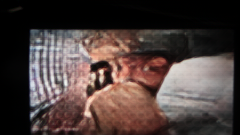}

  & \includegraphics[width=\figsizediffusergen\linewidth,valign=m]{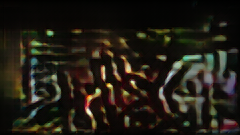}

  & \includegraphics[width=\figsizediffusergen\linewidth,valign=m]{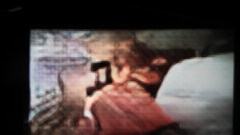}
		\\[\newlinediffusergen]

  \includegraphics[width=\figsizediffusergen\linewidth,valign=m]{figs/benchmark_diffusercam/GROUND_TRUTH/4.png}

  &\includegraphics[width=\figsizediffusergen\linewidth,valign=m]{figs/benchmark_diffusercam/ADMM/100/4.png}
		&\includegraphics[width=\figsizediffusergen\linewidth,valign=m]{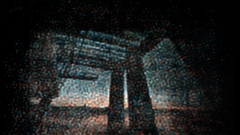}
		& \includegraphics[width=\figsizediffusergen\linewidth,valign=m]{figs/benchmark_diffusercam/hf_tapecam_mirflickr_Unet4M+U5+Unet4M/4.png}

  & \includegraphics[width=\figsizediffusergen\linewidth,valign=m]{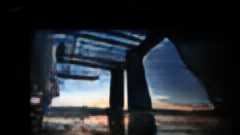}

  & \includegraphics[width=\figsizediffusergen\linewidth,valign=m]{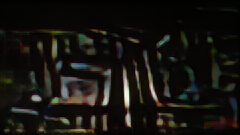}

  & \includegraphics[width=\figsizediffusergen\linewidth,valign=m]{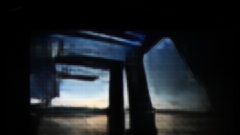}
		\\[\newlinediffusergen]

  \includegraphics[width=\figsizediffusergen\linewidth,valign=m]{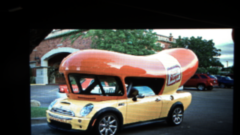}

  &\includegraphics[width=\figsizediffusergen\linewidth,valign=m]{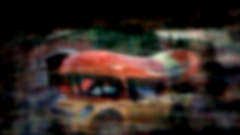}
		&\includegraphics[width=\figsizediffusergen\linewidth,valign=m]{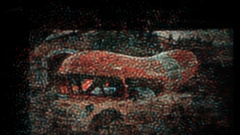}
		& \includegraphics[width=\figsizediffusergen\linewidth,valign=m]{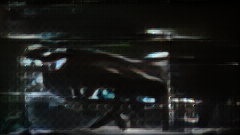}

  & \includegraphics[width=\figsizediffusergen\linewidth,valign=m]{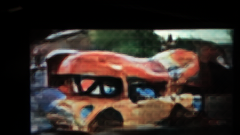}

  & \includegraphics[width=\figsizediffusergen\linewidth,valign=m]{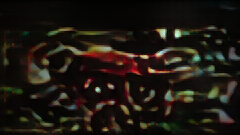}

  & \includegraphics[width=\figsizediffusergen\linewidth,valign=m]{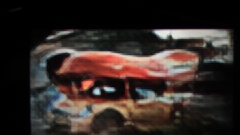}
		\\[\newlinediffusergen]

  \includegraphics[width=\figsizediffusergen\linewidth,valign=m]{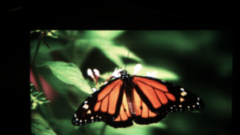}

  &\includegraphics[width=\figsizediffusergen\linewidth,valign=m]{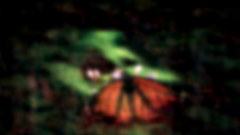}
		&\includegraphics[width=\figsizediffusergen\linewidth,valign=m]{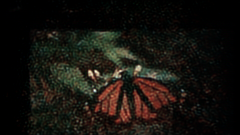}
		& \includegraphics[width=\figsizediffusergen\linewidth,valign=m]{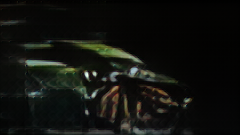}

  & \includegraphics[width=\figsizediffusergen\linewidth,valign=m]{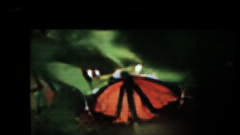}

  & \includegraphics[width=\figsizediffusergen\linewidth,valign=m]{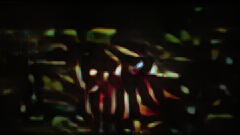}

  & \includegraphics[width=\figsizediffusergen\linewidth,valign=m]{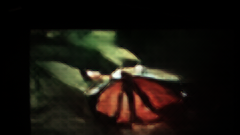}

	\end{tabular}
	\caption{Transfer learning to \textit{DiffuserCam}. Example reconstructions on measured data coming from the \textit{DiffuserCam} test set, without having seen measured lensless data from \textit{DiffuserCam} during training. Fine-tuned models freeze the pre-processor of the original model, and fine-tune the unrolled ADMM parameters and the post-processor on simulations of \textit{DiffuserCam} obtained by convolving ground-truth data with the PSF.}
	\label{fig:diffuser_gen}
\end{figure*}

%% file: sec/6_conclusion.tex
\section{Conclusion}

\noindent In this work, we address the robustness and generalizability of lensless imaging with a modular reconstruction approach, comprised of a pre-processor, camera inversion, a post-processor, and PSF correction (see \cref{fig:pipeline}).
We theoretically show the need for each component due to inevitable model mismatch in lensless imaging systems,
and experimentally demonstrate the benefit of the pre-processor across multiple imaging systems, various reconstruction approaches, and different SNRs.
We also perform the first generalizability study across lensless imaging systems, 
and demonstrate techniques to improve generalizability.
To this end, our modular reconstruction approach allows learnings from one system to transfer to a new one,
in particular the pre-processor component.
This has very practical implications as it can remove the need of collecting a lensed-lensless dataset for a new system,
which is time-consuming and/or may not be possible depending on the application.
Our investigation prioritizes \textit{accessibility} and \textit{reproducibility}.
We open-source datasets collected with inexpensive components: a Raspberry Pi sensor, double-sided tape as a phase mask, and an LCD for our programmable-mask based system -- \textit{DigiCam}.
We also release our measurement software, and reconstruction implementations for the baselines and our modular approach.
Our methods demonstrate improved performance on our low-cost systems and more expensive ones (\textit{DiffuserCam}~\cite{Monakhova:19}).
As future work, we will explore applications of generalizable lensless imaging that would benefit from models that do not need to be retrained with measurements of new PSFs.
For further performance improvements, while our study used a convolutional forward model and the DRUNet architecture~\cite{zhang2021plug} for modular components,
a non-LSI forward model~\cite{Yanny:22,cai2024phocolens} and
different architectures, \eg transformers~\cite{Zamir2021Restormer,Pan:22} or diffusion models~\cite{cai2024phocolens}, can be explored within our modular framework.

%% file: sec/supplementary.tex
\clearpage

\crefalias{section}{appendix}
\counterwithin{figure}{section}
\counterwithin{table}{section}
\counterwithin{equation}{section}

\title{Supplementary Material --\\Towards Robust and Generalizable Lensless Imaging with Modular Learned Reconstruction}

\maketitle

\appendices

\section{Consequences of Model Mismatch to Image Recovery}
\label{app:mismatch}

\noindent Assuming a desired scene is comprised of point sources that are incoherent with each other, a lensless imaging system (with no shift-invariance assumption) can be modeled as a linear matrix-vector multiplication with the system matrix $\bm{H}$:
\begin{align}
    \label{eq:forward_gen_app}
    \bm{y} = \bm{H}\bm{x} + \bm{n},
\end{align}
where $\bm{y}$ and $\bm{x}$ are the vectorized lensless measurement and scene intensity, respectively, and $\bm{n}$ is additive noise.

If we denote our estimate system matrix as $\bm{\hat{H}}=(\bm{H}+\bm{\Delta}_H)$ where the deviation from the true system matrix is $\bm{\Delta}_H$,
our forward model from can be written as:
\begin{align}
    \label{eq:mismatch_forward_app}
    \bm{y} = \bm{H}\bm{x} + \bm{n} = (\bm{\hat{H}} - \bm{\Delta}_H)\bm{x} + \bm{n}.
\end{align}

\subsection{Direct Inversion}

\noindent Assuming the system is invertible and with spectral radius $\rho(\bm{H}) < 1$, using the estimate $\bm{\hat{H}}$ for direct inversion yields~\cite{9546648,9157433}:
\begin{align}
    \bm{\hat{x}} &= \bm{\hat{H}}^{-1} \bm{y} \nonumber \\
    &=(\bm{H}+\bm{\Delta}_H)^{-1} (\bm{H}\bm{x} + \bm{n}) \nonumber \\ 
    &=[\bm{H} (\bm{I}+\bm{H}^{-1}\bm{\Delta}_H)]^{-1} (\bm{H}\bm{x} + \bm{n}) \nonumber \\ 
    &=(\bm{I}+\bm{H}^{-1}\bm{\Delta}_H)^{-1} \bm{H}^{-1} (\bm{H}\bm{x} + \bm{n}) \nonumber \\
    &=(\bm{I}+ \bm{H}^{-1}\bm{\Delta}_H)^{-1} (\bm{x} + \bm{H}^{-1}\bm{n}) \nonumber \\
   &=
   \label{eq:taylor}
   (\bm{I} - \bm{H}^{-1}\bm{\Delta}_H) (\bm{x} + \bm{H}^{-1}\bm{n}) + \mathcal{O}(\| \bm{\Delta}_H\|_F^2)\\
   \label{eq:inversion_terms_app}
   &= \bm{x} - \underbrace{\bm{H}^{-1}\bm{\Delta}_H \bm{x}}_{\text{model mismatch}} + \underbrace{(\bm{I} - \bm{H}^{-1}\bm{\Delta}_H)\bm{H}^{-1}\bm{n}}_{\text{noise amplification}} + \mathcal{O}(\| \bm{\Delta}_H\|_F^2),
\end{align}
where \cref{eq:taylor}  uses the Taylor expansion $(\bm{I}-\bm{X})^{-1} = \bm{I} + \sum_{k=1}^{\infty} \bm{X}^k$ with $\bm{X} = \bm{H}^{-1}\bm{\Delta}_H$.
With \cref{eq:inversion_terms_app}, we see the error terms that arise due to model mismatch in the forward modeling, and how noise can be amplified, particularly if $\bm{H}$ is ill-conditioned.

\subsection{Wiener Filtering}

\noindent Approximating the system as linear shift-invariant (LSI) allows us to write the forward operation as a point-wise multiplication in the frequency domain with a single on-axis point spread function (PSF):
\begin{align}
\label{eq:lsi_forward_app}
    \bm{Y} = \bm{P} \odot \bm{X} + \bm{N},
\end{align}
where $\{\bm{Y}, \bm{P}, \bm{X}, \bm{N}\} \in \mathbb{C}^{N_x \times N_y}$ are 2D Fourier transforms of the measurement, the on-axis PSF, the scene, and the noise respectively, and $\odot$ is point-wise multiplication.
Wiener filtering yields the following estimate:
\begin{align}
    \label{eq:wiener_app}
    \bm{\hat{X}} = \dfrac{\bm{P}^*  \odot \bm{Y}}{ |\bm{P}|^2 + \bm{R}} =  \dfrac{\bm{P}^* \odot (\bm{P} \odot\bm{X} + \bm{N})}{ |\bm{P}|^2 + \bm{R}},
\end{align}
where all operations are point-wise,
the noise $\bm{N}$ is assumed to be independent to $\bm{X}$, and $\bm{R}\in \mathbb{R}^{N_x \times N_y}$ is the \textit{inverse} of the signal-to-noise ratio at each frequency.
$\bm{R}$ is often simplified to a single constant $K$,
which makes \cref{eq:wiener_app} equivalent to least-squares/Tikhonov  regularization~\cite{flatcam} of \cref{eq:opt_gen}, \ie $\mathcal{R}(\cdot) = \|\cdot\|^2_2$ with the appropriate $\lambda$ factor.

If we use a noisy version of the on-axis PSF's Fourier transform, \ie $\bm{\hat{P}} = (\bm{P}+ \bm{\Delta}_P)$, 
our Wiener-filtered estimate of the scene becomes:
\begin{align}
    \bm{\hat{X}}^{\text{noisy}} &= \dfrac{\bm{\hat{P}}^* \odot \bm{Y}}{ |\bm{\hat{P}}|^2 + \bm{R}} \nonumber\\
    &= \dfrac{\bm{P}^* \odot \bm{Y}+ \bm{\Delta}_P^* \odot \bm{Y}}{|\bm{P}|^2 + \bm{R} + |\bm{\Delta}_P|^2 + \bm{\Delta}_P^* \odot \bm{P} + \bm{P}^* \odot \bm{\Delta}_P}.
\end{align}
Using:
\begin{align}
\dfrac{\bm{A}}{\bm{B}+\bm{\Delta}_B} = \dfrac{\bm{A}}{\bm{B}} - \dfrac{\Delta_B \odot \bm{A}}{\bm{B}^2 + \bm{B} \odot \bm{\Delta}_B},
\end{align}
with:
\begin{align}
    \bm{A} &= \bm{P}^* \odot \bm{Y} +\bm{\Delta}_P^* \odot \bm{Y}\\
    \bm{B} &= |\bm{P}|^2 + \bm{R}\\
    \bm{\Delta}_B &= |\bm{\Delta}_P|^2 + \bm{\Delta}_P^* \odot \bm{P} + \bm{P}^*\odot\bm{\Delta}_P,
\end{align}
we obtain:
\begin{align}
    \bm{\hat{X}}^{\text{noisy}} &= \dfrac{\bm{P}^* \odot \bm{Y} +\bm{\Delta}_P^* \odot \bm{Y}}{\bm{B}} - \dfrac{\bm{\Delta}_B \odot \bm{Y} \odot(\bm{P}^* +\bm{\Delta}_P^*)}{\bm{B}^2 + \bm{B} \odot \bm{\Delta}_B} \nonumber\\
    &= \bm{\hat{X}} + \bm{Y} \odot \left[ \dfrac{\bm{\Delta}_P^*}{\bm{B}} - \dfrac{\bm{\Delta}_B \odot (\bm{P}^* + \bm{\Delta}_P^*)}{\bm{B}^2 + \bm{B} \odot \bm{\Delta}_B}  \right]  \nonumber\\
    &=\bm{\hat{X}} + \underbrace{\bm{M} \odot \bm{P} \odot  \bm{X}}_{\text{model mismatch}} + \underbrace{\bm{M} \odot \bm{N}}_{\text{noise amplification}}, \label{eq:noisy_wiener_app}
\end{align}
where:
\begin{align}
    \bm{M} = \dfrac{\bm{\Delta}_P^*}{\bm{B}} - \dfrac{\bm{\Delta}_B \odot (\bm{P}^* + \bm{\Delta}_P^*)}{\bm{B}^2 + \bm{B} \odot \bm{\Delta}_B}.
\end{align}
If there is no model mismatch in the PSF used for Wiener filtering, \ie $\delta_{\bm{f}} = 0$, such that $\bm{B}_2 = 0$, $\bm{M} = 0$, $ \bm{\bm{\hat{X}}_f^{\text{noisy}}} = \bm{\hat{X}}_f $.

\subsection{Gradient Descent}

\noindent A common approach to avoid adverse amplification with $\bm{H}^{-1}$ is to pose the image recovery as a regularized optimization problem:
\begin{align}
\label{eq:opt_gen_app}
   \bm{\hat{x}} = \arg \min_{\bm{x}} \frac{1}{2} ||\bm{H}\bm{x} - \bm{y}||_2^2 + \lambda \mathcal{R}(\bm{x}),
\end{align}
where $\mathcal{R}(\cdot)$ is a regularization function on the estimate image.

Applying gradient descent to solve \cref{eq:opt_gen_app}  (without regularization),
would lead to the following update step (without model mismatch):
\begin{align}
    \label{eq:gradient_step}
    \bm{\hat{x}}^{(k)} = \bm{\hat{x}}^{(k-1)} - \alpha \bm{H}^T(\bm{H}\bm{\hat{x}}^{(k-1)} - \bm{y}).
\end{align}
With model mismatch, we get the following noisy update (assuming no model mismatch in the previous iteration):
\begin{align}
\bm{\hat{x}}^{(k),\text{noisy}} &= \bm{\hat{x}}^{(k-1)} - \alpha (\bm{H}+\bm{\Delta}_H)^T  \left[(\bm{H}+\bm{\Delta}_H)\bm{\hat{x}}^{(k-1)} -\bm{y}\right] \nonumber\\
    &= \bm{\hat{x}}^{(k)} - \alpha \left[ \bm{\Delta}_H^T( \bm{H}\bm{\hat{x}}^{(k-1)} - \bm{y}) + \bm{\hat{H}}^T\bm{\Delta}_H \bm{\hat{x}}^{(k-1)} \right] \nonumber \\
    &= \bm{\hat{x}}^{(k)} + \alpha \left[\underbrace{ \bm{\Delta}_H^T\bm{H} \bm{x} - \delta_{\bm{H}} \bm{\hat{x}}^{(k-1)}}_{\text{model mismatch}}  + \underbrace{\bm{\Delta}_H^T \bm{n}}_{\text{noise amplification}} \right], \label{eq:gradient_descent_mismatch_app}
\end{align}
where \cref{eq:forward_gen_app} is used for $\bm{y}$, and $\delta_{\bm{H}} = \left( \bm{\Delta}_H^T\bm{H} + \bm{\hat{H}}^T \bm{\Delta}_H \right)$.
If there is no model mismatch (\ie $\bm{\Delta}_H = \bm{0}$), the last two terms disappear and $ \bm{\hat{x}}^{(k),\text{noisy}} = \bm{\hat{x}}^{(k)}$.

\subsection{Proximal Gradient Descent}

\noindent Proximal gradient descent applies an operator at each gradient step, \eg shrinkage/soft-thresholding for the fast iterative shrinkage-thresholding algorithm (FISTA)~\cite{beck2009fast}:
\begin{align}
    \bm{\hat{x}}^{(k),\text{noisy}} &= \mathcal{T}_{\beta}
    \left(  \bm{\hat{x}}^{(k)} + \alpha \left[\bm{\Delta}_H^T\bm{H} \bm{x} - \delta_{\bm{H}} \bm{\hat{x}}^{(k-1)}  + \bm{\Delta}_H^T \bm{n} \right] \right),
\end{align}
where the shrinkage operator $\mathcal{T}_{\beta} : \mathbb{R}^n \rightarrow \mathbb{R}^n$ is defined by:
\begin{align}
\mathcal{T}_{\beta}(\bm{x})_i = ( |x_i|- \beta)_{+} \text{sgn}(x_i).
\end{align}
If $\bm{\Delta}_H$ is sufficiently small, the shrinkage operator may discard the unwanted terms, \ie if all element are below $\beta$.
For natural images, we typically promote sparsity in another space, \eg with the TV operator, such that the adjoint of the operator would be applied before applying the shrinkage operator.
In this case, the unwanted terms may not be discarded by the shrinkage operator.

\vspace{1cm}
\subsection{Alternating Direction Method of Multipliers}

\noindent Starting with the ADMM update with model mismatch derived by Zeng \etal~\cite{9546648}, we can expand the terms from the previous iteration in Eq.~15 of~\cite{9546648} that depend on model mismatch:
\begin{align}
    \bm{\hat{x}}^{(k)} &= \left( \bm{W}_1 + \rho_x \delta_{\bm{H}} \right)^{-1} \bm{W}_1 \bm{\hat{x}}^{(k),\text{noisy}}  \nonumber \\
    & \quad - \bm{W}_2 \left( \bm{C}^T \bm{y} + \bm{\epsilon}^{(k-1)} \right) - \bm{W}_3
    \label{eq:original_noisy_admm_app}
\end{align}
where:
\begin{align}
\bm{W}_1 &= \rho_x \bm{\hat{H}}^T \bm{\hat{H}} + \rho_z \bm{C}^T\bm{C} + \rho_y \bm{I}, \\
\bm{W}_2 &= (\bm{W}_1 + \rho_x \delta_{\bm{H}})^{-1} \Delta_{\bm{H}}^T \rho_x (\bm{C}^T\bm{C} + \rho_x \bm{I})^{-1}, \\
\bm{W}_3 &= \left( \bm{W}_1 + \rho_x \delta_{\bm{H}} \right)^{-1} \bm{\hat{H}}^T \rho_x^2 \Delta_{\bm{H}} \bm{\hat{x}}^{(k-1)},
\end{align}
$\{\rho_x, \rho_y, \rho_z\}$ are positive penalty parameters, $\bm{C}$ crops the image to the sensor size~\cite{Antipa:18}, and $\bm{\epsilon}^{(k-1)}$ denotes the combination of terms from the previous iteration's updates that do not depend on model mismatch.
By inserting $(\bm{C}\bm{H}\bm{x} + \bm{n})$ for $\bm{y}$ into \cref{eq:original_noisy_admm_app} and rearranging terms:
\begin{align}
&\bm{\hat{x}}^{(k),\text{noisy}} = \bm{\hat{x}}^{(k)}  + \underbrace{\bm{W}_4 \bm{W}_2\bm{C}^T \bm{n}}_{\text{noise amplification}} \nonumber \\
\label{eq:admm_mismatch_app} 
&\underbrace{+ \bm{W}_1^{-1}\rho_x \delta_{\bm{H}} \bm{\hat{x}}^{(k)} + \bm{W}_4 \bm{W}_2\left(\bm{C}^T\bm{C} \bm{H}\bm{x} + \bm{\epsilon}^{(k-1)} \right) + \bm{W}_4\bm{W}_3}_{\text{model mismatch}}, 
\end{align}
where $\bm{W}_4 = (\bm{I} + \bm{W}_1^{-1} \rho_x \delta_{\bm{H}})$.
If there is no model mismatch (\ie $\bm{\Delta}_H = \bm{0}$), $\bm{W}_2 = \bm{0}$, $\bm{W}_3 = \bm{0}$, $\bm{W}_4 = \bm{I}$, and $\delta_{\bm{H}} = \bm{0}$ such that \cref{eq:admm_mismatch_app} simplifies to $\bm{\hat{x}}^{(k),\text{noisy}} =   \bm{\hat{x}}^{(k)}$.

\section{Pre- and Post-Processor Architecture}
\label{sec:drunet}

\begin{figure}[t!]
	\centering
	\includegraphics[width=1.0\linewidth]{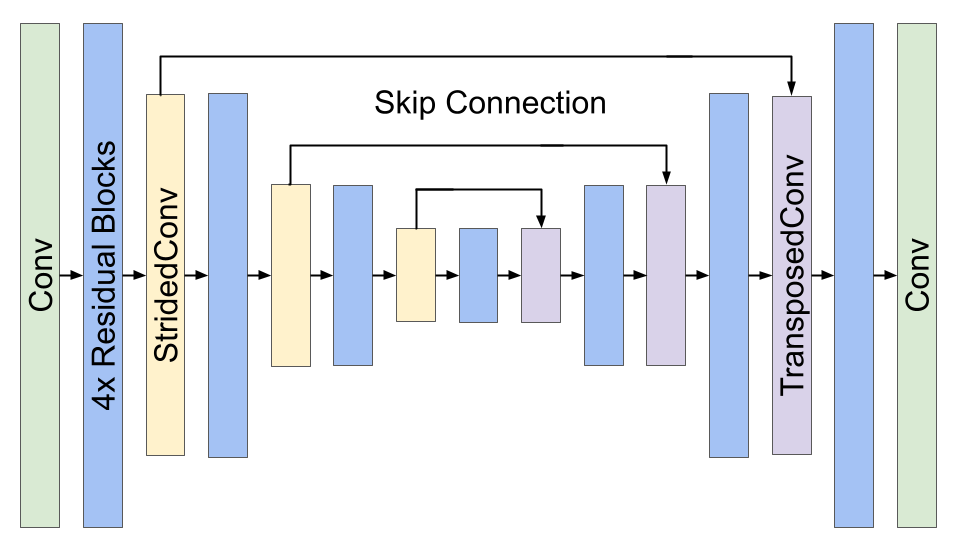}
	\caption{Denoising residual U-Net (DRUNet) architecture. where the sequence of operations is identical to the architecture proposed in~\cite{zhang2021plug}: a U-Net architecture with four scales and sandwiched by 2D convolutional layers (\textit{Conv}) with no activation function.
		Each scale has an identity skip connection between a $(2\times2)$ strided-convolution downscaling block (\textit{StridedConv}) and a corresponding $(2\times2)$ transposed-convolution upscaling block (\textit{TransposedConv}). Each residual blocks uses two \textit{Conv} layers, a \textit{ReLU} activation, a skip connection, and no batch normalization.}
	\label{fig:drunet}
\end{figure}

\noindent For the pre- and post-processor architectures, we use a denoising residual U-Net (DRUNet) architecture  that has been shown to be very effective for denoising, deblurring, and super-resolution tasks~\cite{zhang2021plug}.
The architecture of DRUNet is shown in \cref{fig:drunet}.

For unrolled ADMM with model mismatch compensation, before going through the upsampling residual blocks of the post-processor (see \cref{fig:drunet}), the output of the compensation branch is concatenated to the last \textit{StridedConv} output and passed through a 2D convolutional layer whose number of output channels is equivalent to the number of channels of the post-processor's fourth scale, \eg 256 for $P_8$, and then passed through a \textit{ReLU} activation.

\section{Visualization of Camera Inversion Approaches}
\label{app:inversion}

\begin{figure*}[t!]
    \centering
    \begin{subfigure}{0.48\linewidth}
		\centering
        \includegraphics[width=0.99\linewidth]{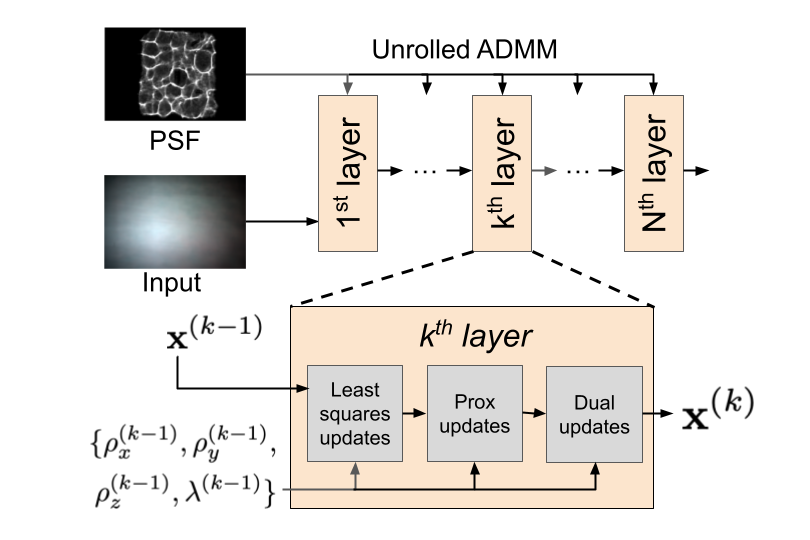} 
		\caption{Unrolled ADMM~\cite{Monakhova:19}.}
		\label{fig:unrolled_admm}
	\end{subfigure}
 \begin{subfigure}{0.5\linewidth}
		\centering
        \includegraphics[width=\linewidth]{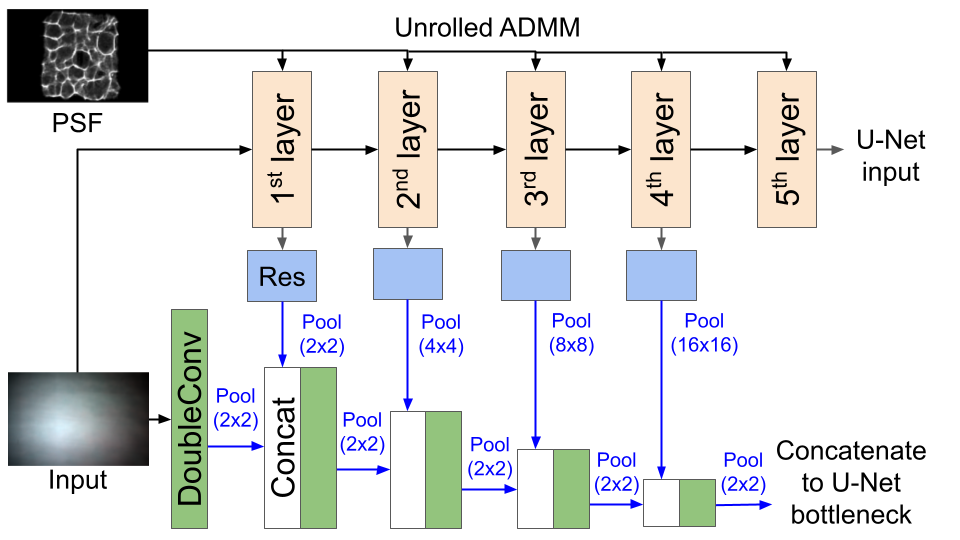} 
		\caption{Unrolled ADMM with model mismatch compensation network~\cite{9546648}.}
		\label{fig:compensation_branch}
	\end{subfigure}
    \hspace{2cm}
	\begin{subfigure}{0.48\linewidth}
		\centering
		\includegraphics[width=0.99\linewidth]{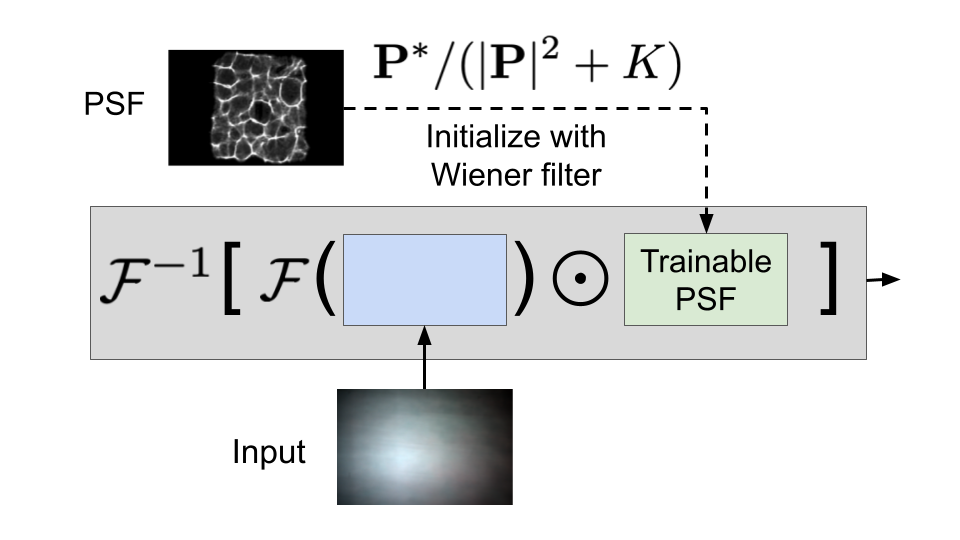} 
		\caption{Trainable inversion~\cite{9239993}.}
		\label{fig:trainable_inv}
	\end{subfigure}
	\begin{subfigure}{0.5\linewidth}
		\centering
		\includegraphics[width=0.99\linewidth]{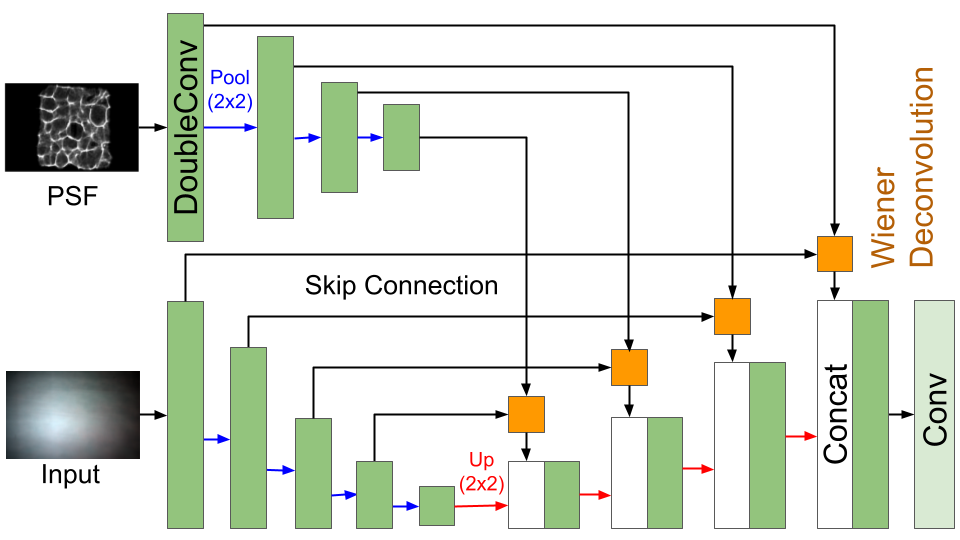} 
		\caption{Multi-Wiener deconvolution network with PSF correction~\cite{Li:23}.}
		\label{fig:multiwiener}
	\end{subfigure}
	\caption{Camera inversion approaches considered in this work. The input is either the raw measurement or the output of the pre-processor, while the output can be fed to a post-processor.
 }
	\label{fig:camera_inv_arch}
\end{figure*}

\noindent \cref{fig:camera_inv_arch} visualizes all the camera inversion approaches.
In \cref{fig:trainable_inv}, $\mathcal{F}$ and $\mathcal{F}^{-1}$ refer to the 2D Fourier transform and its inverse respectively.
In \cref{fig:compensation_branch} and \cref{fig:multiwiener}, \textit{DoubleConv} corresponds to two 2D convolutional layers each followed by batch normalization and a \textit{ReLU} activation, \textit{Conv} corresponds to a 2D convolutional layer followed by a \textit{ReLU} activation, \textit{Res} corresponds to \textit{DoubleConv} with a skip connection before the final \textit{ReLU} activation, and \textit{Pool} refers to max-pooling. All convolutional layers use $(3\times3)$ kernels.
For MMCN, before going through the bottleneck residual blocks of the post-processor (see \cref{fig:drunet}), the output of the compensation branch is concatenated to the last \textit{StridedConv} output and passed through a 2D convolutional layer whose number of output channels is equivalent to the number of channels of the post-processor's fourth scale, \eg 256 for $P_8$, and then passed through a \textit{ReLU} activation.
\cref{fig:multiwiener} shows the architecture for a multi-Wiener deconvolution network with PSF correction (MWDN)~\cite{Li:23}.
As it already uses convolutional layers before and after Wiener filtering, we do not incorporate pre- and post-processors.

\section{Point Spread Function Modeling}
\label{app:psf_modeling}

\begin{figure}[t!]
	\centering
	\includegraphics[width=1.0\linewidth]{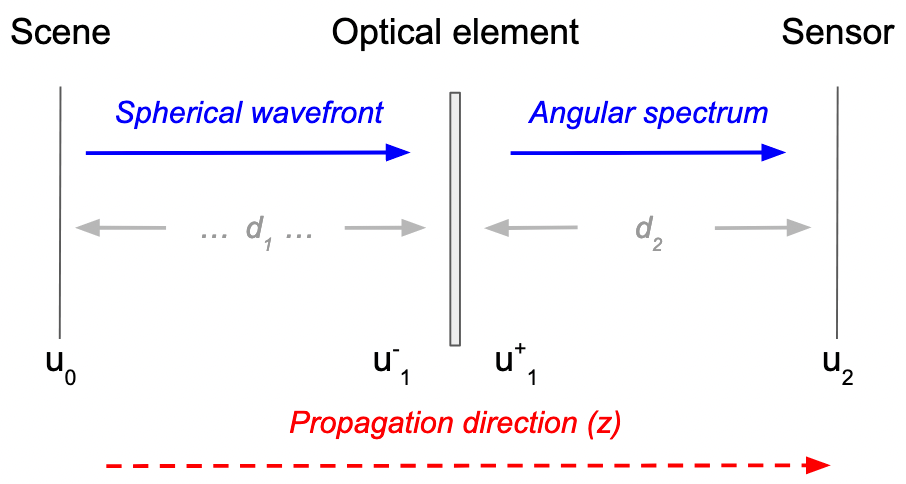}
	\caption{Modeling of propagation to simulate the point spread function (not drawn to scale).}
	\label{fig:propagation_model}
\end{figure}

\noindent We model the PSF similar to~\cite{sitzmann2018e2e}, \ie as spherical waves up to the optical element followed by free-space propagation to the sensor, as shown in \cref{fig:propagation_model}.
The wave field at the sensor for a given wavelength $\lambda$ and for a point source at a distance $d_1$ from the optical element, which is at a distance $d_2$ from the sensor, can be written as:
\begin{align}
	\label{eq:wavefield_app}
	&u_2(\bm{r}; d_1, d_2, \lambda) = \nonumber \\ 
	&\mathcal{F}^{-1}\Big(\mathcal{F} \Big( m(\bm{r}; \lambda) \underbrace{e^{j \frac{2\pi}{\lambda} \sqrt{\|\bm{r}\|_2^2 +  d_1^2}}}_{\text{spherical waves}}
	\Big) \times h(\bm{u}; z=d_2, \lambda) \Big),
\end{align}
where 
$ h(\bm{u}; z, \lambda)$ is the free-space propagation frequency response, and
$\bm{u} \in \mathbb{R}^2$ are the spatial frequencies of $\bm{r} \in \mathbb{R}^2$.
For the free-space propagation kernel, we use bandlimited angular spectrum (BLAS)~\cite{Matsushima2009}:
\begin{align}
	\label{eq:freespace}
	h(\bm{u}; z=d_2, \lambda ) = e^ {j \frac{2 \pi}{\lambda} z \sqrt{1 - \|\lambda \bm{u}\|_2^2} } \,\text{rect}_{\text{2d}}\Big(\frac{\bm{u}}{2\bm{u}_{\text{limit}}}\Big),
\end{align}
where $ \text{rect}_{\text{2d}} $ is a 2D rectangular function for bandlimiting by the frequencies $\bm{u}_{\text{limit}} = \sqrt{(z / \bm{S} )^2  + 1} / \lambda$
and $ \bm{S} \in \mathbb{R}^2 $ are the physical dimensions of the propagation region (in our case the physical dimensions of the sensor).

\section{Mask Modeling of \textit{DigiCam}}
\label{app:mask_modeling}

\begin{figure}[t!]
		\centering
		\includegraphics[width=1.0\linewidth]{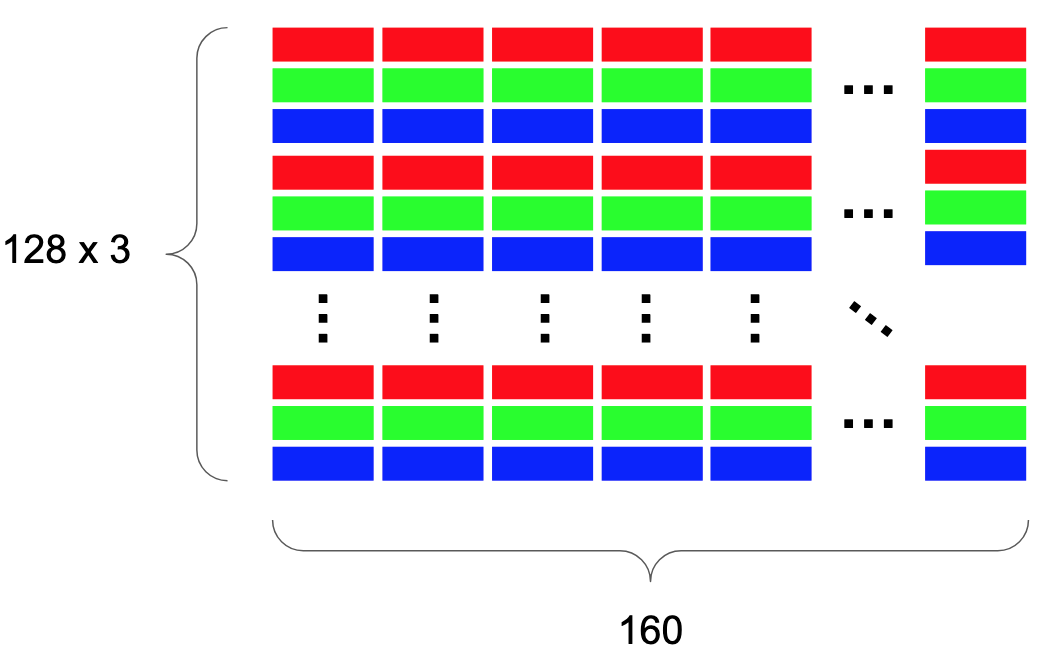}
	\caption{Pixel layout of the ST7735R component~\cite{adafruitlcd}: red, green, blue color filter arrangement.}
	\label{fig:pixel_layout}
\end{figure}

\begin{figure*}[t!]
    \centering
    \begin{subfigure}{0.23\linewidth}
		\centering
\includegraphics[width=0.99\linewidth]{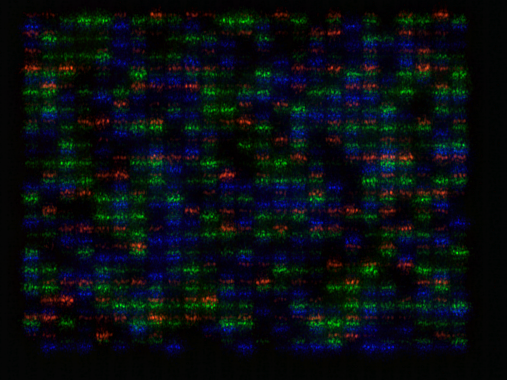} 
		\caption{Measured PSF with a white LED at \SI{30}{\centi\meter}.}
		\label{fig:digicam_celeba_meas}
	\end{subfigure}
 \begin{subfigure}{0.23\linewidth}
		\centering
\includegraphics[width=0.99\linewidth]{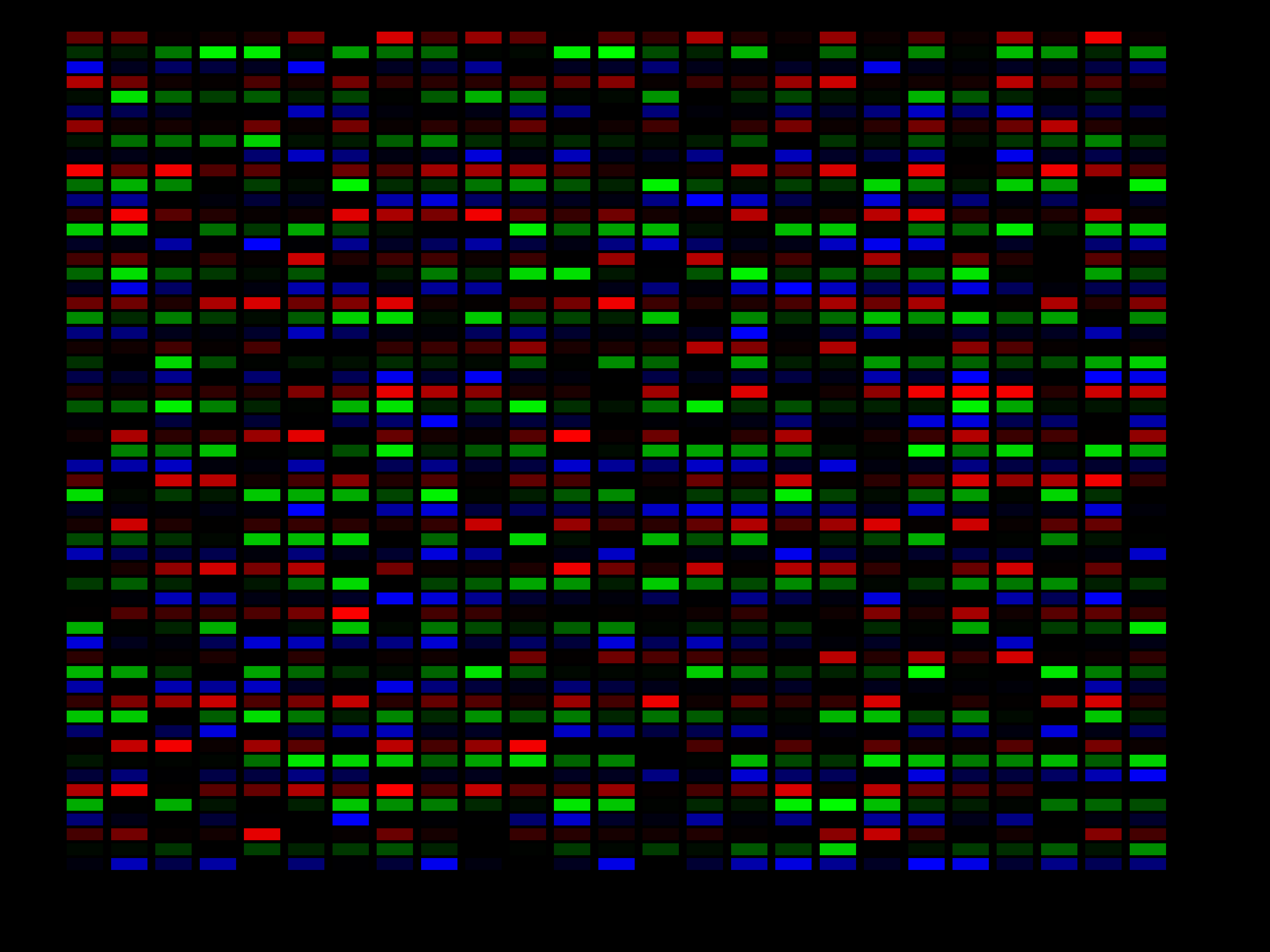} 
		\caption{Simulated PSF without wave propagation.}
		\label{fig:digicam_celeba_nowave}
	\end{subfigure}
 \begin{subfigure}{0.23\linewidth}
		\centering
\includegraphics[width=0.99\linewidth]{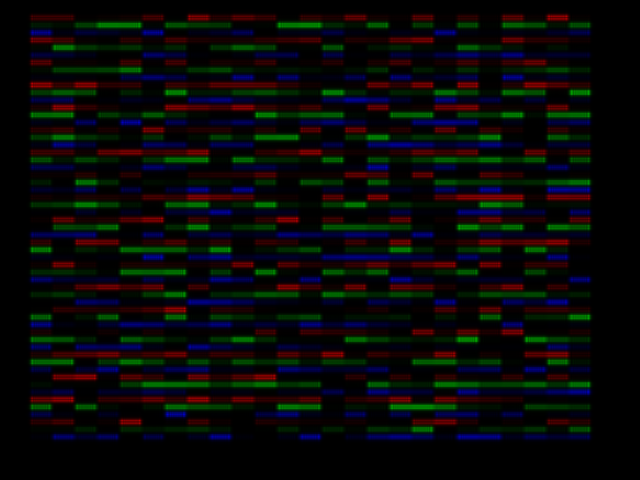} 
		\caption{Simulated PSF without deadspace.}
		\label{fig:digicam_celeba_wave_nodead}
	\end{subfigure}
 \begin{subfigure}{0.23\linewidth}
		\centering
\includegraphics[width=0.99\linewidth]{figs/fig3_digicam_celeba.png} 
		\caption{Simulated PSF with wave propagation and deadspace.}
		\label{fig:digicam_celeba_wave}
	\end{subfigure}
  \caption{Comparing measured and simulation point spread functions (PSFs) of \textit{DigiCam}.}
	\label{fig:compared_psfs_crop_app}
\end{figure*}

\noindent The LCD component used for \textit{DigiCam} has an interleaved pattern of red, blue, and green sub-pixels, 
as shown in \cref{fig:pixel_layout}.
A programmable mask can be modeled as a superposition of $K$ apertures for each adjustable pixel in $ \bm{r} \in \mathbb{R}^2 $:
\begin{align}
	\label{eq:mask_gen_main}
	m(\bm{r}) = \sum_{k=1}^K w_{k} \hspace{0.08cm} a(\bm{r} - \bm{r}_k),
\end{align}
where the complex-valued weights $\{w_{k}\}_{k=1}^K$ satisfy $|w_{k}| \leq 1$, the coordinates $ \{(\bm{r}_k)\}_{k=1}^{K} $ are the centers of each pixel, and the aperture function $a(\cdot)$ of each pixel is assumed to be identical. 
For a mask with a color filter, there is an additional dependence on the wavelength $\lambda$:
\begin{align}
	\label{eq:mask_gen_main_wv}
	m(\bm{r}; \lambda) &= \sum_{c\in\mathcal{C}} \gamma_c(\lambda)  \, \sum_{k_c \in K_c}
	w_{k_c}
	a(\bm{r} -\bm{r}_{k_c}),
\end{align}
where $\gamma_c$ is the wavelength sensitivity function for a specific color filter $c$, and $K_c$ is the set of pixels corresponding to $c$. 
\cref{eq:mask_gen_main_wv} accounts for the pixel pitch and deadspace of the mask by setting the appropriate centers $ \{(\bm{r}_{k_c})\}_{k_c=1}^{K_c} $.
An alternative approach to account for pixel pitch is to modify the wave propagation model to include higher-order diffraction and attenuation~\cite{Gopakumar:21}, but this approach does not account for the deadspace.

For our component~\cite{adafruitlcd}, the pixel value weights $w_{k_c}$ are real-valued, as the LCD only modulates amplitude.
Moreover, we do not have the ground-truth color functions $\gamma_c$, but since our LCD and sensor both have color filters $ c\in\{R,G,B\} $, we compute the mask function as:
\begin{align}
	\label{eq:mask_simple_app}
	m(\bm{r}; \lambda=\lambda_c) &= \sum_{k_c \in K_c}
	w_{k_c}
	a(\bm{r} -\bm{r}_{k_c}), \quad c\in\{R,G,B\},
\end{align}
with a narrowband around the RGB wavelengths. 
Furthermore, each aperture function $a(\cdot)$ is modeled as a rectangle of size $\SI{0.06}{\milli\meter}\times\SI{0.18}{\milli\meter}$ (the dimensions of each sub-pixel).

\section{Comparison Between Simulated and Measured Point Spread Functions}
\label{app:compare_psf}

\newcommand{\figsizecelebadmm}{0.115}
\begin{figure*}[t!]
	\centering
	\renewcommand{\arraystretch}{1} 
	\setlength{\tabcolsep}{0.2em} 
	\begin{tabular}{cccccccc}
    \multicolumn{2}{c}{Measured PSF (\cref{fig:digicam_celeba_meas})} 
    &
    \multicolumn{2}{c}{\makecell{Simulated PSF without\\wave propagation (\cref{fig:digicam_celeba_nowave})}}
    &
    \multicolumn{2}{c}{\makecell{Simulated PSF without\\deadspace (\cref{fig:digicam_celeba_wave_nodead})}}
    &
 \multicolumn{2}{c}{\makecell{Simulated PSF with\\wave propagation (\cref{fig:digicam_celeba_wave})}}
    \\ 
    \cmidrule(r){1-2} \cmidrule(r){3-4} \cmidrule(r){5-6} \cmidrule(r){7-8} 
    \includegraphics[width=\figsizecelebadmm\linewidth,valign=m]{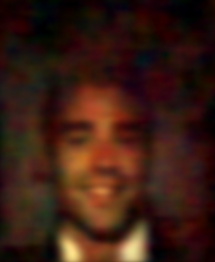}
		&
        \includegraphics[width=\figsizecelebadmm\linewidth,valign=m]{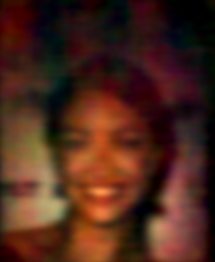} &
        \includegraphics[width=\figsizecelebadmm\linewidth,valign=m]{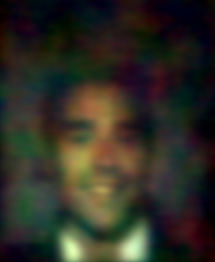}
        &
        \includegraphics[width=\figsizecelebadmm\linewidth,valign=m]{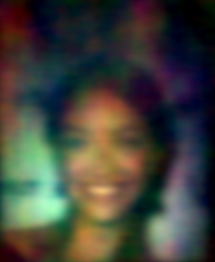}
        &

        \includegraphics[width=\figsizecelebadmm\linewidth,valign=m]{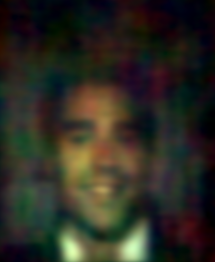}
        &
        \includegraphics[width=\figsizecelebadmm\linewidth,valign=m]{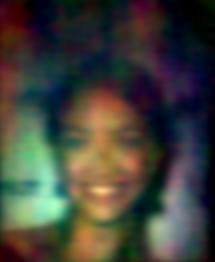}
        &
        
\includegraphics[width=\figsizecelebadmm\linewidth,valign=m]{figs/benchmark_digicam_celeba/ADMM/100/4.png}
		&  \includegraphics[width=\figsizecelebadmm\linewidth,valign=m]{figs/benchmark_digicam_celeba/ADMM/100/9.png}
		\\
	\end{tabular}
	\caption{\textit{ADMM100} reconstructions of measured data with simulated and measured PSFs of \textit{DigiCam}. Ground-truth data can be seen in \cref{fig:celeba_ground_truth}.}
	\label{fig:sim_vs_meas_recon}
\end{figure*}

\begin{figure}[t!]
		\centering
\includegraphics[width=0.25\linewidth]{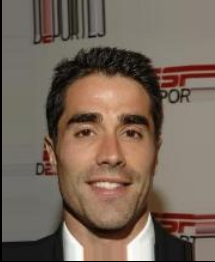}
\includegraphics[width=0.25\linewidth]{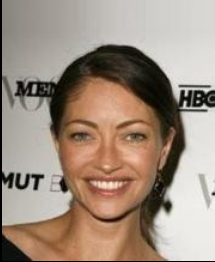}
	\caption{Ground-truth CelebA data~\cite{digicam_celeba}.}
	\label{fig:celeba_ground_truth}
\end{figure}

\begin{table}[!t]
	\renewcommand{\arraystretch}{1.3}
	\caption{Average image quality metrics to compare using simulated PSF variants against a measured PSF for image recovery on the test set of \textit{DigiCam-CelebA} with 100 iterations of ADMM.}
	\label{tab:sim_v_meas}
	\centering
	\begin{tabular}{c|c|c|c}
		\hline
		 & PSNR $\uparrow$ & SSIM $\uparrow$ & LPIPS $\downarrow$ \\
		\hline\hline
		Measured PSF (\cref{fig:digicam_celeba_meas}) &   9.38 & 0.294 & \textbf{0.695}  \\
  \hline
  \makecell{Simulated PSF without\\wave propagation (\cref{fig:digicam_celeba_nowave})} & 10.1 & 0.352 & 0.737 \\
  \hline
  \makecell{Simulated PSF without\\deadspace(\cref{fig:digicam_celeba_wave_nodead})} & 10.0 & 0.345 & 0.730 \\
  \hline
  \makecell{Simulated PSF with\\wave propagation  and\\deadspace(\cref{fig:digicam_celeba_wave})} & \textbf{10.2} & \textbf{0.356} & 0.739  \\
  \hline
	\end{tabular}
\end{table}

\noindent \cref{fig:compared_psfs_crop_app} compares a PSF measured with a white LED (\cref{fig:digicam_celeba_meas}) with PSFs simulated with different approaches:
\begin{itemize}
    \item \textit{Without wave propagation} (\cref{fig:digicam_celeba_nowave}): Simply using \cref{eq:mask_simple_app} as the PSF.
    \item \textit{Without deadspace} (\cref{fig:digicam_celeba_nowave}): 
    The mask is modeled as a single aperture with dimensions $(Mp, Np)$ where $(M,N)$ is the dimensions of the programmable array and $p$ is the pixel pitch.
    \item \textit{With wave propagation and deadspace} (\cref{fig:digicam_celeba_wave}): When forming the mask function with \cref{eq:mask_simple_app}, the pixel centers $ \{(\bm{r}_k)\}_{k=1}^{K} $ account for the aperture of each sub-pixel and the pixel pitch, such that there is a \textit{deadspace} between pixels that is occluding, \ie due to the mask's circuitry.
\end{itemize}
We can observe that incorporating deadspace (\ie \cref{fig:digicam_celeba_nowave,fig:digicam_celeba_wave}) more closely resembles the structure of the measured PSF in \cref{fig:digicam_celeba_meas}.

For PSF simulation, we are ultimately interested in how well a PSF describes the forward model to obtain a high-quality reconstruction, 
\ie a PSF that faithfully describes the forward model in \cref{eq:opt_gen_app}.
To this end, we compare each PSF when used to reconstruct images from the \textit{DigiCam-CelebA} dataset~\cite{digicam_celeba} with 100 iterations of ADMM.
\cref{tab:sim_v_meas} compares the average image quality metrics on the test set (3900 files).
All the simulated PSFs yield similar image quality metrics,
while using the measured PSF is worse with regards to PSNR/SSIM but better in LPIPS.

\cref{fig:sim_vs_meas_recon} shows example outputs. All simulated PSFs yield reconstructions that look very similar.
While the measured PSF yields a reddish-reconstruction,
the overall quality is very similar to those of the simulated PSFs.
In both cases, the reddish/greenish tint can be removed with white-balancing (which can also be learned by the post-processor). 

For the \textit{DigiCam-Multi} dataset~\cite{digicam_multi} that consists of 100 different masks patterns,
we rely on simulated PSFs to avoid the hassle of measuring 100 PSFs.
To this end, we use wave propagation and deadspace as it is more realistic.


\section{Intermediate Outputs}
\label{app:intermediate}

\noindent\cref{fig:intermediate_diffusercam} shows intermediate outputs for various models trained on the \textit{DiffuserCam} dataset~\cite{Monakhova:19}.
When only using a pre-processor ($\text{Pre}_8$+LeADMM5),
we can observe more consistent coloring but the final outputs lack the perceptual enhancements that a post-processor can perform after the camera inversion.
Using both a pre-processor and a post-processor achieves the best results (Table III in main paper), 
but the intermediate outputs (\eg camera inversion output) may be less interpretable (last two rows of \cref{fig:intermediate_diffusercam}).

To improve the interpretability of intermediate outputs,
an auxiliary loss can be used from the camera inversion output during training~\cite{Perron2023}:
\begin{align}
        \mathscr{L}_{\text{res}}\left(\bm{x},\bm{\hat{x}},\bm{\hat{x}}_{\text{inv}}\right) = \mathscr{L}\left(\bm{x},\bm{\hat{x}}\right) + \alpha \hspace{0.2em} \mathscr{L}\left(\bm{x},\bm{\hat{x}}_{\text{inv}}\right).
\end{align}
where the loss $\mathscr{L}\left(\bm{x},\bm{\hat{x}}\right)$ can be a combined MSE and LPIPS loss (\ie \cref{eq:loss_mse_lpips}), $\bm{\hat{x}}$ is the output of the post-processor, 
$\bm{\hat{x}}_{\text{inv}}$ is the output of camera inversion, and $\alpha$ weights the amount of auxiliary loss.
Higher values of $\alpha$ can lead to more consistent coloring at the camera inversion output,
but to slightly worse image quality metrics~\cite{Perron2023}.

\cref{fig:intermediate_diffusercam_noise,fig:intermediate_diffusercam_noisy_psf} shows intermediate outputs for our robustness experiments in the main paper:
simulated shot noise in the measurement (\cref{sec:shot_noise_exp}) and mismatch in the PSF (\cref{sec:mismatch_exp}).
While the intermediate outputs of approaches that use a pre-processor are significantly different with respect to coloring,
the robustness to measurement noise and model mismatch is much better (see Tables IV and V in the main paper).
As mentioned above,
using an auxiliary loss can lead to more consistent coloring at the camera inversion output but at the expense of slightly worse image quality metrics.

\cref{fig:intermediate_tapecam,fig:intermediate_tapecam} show intermediate outputs for \textit{TapeCam} and \textit{DigiCam}.
We observe similar behavior as with \textit{DiffuserCam}:
significant discoloring at the camera inversion output when both pre- and post-processor are used.

\newcommand{\figsizeinter}{0.19}
\newcommand{\newlineinter}{32pt}
\begin{figure*}[t!]
\centering
	\begingroup
	\renewcommand{\arraystretch}{1} 
	\setlength{\tabcolsep}{0.08em} 
	\begin{tabular}{c cccc}
		  & Lensless measurement & PSF for inversion & Camera inversion output & Final output \\
    
\makecell{LeADMM5\\+$\text{Post}_8$~\cite{Monakhova:19}} 
  & \includegraphics[width=\figsizeinter\linewidth,valign=m]{figs/benchmark_diffusercam/LENSLESS/1.png}
  & \includegraphics[width=\figsizeinter\linewidth,valign=m]{figs/fig3_diffusercam_psf.png}
   & \includegraphics[width=\figsizeinter\linewidth,valign=m]{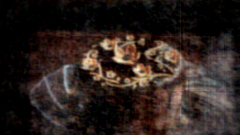}
  & \includegraphics[width=\figsizeinter\linewidth,valign=m]{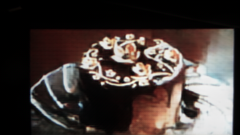}
\\[\newlineinter]

& Pre-processor output &  &  &  \\

\makecell{$\text{Pre}_8$+LeADMM5} 
  & \includegraphics[width=\figsizeinter\linewidth,valign=m]{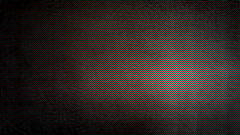}  
  & 
\includegraphics[width=\figsizeinter\linewidth,valign=m]{figs/fig3_diffusercam_psf.png}
  & 
\includegraphics[width=\figsizeinter\linewidth,valign=m]{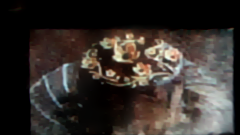} 
  & 
\\[\newlineinter]

\makecell{$\text{Pre}_4$+LeADMM5\\+$\text{Post}_4$} 
  & \includegraphics[width=\figsizeinter\linewidth,valign=m]{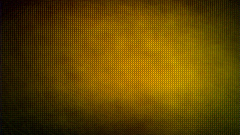}  
  & 
\includegraphics[width=\figsizeinter\linewidth,valign=m]{figs/fig3_diffusercam_psf.png}
  & 
\includegraphics[width=\figsizeinter\linewidth,valign=m]{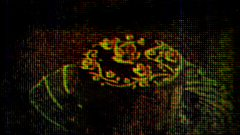} 
  & \includegraphics[width=\figsizeinter\linewidth,valign=m]{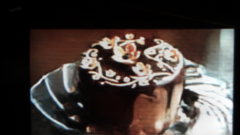} 
\\[\newlineinter]
& & Corrected PSF &  &  \\
\makecell{$\text{Pre}_4$+LeADMM5\\+$\text{Post}_4$ (PSF correction)} 
  & \includegraphics[width=\figsizeinter\linewidth,valign=m]{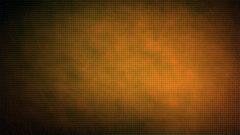} 
  & 
\includegraphics[width=\figsizeinter\linewidth,valign=m]{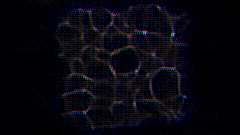}
  & 
\includegraphics[width=\figsizeinter\linewidth,valign=m]{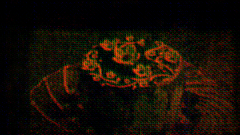} 
  & 
  \includegraphics[width=\figsizeinter\linewidth,valign=m]{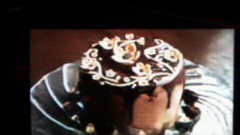}
\\
	\end{tabular}
	\endgroup
	\caption{Intermediate outputs for \textit{DiffuserCam}.}
  \label{fig:intermediate_diffusercam}
\end{figure*}

\newcommand{\figsizeinternoisy}{0.135}
\newcommand{\newlineinternoisy}{15pt}
\begin{figure*}[t!]
\centering
	\begingroup
	\renewcommand{\arraystretch}{1} 
	\setlength{\tabcolsep}{0.08em} 
	\begin{tabular}{c cccc}
		  & Lensless measurement & Camera inversion output & Final output \\
    
\makecell{LeADMM5\\+$\text{Post}_8$~\cite{Monakhova:19}} 
  & 
  \includegraphics[width=\figsizeinternoisy\linewidth,valign=m]{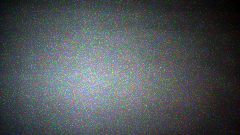}
\includegraphics[width=\figsizeinternoisy\linewidth,valign=m]{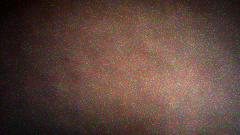}
   & 
\includegraphics[width=\figsizeinternoisy\linewidth,valign=m]{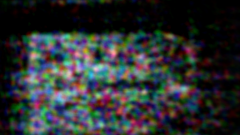}
\includegraphics[width=\figsizeinternoisy\linewidth,valign=m]{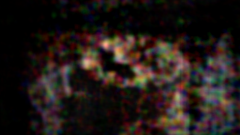}
  & 
\includegraphics[width=\figsizeinternoisy\linewidth,valign=m]{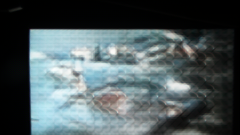}
\includegraphics[width=\figsizeinternoisy\linewidth,valign=m]{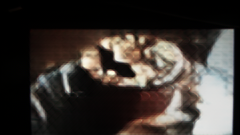}
\\[\newlineinternoisy]

& Pre-processor output &  &  &  \\

\makecell{$\text{Pre}_4$\\+LeADMM5\\+$\text{Post}_4$} 
  & 
\includegraphics[width=\figsizeinternoisy\linewidth,valign=m]{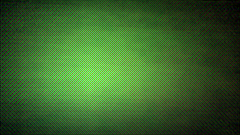} \includegraphics[width=\figsizeinternoisy\linewidth,valign=m]{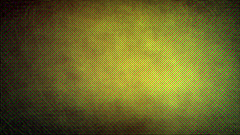}  
  & 
\includegraphics[width=\figsizeinternoisy\linewidth,valign=m]{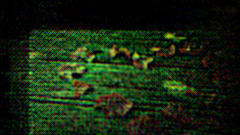} 
\includegraphics[width=\figsizeinternoisy\linewidth,valign=m]{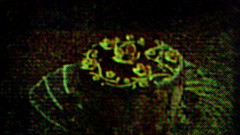} 
  &
\includegraphics[width=\figsizeinternoisy\linewidth,valign=m]{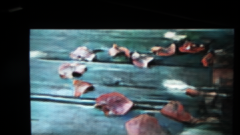} 
\includegraphics[width=\figsizeinternoisy\linewidth,valign=m]{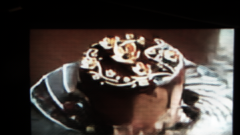} 
\\
	\end{tabular}
	\endgroup
	\caption{Intermediate outputs for \textit{DiffuserCam} in the presence of digitally-added Poisson noise with an SNR of \SI{10}{\decibel}.}
  \label{fig:intermediate_diffusercam_noise}
\end{figure*}

\begin{figure*}[t!]
\centering
	\begingroup
	\renewcommand{\arraystretch}{1} 
	\setlength{\tabcolsep}{0.08em} 
	\begin{tabular}{c cccc}
		  & Lensless measurement & Camera inversion output & Final output \\
    
\makecell{LeADMM5\\+$\text{Post}_8$~\cite{Monakhova:19}} 
  & 
  \includegraphics[width=\figsizeinternoisy\linewidth,valign=m]{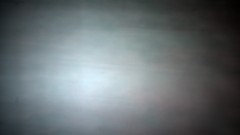}
\includegraphics[width=\figsizeinternoisy\linewidth,valign=m]{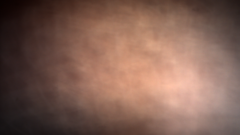}
   & 
\includegraphics[width=\figsizeinternoisy\linewidth,valign=m]{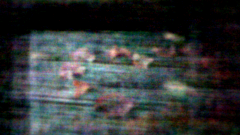}
\includegraphics[width=\figsizeinternoisy\linewidth,valign=m]{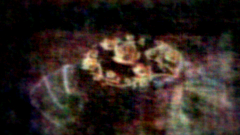}
  & 
\includegraphics[width=\figsizeinternoisy\linewidth,valign=m]{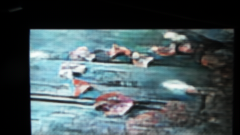}
\includegraphics[width=\figsizeinternoisy\linewidth,valign=m]{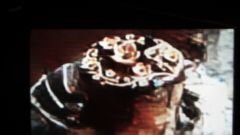}
\\[\newlineinternoisy]

& Pre-processor output &  &  &  \\

\makecell{$\text{Pre}_4$\\+LeADMM5\\+$\text{Post}_4$} 
  & 
  \includegraphics[width=\figsizeinternoisy\linewidth,valign=m]{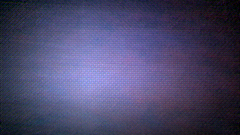}
\includegraphics[width=\figsizeinternoisy\linewidth,valign=m]{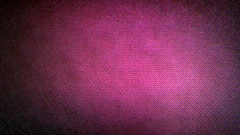}
   & 
\includegraphics[width=\figsizeinternoisy\linewidth,valign=m]{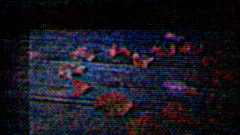}
\includegraphics[width=\figsizeinternoisy\linewidth,valign=m]{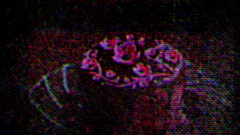}
  & 
\includegraphics[width=\figsizeinternoisy\linewidth,valign=m]{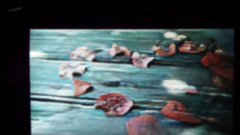}
\includegraphics[width=\figsizeinternoisy\linewidth,valign=m]{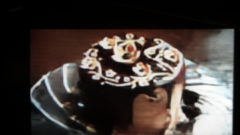}
\\[\newlineinternoisy]

\makecell{$\text{Pre}_4$+LeADMM5\\+$\text{Post}_4$ (PSF corr.)} 
  & 
\includegraphics[width=\figsizeinternoisy\linewidth,valign=m]{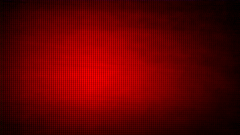}
\includegraphics[width=\figsizeinternoisy\linewidth,valign=m]{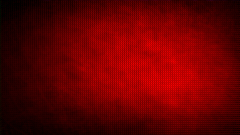}
  & 
\includegraphics[width=\figsizeinternoisy\linewidth,valign=m]{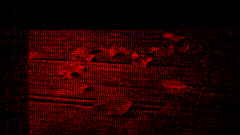}
\includegraphics[width=\figsizeinternoisy\linewidth,valign=m]{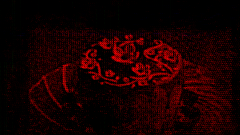}
  &
\includegraphics[width=\figsizeinternoisy\linewidth,valign=m]{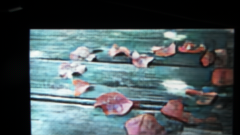}
\includegraphics[width=\figsizeinternoisy\linewidth,valign=m]{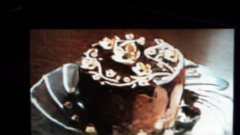}
\\
	\end{tabular}
	\endgroup
	\caption{Intermediate outputs for \textit{DiffuserCam} when model mismatch is added to the PSF (Gaussian noise with an SNR of \SI{0}{\decibel}).}
\label{fig:intermediate_diffusercam_noisy_psf}
\end{figure*}

\newcommand{\figsizeinterrpi}{0.17}
\begin{figure*}[t!]
\centering
	\begingroup
	\renewcommand{\arraystretch}{1} 
	\setlength{\tabcolsep}{0.08em} 
	\begin{tabular}{c cccc}
	 & Lensless measurement &  PSF for inversion & Camera inversion output & Final output (cropped) \\

    \makecell{LeADMM5\\+$\text{Post}_8$~\cite{Monakhova:19}} 
  & 
\includegraphics[width=\figsizeinterrpi\linewidth,valign=m]{figs/benchmark_tapecam/LENSLESS/2.png}

  & 
  \includegraphics[width=\figsizeinterrpi\linewidth,valign=m]{figs/fig3_tapecam_psf.png}
   & 
   \includegraphics[width=\figsizeinterrpi\linewidth,valign=m]{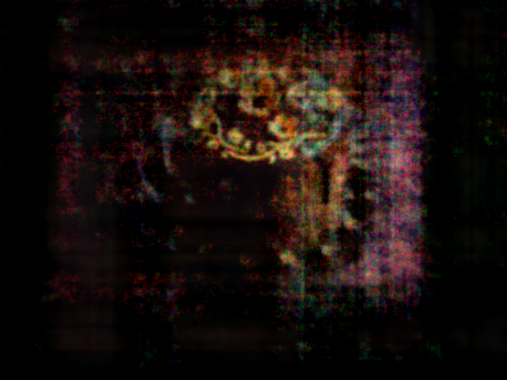}
  & \includegraphics[width=\figsizeinterrpi\linewidth,valign=m]{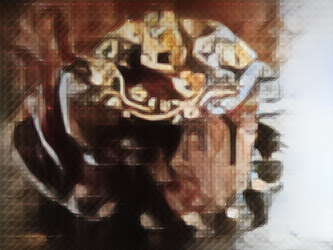}
\\[\newlineinter]

& Pre-processor output & &  &  \\
\makecell{$\text{Pre}_8$+LeADMM5} 
  & \includegraphics[width=\figsizeinterrpi\linewidth,valign=m]{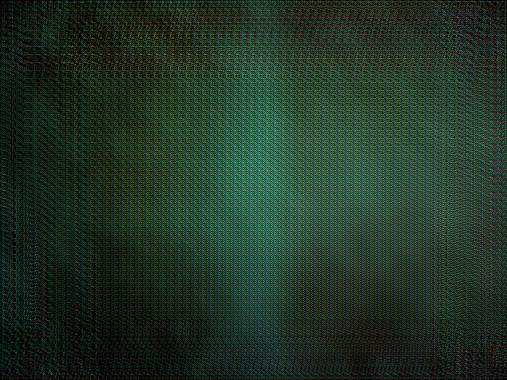}  
  & 
\includegraphics[width=\figsizeinterrpi\linewidth,valign=m]{figs/fig3_tapecam_psf.png}
  & 
\includegraphics[width=\figsizeinterrpi\linewidth,valign=m]{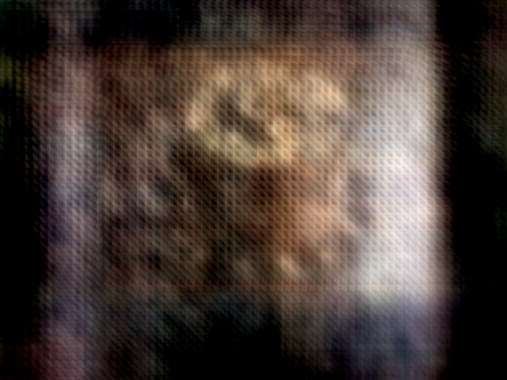} 
  & 
\\[\newlineinter]

\makecell{$\text{Pre}_4$+LeADMM5\\+$\text{Post}_4$} 
  & 

\includegraphics[width=\figsizeinterrpi\linewidth,valign=m]{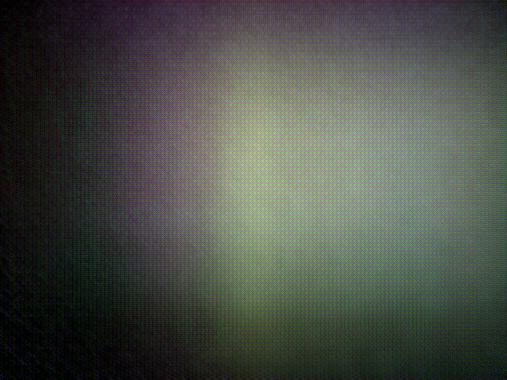}  
  & 
\includegraphics[width=\figsizeinterrpi\linewidth,valign=m]{figs/fig3_tapecam_psf.png}
  & 
\includegraphics[width=\figsizeinterrpi\linewidth,valign=m]{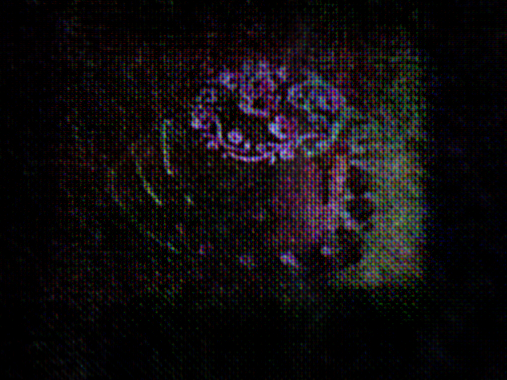}
  & \includegraphics[width=\figsizeinterrpi\linewidth,valign=m]{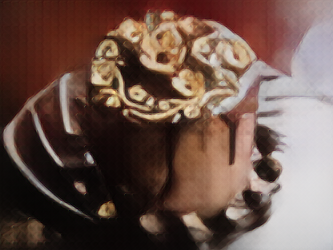}
\\[\newlineinter]
& & Corrected PSF &  &  \\
\makecell{$\text{Pre}_4$+LeADMM5\\+$\text{Post}_4$ (PSF correction)} 
  & \includegraphics[width=\figsizeinterrpi\linewidth,valign=m]{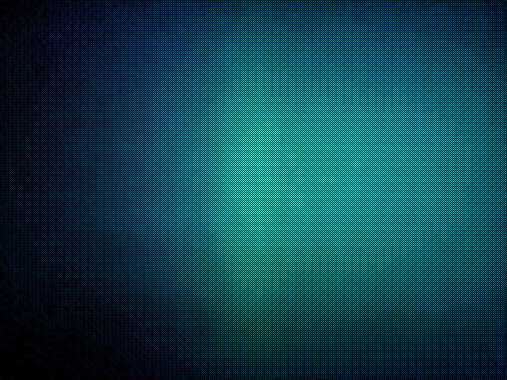} 
  & 
\includegraphics[width=\figsizeinterrpi\linewidth,valign=m]{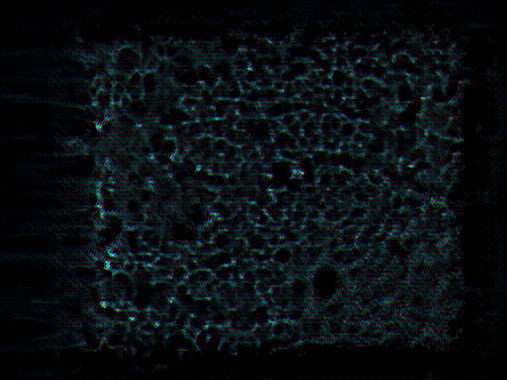}
  & 
\includegraphics[width=\figsizeinterrpi\linewidth,valign=m]{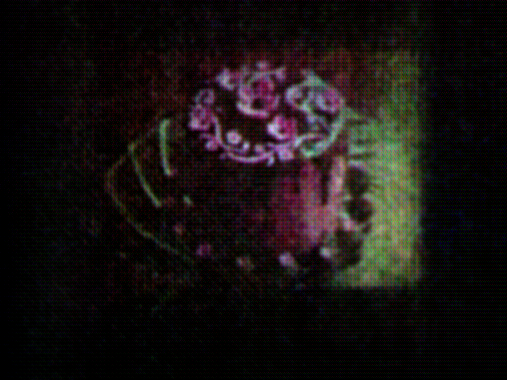} 
  & 
  \includegraphics[width=\figsizeinterrpi\linewidth,valign=m]{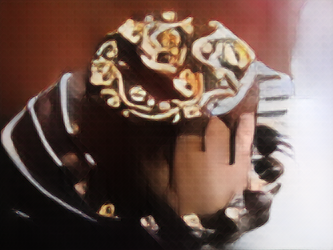}
\\
	\end{tabular}
	\endgroup
	\caption{Intermediate outputs for \textit{TapeCam}.}
  \label{fig:intermediate_tapecam}
\end{figure*}

\begin{figure*}[t!]
\centering
	\begingroup
	\renewcommand{\arraystretch}{1} 
	\setlength{\tabcolsep}{0.08em} 
	\begin{tabular}{c cccc}
		  & Lensless measurement & PSF for inversion & Camera inversion output & Final output (cropped) \\
    
\makecell{LeADMM5\\+$\text{Post}_8$~\cite{Monakhova:19}} 
  & \includegraphics[width=\figsizeinterrpi\linewidth,valign=m]{figs/benchmark_digicam_mirflickr/LENSLESS/2.png}
  & \includegraphics[width=\figsizeinterrpi\linewidth,valign=m]{figs/fig3_DigiCam-Mirflickr-SingleMask-25K_psf.png}
   & \includegraphics[width=\figsizeinterrpi\linewidth,valign=m]{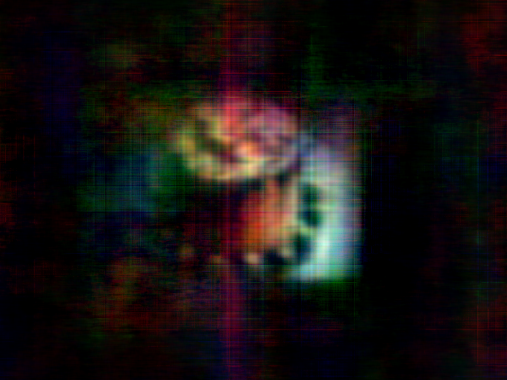}
  & \includegraphics[width=\figsizeinterrpi\linewidth,valign=m]{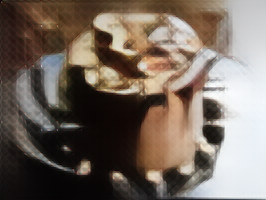}
\\[\newlineinter]

& Pre-processor output &  &  &  \\
\makecell{$\text{Pre}_8$+LeADMM5} 
  & \includegraphics[width=\figsizeinterrpi\linewidth,valign=m]{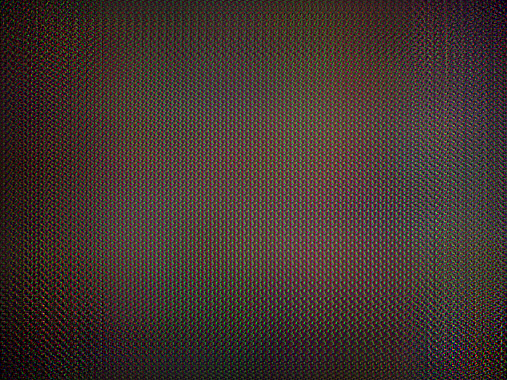}  
  & 
\includegraphics[width=\figsizeinterrpi\linewidth,valign=m]{figs/fig3_DigiCam-Mirflickr-SingleMask-25K_psf.png}
  & 
\includegraphics[width=\figsizeinterrpi\linewidth,valign=m]{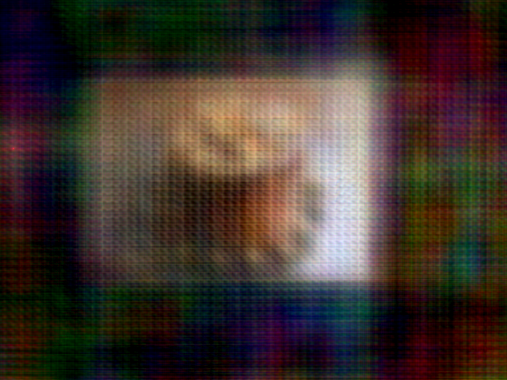} 
  & 
\\[\newlineinter]

\makecell{$\text{Pre}_4$+LeADMM5\\+$\text{Post}_4$} 
  & \includegraphics[width=\figsizeinterrpi\linewidth,valign=m]{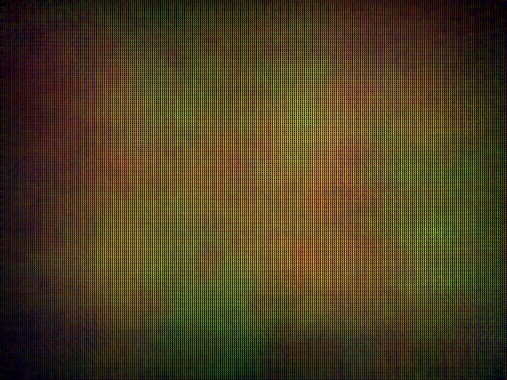}  
  & 
\includegraphics[width=\figsizeinterrpi\linewidth,valign=m]{figs/fig3_DigiCam-Mirflickr-SingleMask-25K_psf.png}
  & 
\includegraphics[width=\figsizeinterrpi\linewidth,valign=m]{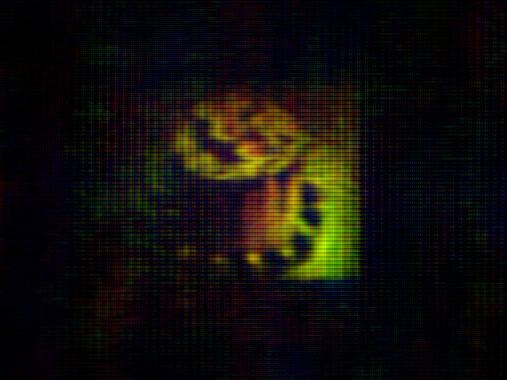}
  & \includegraphics[width=\figsizeinterrpi\linewidth,valign=m]{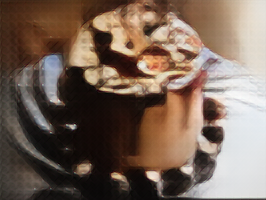}
\\[\newlineinter]
& & Corrected PSF &  &  \\
\makecell{$\text{Pre}_4$+LeADMM5\\+$\text{Post}_4$ (PSF correction)} 
  & \includegraphics[width=\figsizeinterrpi\linewidth,valign=m]{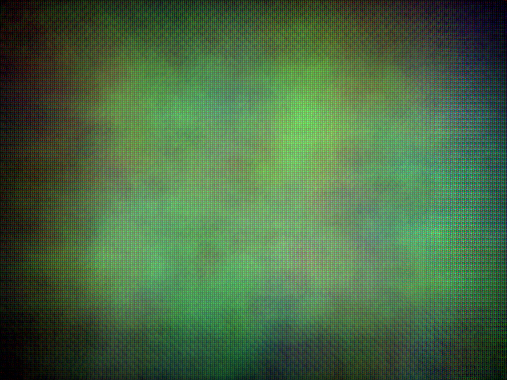} 
  & 
\includegraphics[width=\figsizeinterrpi\linewidth,valign=m]{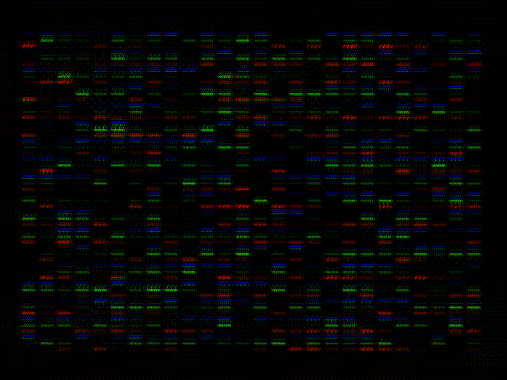}
  & 
\includegraphics[width=\figsizeinterrpi\linewidth,valign=m]{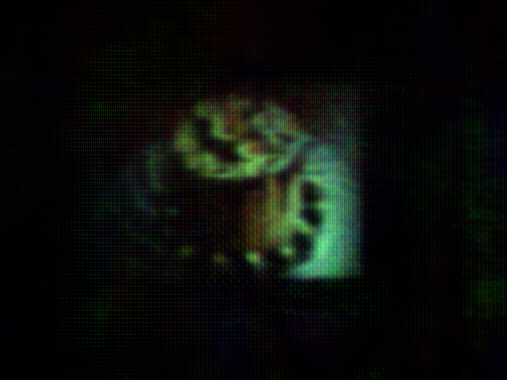} 
  & 
  \includegraphics[width=\figsizeinterrpi\linewidth,valign=m]{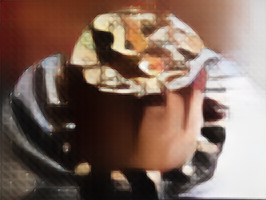}
\\
	\end{tabular}
	\endgroup
	\caption{Intermediate outputs for \textit{DigiCam}.}
  \label{fig:intermediate_digicam}
\end{figure*}

\section{Benchmark Generalizability to PSF Changes with PSF Correction Models}
\label{app:gen_benchmark_psf_corr}

\begin{figure*}[t!]
\centering
	\renewcommand{\arraystretch}{1} 
	\setlength{\tabcolsep}{0.1em} 
	\begin{tabular}{c|cc|cc|cc||cc|cc}
    \makecell{\textit{Train set} $\rightarrow$\\\textit{Test set} $\downarrow$}
    &   
    \multicolumn{2}{c|}{DiffuserCam}
    & \multicolumn{2}{c|}{TapeCam}
    & \multicolumn{2}{c||}{DigiCam-Single}
    & \multicolumn{2}{c|}{\makecell{ADMM100\\(no training)}}
    & 
    \multicolumn{2}{c}{Ground-truth}
    \\
\hline 
\makecell{DiffuserCam}
&\includegraphics[width=\figsizegentrans\linewidth,valign=m]{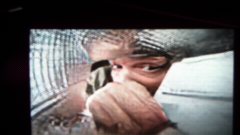}
&\includegraphics[width=\figsizegentrans\linewidth,valign=m]{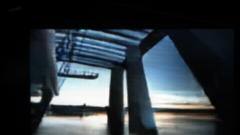}
&\includegraphics[width=\figsizegentrans\linewidth,valign=m]{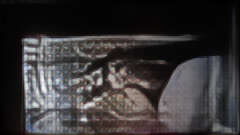}
&\includegraphics[width=\figsizegentrans\linewidth,valign=m]{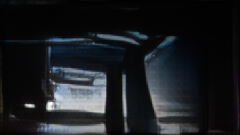}
&\includegraphics[width=\figsizegentrans\linewidth,valign=m]{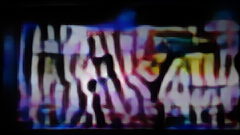}
&\includegraphics[width=\figsizegentrans\linewidth,valign=m]{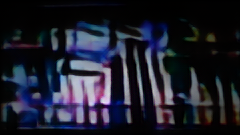}
&\includegraphics[width=\figsizegentrans\linewidth,valign=m]{figs/benchmark_diffusercam/ADMM/100/3.png}
 
&\includegraphics[width=\figsizegentrans\linewidth,valign=m]{figs/benchmark_diffusercam/ADMM/100/4.png}

&\includegraphics[width=\figsizegentrans\linewidth,valign=m]{figs/benchmark_diffusercam/GROUND_TRUTH/3.png}
&\includegraphics[width=\figsizegentrans\linewidth,valign=m]{figs/benchmark_diffusercam/GROUND_TRUTH/4.png}
    \\[6pt]
\hline 
\makecell{TapeCam}
&\includegraphics[width=\figsizegentrans\linewidth,valign=m]{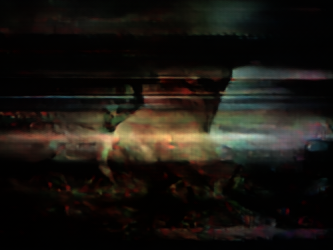}
&\includegraphics[width=\figsizegentrans\linewidth,valign=m]{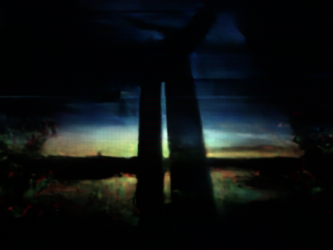}
&\includegraphics[width=\figsizegentrans\linewidth,valign=m]{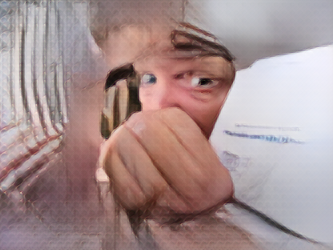}
&\includegraphics[width=\figsizegentrans\linewidth,valign=m]{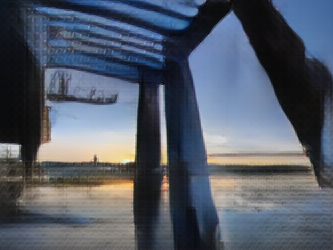}
&\includegraphics[width=\figsizegentrans\linewidth,valign=m]{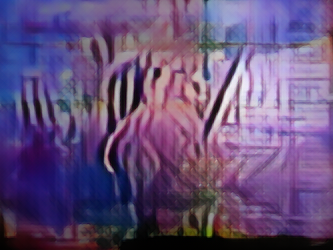}
&\includegraphics[width=\figsizegentrans\linewidth,valign=m]{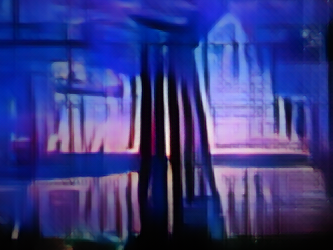}
&\includegraphics[width=\figsizegentrans\linewidth,valign=m]{figs/benchmark_tapecam/ADMM/100/4.png}
&\includegraphics[width=\figsizegentrans\linewidth,valign=m]{figs/benchmark_tapecam/ADMM/100/5.png}

&\includegraphics[width=\figsizegentrans\linewidth,valign=m]{figs/benchmark_tapecam/GROUND_TRUTH/4.png}
&\includegraphics[width=\figsizegentrans\linewidth,valign=m]{figs/benchmark_tapecam/GROUND_TRUTH/5.png}
\\[10pt]
\hline
\makecell{DigiCam\\-Single}
&\includegraphics[width=\figsizegentrans\linewidth,valign=m]{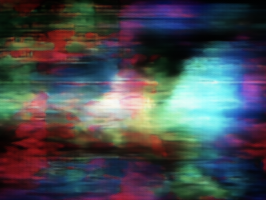}
&\includegraphics[width=\figsizegentrans\linewidth,valign=m]{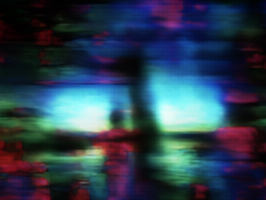}
&\includegraphics[width=\figsizegentrans\linewidth,valign=m]{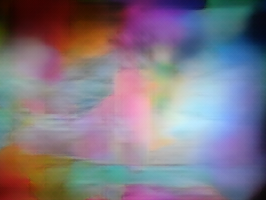}
&\includegraphics[width=\figsizegentrans\linewidth,valign=m]{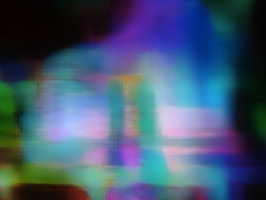}
&\includegraphics[width=\figsizegentrans\linewidth,valign=m]{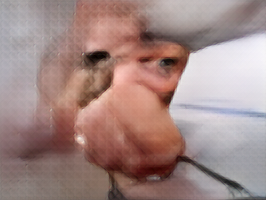}
&\includegraphics[width=\figsizegentrans\linewidth,valign=m]{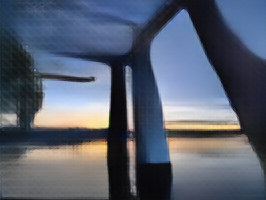}
&\includegraphics[width=\figsizegentrans\linewidth,valign=m]{figs/benchmark_digicam_mirflickr/ADMM/100/4.png}
&\includegraphics[width=\figsizegentrans\linewidth,valign=m]{figs/benchmark_digicam_mirflickr/ADMM/100/5.png}

&\includegraphics[width=\figsizegentrans\linewidth,valign=m]{figs/benchmark_digicam_mirflickr/GROUND_TRUTH/4.png}
&\includegraphics[width=\figsizegentrans\linewidth,valign=m]{figs/benchmark_digicam_mirflickr/GROUND_TRUTH/5.png}
\\[10pt]
\hline
\hline
\makecell{DigiCam\\-Multi}
&\includegraphics[width=\figsizegentrans\linewidth,valign=m]{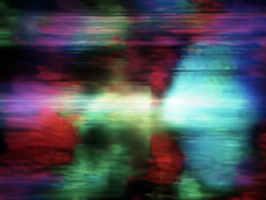}
&\includegraphics[width=\figsizegentrans\linewidth,valign=m]{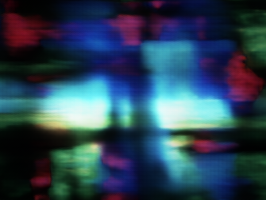}
&\includegraphics[width=\figsizegentrans\linewidth,valign=m]{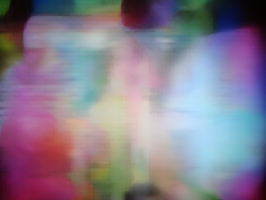}
&\includegraphics[width=\figsizegentrans\linewidth,valign=m]{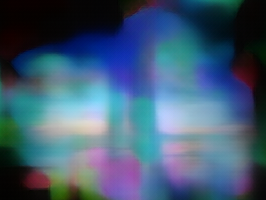}
&\includegraphics[width=\figsizegentrans\linewidth,valign=m]{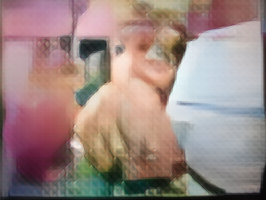}
&\includegraphics[width=\figsizegentrans\linewidth,valign=m]{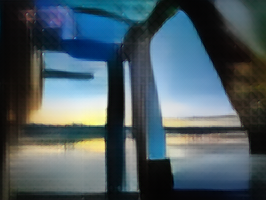}
&\includegraphics[width=\figsizegentrans\linewidth,valign=m]{figs/benchmark_digicam_multimask/ADMM/100/4.png}
&\includegraphics[width=\figsizegentrans\linewidth,valign=m]{figs/benchmark_digicam_multimask/ADMM/100/5.png}

&\includegraphics[width=\figsizegentrans\linewidth,valign=m]{figs/benchmark_digicam_multimask/GROUND_TRUTH/4.png}
& \hspace{-0.5em}\includegraphics[width=\figsizegentrans\linewidth,valign=m]{figs/benchmark_digicam_multimask/GROUND_TRUTH/5.png}
\\
	\end{tabular}
	\caption{Example outputs of ($\textit{Pre}_{4}$\textit{+LeADMM5+}$\textit{Post}_{4}$) with PSF correction trained on the system/dataset indicated along the columns, and evaluated on the system/dataset indicated along the rows.}
\label{fig:exp2_visual_comparison_psfNN}
\end{figure*}

\noindent\cref{fig:exp2_visual_comparison_psfNN} performs a similar evaluation as that of \cref{fig:exp2_visual_comparison},
namely evaluating a model
trained on measurements from one system on measurements from other systems.
However, in \cref{fig:exp2_visual_comparison_psfNN} 
the evaluated models are ($\textit{Pre}_{4}$\textit{+LeADMM5+}$\textit{Post}_{4}$) \textit{with PSF correction}, 
\ie by inputting the PSF to a DRUNet with (4, 8, 16, 32) feature representation channels (128K parameters) prior to camera inversion.
The pre-processor is slightly decreased to (32, 64, 112, 128) channels (3.9M parameters) to maintain an approximately equivalent number of parameters as the model without PSF correction.
Even with PSF correction, we observe similar behavior as in Fig.~8 of the main paper:
image recovery approaches trained on measurements from a single system fail to generalize to measurements of other systems.

